\documentclass[10pt,english]{article}
\usepackage[]{graphicx,amssymb,amsmath,bm}

\usepackage[colorlinks=true, citecolor=blue]{hyperref}

\title{Motion in classical field theories and the foundations of the self-force problem}

\author{Abraham I. Harte\footnote{Email: \texttt{harte@aei.mpg.de}} \\
        Albert-Einstein-Institut\\
        Max-Planck-Institut f\"{u}r Gravitationsphysik\\
        Am M\"{u}hlenberg 1, 14476 Golm, Germany}

\date{\today}

\newcommand{\rmd}{\mathrm{d}}
\newcommand{\rmD}{\mathrm{D}}

\newcommand{\bx}{\bm{x}}

\newcommand{\bz}{\bm{z}}
\newcommand{\bs}{\bm{S}}
\newcommand{\bp}{\bm{p}}
\newcommand{\bnabla}{\bm{\nabla}}
\newcommand{\fB}{\mathfrak{B}}
\newcommand{\fW}{\mathfrak{W}}

\newcommand{\Lie}{\mathcal{L}}
\newcommand{\phiS}{ \phi_\mathrm{S} }

\begin{document}
\maketitle

\begin{abstract}

This article serves as a pedagogical introduction to the problem of motion in classical field theories. The primary focus is on self-interaction: How does an object's own field affect its motion? General laws governing the self-force and self-torque are derived using simple, non-perturbative arguments. The relevant concepts are developed gradually by considering motion in a series of increasingly complicated theories. Newtonian gravity is discussed first, then Klein-Gordon theory, electromagnetism, and finally general relativity. Linear and angular momenta as well as centers of mass are defined in each of these cases. Multipole expansions for the force and torque are derived to all orders for arbitrarily self-interacting extended objects. These expansions are found to be structurally identical to the laws of motion satisfied by extended test bodies, except that all relevant fields are replaced by effective versions which exclude the self-fields in a particular sense. Regularization methods traditionally associated with self-interacting point particles arise as straightforward perturbative limits of these (more fundamental) results. Additionally, generic mechanisms are discussed which dynamically shift --- i.e., renormalize --- the apparent multipole moments associated with self-interacting extended bodies. Although this is primarily a synthesis of earlier work, several new results and interpretations are included as well.

\end{abstract}

\section{Introduction}

How are charges accelerated by electromagnetic fields? How do masses fall in curved spacetimes? Such questions can be answered in many different ways. Consider, for example, the Newtonian $n$-body problem. This is typically posed as a system of ordinary differential equations which describe the motions of $n$ points in $\mathbb{R}^3$. Besides its position, each point is characterized only by its mass. This is a considerable abstraction from the extended stars or planets whose motions the $n$-body problem is physically intended to describe, and the internal density distributions, velocity fields, and temperatures of each body might be governed by complicated sets of nonlinear partial differential equations. From one point of view, it is the solutions to these continuum equations which represent ``the motion'' of each mass.

This is not, however, the approach which is typically adopted in celestial mechanics. In that context, one instead focuses only on a body's center of mass (and perhaps its spin angular momentum): observables which describe motion ``in the large.'' It is a central result of Newtonian gravity that much of the dynamics of these observables can be understood without detailed knowledge of each body's internal structure. This is why the extended stars and associated partial differential equations of the ``physical $n$-body problem'' can often be modeled as discrete points satisfying a simple set of ordinary differential equations --- an enormous simplification. 

This work is intended as an introduction to techniques which have recently been developed \cite{HarteSyms, HarteScalar, HarteEM, HarteBrokenSyms, HarteGR} to similarly simplify problems of motion in a wide variety of contexts. Is it possible, for example, to describe extended masses in general relativity using appropriately-defined centers of mass? Do these mass centers obey simple laws of motion? Of course, the same questions may also be asked for charged matter coupled to electromagnetic fields. In simple cases, appropriate laws of motion are well-known in both electromagnetism and general relativity; sufficiently small test charges accelerate via the Lorentz force law and sufficiently small test masses fall on geodesics. Test body motion is also understood in cases where a body's higher multipole moments cannot be neglected \cite{Dix74}. 

Although it has historically been difficult to relax the test body assumption in relativistic theories, several important cases have nevertheless been understood \cite{ErberReview, Jackson, Spohn, PoissonRev, BarackReview, BlanchetReview, FutamaseReview, Kopeikin}. The majority of this work has been intrinsically perturbative. It makes detailed assumptions about the systems to be studied and uses these assumptions at all stages in the analysis. Concepts like the mass and momentum of individual objects typically arise as purely perturbative structures with no clear connection to the full theory. This review takes a different approach. Although approximations may be needed to understand specific applications, we adopt the point of view that approximating exact concepts is preferable to considering structures which emerge only as artifacts of a particular approximation scheme. We therefore focus on non-perturbative descriptions of motion. Somewhat surprisingly, considerable progress can be made from this perspective. Indeed, applying perturbation theory ``too early'' serves mainly to increase the computational burden and to obscure the underlying physics. 

The first step in our program is to define exact linear and angular momenta for arbitrary extended objects\footnote{Classical point particles are sometimes discussed as though they were the fundamental building blocks of all classical matter. This viewpoint is severely problematic on both mathematical and physical grounds, and is rejected here. That said, appropriately-regularized point particles do arise \textit{as mathematical structures} obtained from certain well-defined limits involving families of extended bodies. All mention of point particles here is to be understood in this (effective) sense.}. It is these momenta which are used to characterize an object's motion. Their evolution equations are derived without placing any significant constraints on an object's shape, composition, or degree of rigidity. Despite this generality, the methods used here are very easy to apply once the main concepts have been understood. Almost all difficulties lie in finding appropriate definitions and interpreting relations between those definitions. Complicated calculations are not required\footnote{As is typical throughout physics, simple underlying principles do not imply simple applications to explicit problems. Applications typically do require significant computations.}. Nevertheless, many of the concepts used here are likely to be unfamiliar. Considerable effort has therefore been devoted to explaining these concepts slowly and carefully by applying them in a series of increasingly complicated contexts.  

The prototype for all of our discussion is Newtonian celestial mechanics. The laws of motion for this theory are therefore reviewed in Section \ref{Sect:Newton}. This serves two purposes. First, Section \ref{Sect:NewtonianOld} uses standard techniques to remind the reader which ideas are important and why they are true. What, for example, is a self-field? Why do the self-force and self-torque vanish in Newtonian gravity? The methods used to discuss these questions make essential use of the vector space structure of Euclidean space, and cannot be generalized to curved spacetimes. 

Section \ref{Sect:NewtonianNew} introduces techniques which remove this objection. The Newtonian problem is reformulated such that all references to the detailed properties of Euclidean space are eliminated. Indeed, all that is required at this stage is a Riemannian space which admits a maximal set of Killing vector fields. Using this, a ``generalized momentum'' is introduced which serves to describe a body's large-scale behavior. The ordinary linear and angular momenta arise as two aspects of this more fundamental structure, and are therefore equivalent. Many computations simplify, however, when expressed in terms of the generalized momentum rather than its linear and angular components. Employing it, Newtonian self-forces and self-torques are seen to vanish using a one-line computation which employs only the symmetry of an appropriate Green function. That symmetry is physically related to Newton's third law.

From this perspective, certain generalizations of Newtonian gravity may be considered with almost no additional effort. The standard Euclidean background space can, for example, be replaced by one which is spherical or hyperbolic. The usual laws of motion still hold in these cases, except for the addition of Mathisson-Papapetrou spin-curvature couplings. Such terms arise kinematically even in these non-relativistic problems, and are shown to have a simple geometrical interpretation. 

Fully relativistic motion is first discussed in Section \ref{Sect:SR}, although only in flat or otherwise maximally-symmetric backgrounds. We first consider the motion of matter coupled to a linear scalar field. It is shown that a non-perturbative self-field can be defined in this context which is only a slight generalization of its Newtonian analog. Unlike in the non-relativistic case, however, forces and torques exerted by relativistic self-fields do not necessarily vanish. They instead act to renormalize\footnote{It is common in the literature to use the words renormalization and regularization interchangeably, both implying the removal of unwanted infinities. This is not the usage here. Renormalization is intended in this review essentially as a synonym for ``dynamical shift.'' These shifts need not be infinite. Regularizations, by contrast, always refer to rules for handling singular behavior. Almost all discussion here focuses on finite renormalizations. Regularizations arise only in certain limiting cases.} an object's linear and angular momenta. This effect is finite and non-perturbative. Physically, it represents the inertia associated with an object's self-field. Mathematically, it is related to the hyperbolicity of the underlying field equations. Similar effects apply generically for all matter coupled to long-range hyperbolic fields.

Next, Section \ref{Sect:Curved} considers motion in fully generic curved spacetimes. The Killing vectors used to define momenta in simpler cases must then be replaced by an appropriate set of ``generalized Killing fields.'' This is accomplished in Section \ref{Sect:GKF}. The scalar problem of motion is analyzed first in this more general context, where a new type of renormalization is found to occur which affects the quadrupole and higher multipole moments of a body's stress-energy tensor, and may be viewed as a consequence of the ``passive gravitational mass distribution'' of an object's self-field. Matter coupled to electromagnetic fields in generic background spacetimes may be understood similarly, and is discussed in Section \ref{Sect:emSR}. Finally, Section \ref{Sect: GR} considers motion in general relativity.

\subsection*{Notation}

The sign conventions used here are those of Wald \cite{Wald}. Metrics have positive signature. The Riemann tensor satisfies $2 \nabla_{[a} \nabla_{b]} \omega_{c} = R_{abc}{}^{d} \omega_{d}$ for any 1-form $\omega_{a}$, and the Ricci tensor is defined by $R_{ab} = R_{acb}{}^{c}$. Abstract indices are denoted using letters $a,b,\ldots$ from beginning of the Latin alphabet, while $i,j, \ldots$ represent coordinate components. Boldface symbols are used to denote Euclidean vectors and tensors. Units are chosen such that $c=G=1$.

\section{Newtonian gravity}
\label{Sect:Newton}

Consider a Newtonian test body immersed in a gravitational potential $\phi(\bx,t)$. If such a body is sufficiently small, it is well-known that its center of mass location $\bm{\gamma}_t$ at time $t$ evolves via
\begin{equation}
    \ddot{\bm{\gamma}}_t = - \bnabla \phi (\bm{\gamma}_t,t).
    \label{AccelNewtTest}
\end{equation}
This is not the correct equation of motion for (non-spherical) objects with significant self-gravity. Relaxing the test body assumption while still imposing an appropriate smallness condition instead results in
\begin{equation}
    \ddot{\bm{\gamma}}_t = - \bnabla \hat{\phi} (\bm{\gamma}_t,t),
    \label{AccelNewtFull}
\end{equation}
where $\hat{\phi}$ denotes that part of the potential which is determined only by masses external to the body of interest. Comparison of these two equations shows that the field $\bnabla \hat{\phi}$ which accelerates a large mass can differ\footnote{Although the gradients of $\phi$ and $\hat{\phi}$ coincide at the center of a spherically-symmetric mass, they can be quite different in general. Consider, for example, a barbell constructed by joining two unequal spheres with a massless strut.} from the field $\bnabla \phi$ which would be inferred by measuring the accelerations of nearby test particles. Although it is not often emphasized, this is a standard result in Newtonian gravity. The well-known laws of motion which describe ``Newtonian point masses'' are, for example, equivalent to \eqref{AccelNewtFull}, not \eqref{AccelNewtTest}.

A central goal of this review is to explain how similar results hold in more complicated relativistic theories. In all cases, the laws of motion are structurally identical to those associated with test bodies. The fields which appear in those laws of motion are not, however, the physical ones. Effective fields appear instead, their details depending on appropriate notions of self-interaction. Once the precise nature of the effective field has been determined in a particular theory, ``point particle limits'' and related approximations follow very easily. 

Many of the difficulties encountered in the relativistic theory motion already appear in the Newtonian problem (where they can be so simple as to easily pass by unnoticed). It is therefore instructive to open this review by carefully discussing the Newtonian theory of motion. Section \ref{Sect:NewtonianOld} accomplishes this using essentially standard arguments. Concepts such as the self-field and self-force are emphasized, as well as their connections to physical principles like Newton's third law. Similar discussions may be found in, e.g., \cite{Kopeikin, DamourReview, Dix79}. Unfortunately, the familiar techniques used in Section \ref{Sect:NewtonianOld} cannot be readily applied to more complicated theories. Section \ref{Sect:NewtonianNew} therefore uses Newtonian gravity as a familiar setting with which to introduce a different, more geometrical, approach. The resulting formulation, some of which originally appeared in \cite{HarteScalar, HarteGR}, does generalize. It is used throughout the remainder of this review.

\subsection{Newtonian celestial mechanics}
\label{Sect:NewtonianOld}

Consider an extended body residing inside a finite (and possibly time-dependent) region of space $\fB_t \subset \mathbb{R}^3$ which contains no other matter. The mass density $\rho$ and momentum density\footnote{In simple cases, $\bm{v}$ represents a velocity field in the standard sense. More generally, it might be only an effective construction. This occurs, for example, if a body is composed of multiple interpenetrating fluids.} $\rho \bm{v}$ of this body are both assumed to be smooth. Local mass and momentum conservation then imply that \cite{Dix79, MarsdenHughes, TruesdellNoll}
\begin{align}
    \frac{ \partial \rho }{ \partial t} + \bnabla \cdot ( \rho \bm{v}) = 0 
    \label{MassCons}
\end{align}
and
\begin{align}
    \frac{\partial}{\partial t} (\rho \bm{v} ) + \bnabla \cdot (\rho \bm{v} \otimes \bm{v} - \bm{\tau} ) = \bm{f}.
        \label{CauchyMomentum}
\end{align}
The Cauchy stress tensor $\bm{\tau}$ describes how matter interacts via contact forces, while the force density $\bm{f}$ describes longer-range interactions (also known as body forces). If the only long-range forces are gravitational, there exists a potential $\phi$ such that\footnote{This follows from noting that any sufficiently small piece of matter with finite density responds to gravitational forces as though it were a test body.}
\begin{equation}
    \bm{f} = - \rho \bnabla \phi.
    \label{NewtForce}
\end{equation}
Inside $\fB_t$, the potential must satisfy Poisson's equation
\begin{equation}
    \nabla^2 \phi = 4 \pi \rho.
    \label{Poisson}
\end{equation}
The influence of masses external to $\fB_t$ may be encoded using, e.g., $\phi$ or its normal derivative on the boundary $\partial \fB_t$.  

Equations \eqref{MassCons}-\eqref{Poisson} are very general. They are not, however, complete. Imposing appropriate boundary conditions, Poisson's equation determines $\phi$ in terms of $\rho$, mass conservation evolves $\rho$ using $\bm{v}$, and momentum conservation evolves $\bm{v}$ using $\bm{\tau}$. The stress tensor cannot, however, be determined without additional assumptions. Its evolution is not universal. Stresses depend on an object's detailed composition, reflecting the trivial fact that different types of materials move differently. This is but a minor obstacle in celestial mechanics, and no particular form for $\bm{\tau}$ is assumed here.

Observables which describe a body's ``large-scale'' motion may be obtained by integrating the conservation laws \eqref{MassCons} and \eqref{CauchyMomentum}. This results in the total mass $m$, linear momentum $\bp(t)$, and angular momentum $\bs(\bz_t, t)$:
\begin{equation}
    m := \int_{\fB_t} \!\! \rho \rmd^3 \bm{x}, \quad \bm{p} := \int_{\fB_t} \!\! \rho \bm{v} \rmd^3 \bm{x}, \quad \bs := \int_{\fB_t} \!\! \rho (\bx - \bz_t) \times \bm{v}\rmd^3 \bm{x}.
    \label{pDef}
\end{equation}
The general philosophy of celestial mechanics is to focus on $\bp$ and $\bs$ while ignoring $\rho$ and $\bm{v}$ as much as possible. The vast majority of information concerning an object's internal structure is set aside; only its momenta matter. Evolution equations for these momenta are easily obtained from \eqref{MassCons} and \eqref{CauchyMomentum}, which show that the mass remains constant and that
\begin{equation}
    \dot{\bp} = - \int_{\fB_t} \!\! \rho \bnabla \phi \rmd^3 \bx, \qquad \dot{ \bs } = - \int_{\fB_t} \!\! \rho (\bx-\bz_t) \times \bnabla \phi \rmd^3 \bx - \dot{\bz}_t \times \bp.
    \label{NewtForceTot}
\end{equation}
The gravitational force and torque acting on an extended body therefore depend on its mass distribution, its internal gravitational potential, and a ``choice of origin'' parametrized by $\bz_t$. 

One reason for considering $\bm{p}$ is its close relation to the center of mass position $\bm{\gamma}_t$. This can be defined by
\begin{equation}
    \bm{\gamma}_t := \frac{1}{m} \int_{\fB_t} \bx \rho(\bx,t) \rmd^3 \bx,
    \label{CMNewt2}
\end{equation}
or equivalently by demanding that a body's mass dipole moment vanish when evaluated about $\bm{\gamma}_t$:
\begin{equation}
    \int_{\fB_t} \!\! (\bx - \bm{\gamma}_t) \rho(\bx,t)  \rmd^3 \bx = 0.
    \label{CMNewt}
\end{equation}
Regardless, it follows from \eqref{MassCons} and \eqref{CMNewt2} that the center of mass velocity must satisfy
\begin{equation}
    m \dot{\bm{\gamma}}_t = \bm{p}. 
    \label{MomVelNewt}
\end{equation}
Note that this is a derived result, not a definition; $m$, $\bm{\gamma}_t$ and $\bm{p}$ are defined in terms of $\rho$ and $\bm{v}$ via \eqref{pDef} and \eqref{CMNewt2}. In more complicated theories, the center of mass velocity need not be parallel to the momentum. 

Once $\bm{\gamma}_t$ has been defined, its time evolution is easily found by combining \eqref{NewtForceTot} and \eqref{MomVelNewt} to yield
\begin{equation}
 m \ddot{\bm{\gamma}}_t = - \int_{\fB_t} \!\! \rho \bnabla \phi \rmd^3 x.
 \label{CMevolveNewt}
\end{equation}
Evaluating $\bs(\bz_t,t)$ with $\bz_t = \bm{\gamma}_t$ isolates the spin component of the angular momentum from the orbital component which can appear more generally, resulting in
\begin{equation}
    \dot{\bs} = - \int_{\fB_t} \!\! \rho (\bx-\bm{\gamma}_t) \times \bnabla \phi \rmd^3 \bx.
    \label{SpinEvolveNewt}
\end{equation}
In astrophysical applications, $\rho$ and $\bm{\nabla} \phi$ are typically only very coarsely constrained by observations. Integral expressions like \eqref{CMevolveNewt} and \eqref{SpinEvolveNewt} are therefore unsuitable for applications. They must first be simplified. 

Such simplifications are immediate if $\bnabla \phi$ varies negligibly throughout $\fB_t$, as can occur if the mass in question is a test body whose dimensions are small compared with the distances to all other masses in the universe. In these cases, it follows from \eqref{CMevolveNewt} and \eqref{SpinEvolveNewt} that the center of mass acceleration satisfies \eqref{AccelNewtTest} and that $\dot{\bs} = 0$. Simplifying the laws of motion in more general contexts requires understanding the influence of an object's own gravitational field.

A precise definition for the self-field may be obtained via a two-point function (or propagator) $G(\bx,\bx')$ describing ``the gravitational potential at $\bx$ per unit mass at $\bx'$.'' Any potential constructed from such a propagator can reasonably be called a self-field only if it is a Green function for the Poisson equation:
\begin{equation}
    \nabla^2 G(\bx,\bx') = 4 \pi \delta^3(\bx- \bx').
    \label{GLaplace}
\end{equation}
There are, of course, many possible Green functions. A particular one may be singled out by demanding that self-fields described by $G$ be compatible with Newton's third law. Consider two distinct points $\bx, \bx' \in \fB_t$. It is then natural to interpret ``the force on mass at $\bx$ due to mass at $\bx'$'' to mean
\begin{equation}
    - \rho (\bm{x},t) \bnabla \left[ G(\bx,\bx') \rho(\bx',t) \rmd^3 \bx'  \right] \rmd^3 \bx .
    \label{dForce}
\end{equation}
The weak form of Newton's third law states that the force at $\bx$ due to $\bx'$ must be equal and opposite to the force at $\bx'$ due to $\bx$, implying that
\begin{equation}
    \bnabla G(\bx,\bx') = - \bnabla' G(\bx',\bx).
    \label{GradG}
\end{equation}
The strong form of Newton's third law instead requires that the force at $\bx$ due to $\bx'$ point along the line which connects these two points. Imposing this,
\begin{equation}
    \bnabla G(\bx,\bx') \propto \bx - \bx'.
    \label{GradG2}
\end{equation}
Any $G(\bx,\bx')$ which is compatible with the strong form of Newton's third law can therefore depend only on the distance $|\bx - \bx'|$ between its arguments. Up to an irrelevant additive constant, it follows from \eqref{GLaplace} that
\begin{equation}
    G(\bx,\bx') = G(\bx',\bx) = - \frac{1}{|\bx - \bx'|} ,
    \label{GreenNewt}
\end{equation}
and the total self-field is
\begin{equation}
    \phiS(\bx,t) := \int_{\fB_t} \rho(\bx',t) G(\bx,\bx') \rmd^3 \bx' = - \int_{\fB_t} \frac{\rho(\bx',t)}{|\bx-\bx'|} \rmd^3 \bx' .
    \label{SelfFieldNewt}
\end{equation}
The physical field $\phi$ may be viewed as the sum of the self-field $\phiS$ and an appropriate remainder $\hat{\phi}$:
\begin{equation}
    \hat{\phi} := \phi - \phiS.
\end{equation}
It follows from \eqref{Poisson}, \eqref{GLaplace}, and \eqref{SelfFieldNewt} that the effective potential $\hat{\phi}$ satisfies the vacuum field equation $\nabla^2 \hat{\phi} = 0$ throughout $\fB_t$.

Now consider the total force exerted by $\phiS$, the ``self-force.'' Noting \eqref{NewtForceTot}, it is natural to let this refer to
\begin{equation}
    \bm{F}_\mathrm{S} := - \int_{\fB_t} \!\! \rho \bm{\nabla} \phiS \rmd^3 \bx.
\end{equation}
Substituting \eqref{SelfFieldNewt} into this expression results in an integral over the product space $\fB_t \times \fB_t$:
\begin{equation}
    \bm{F}_\mathrm{S} = 
     -  \int_{\fB_t \times \fB_t} \!\!\! \rho(\bx,t) \rho(\bx',t)  \bnabla G(\bx,\bx') \rmd^3 \bx \rmd^3 \bx' .
\end{equation}
Recalling \eqref{GradG} or \eqref{GreenNewt}, the integrand is antisymmetric under interchange of $\bx$ and $\bx'$. The Newtonian self-force therefore vanishes. This is an exact result. It holds for all compact mass distributions. Whatever the shape a particular body happens to be in $\fB_t$, the self-force vanishes because that shape is trivially symmetric when copied into $\fB_t \times \fB_t$. A similar argument may be used to show that the self-torque, the net torque exerted by $\phiS$, vanishes as well.

The main point of this discussion is that the net gravitational force exerted on any isolated extended mass satisfies
\begin{equation}
    \dot{\bp} = - \int_{\fB_t} \!\! \rho \bm{\nabla} (\hat{\phi} + \phiS) \rmd^3 \bx = - \int_{\fB_t} \!\! \rho \bm{\nabla} \hat{\phi} \rmd^3 \bx.
    \label{CMevolveNewt2}
\end{equation}
Not necessarily choosing $\bz_t$ to be the center of mass, the equivalent evolution equation for $\bs(\bz_t,t)$ is
\begin{equation}
    \dot{\bs} = - \int_{\fB_t} \!\! \rho (\bx-\bz_t) \times \bnabla \hat{\phi} \rmd^3 \bx - \dot{\bz}_t \times \bm{p} .
    \label{sDotNewtOld}
\end{equation}
The vanishing self-force and self-torque therefore allows $\phi$ to be replaced by $\hat{\phi}$ in the evolution equations for both the linear and angular momenta. Although forces and torques may be computed using either the physical field $\phi$ or the fictitious effective field $\hat{\phi}$, the latter computation is often simpler. In most cases of practical interest, $\bnabla \hat{\phi}$ varies far more slowly in $\fB_t$ than does $\bnabla \phi$. The integral involving $\bnabla \hat{\phi}$ can therefore be amenable to approximation when the (otherwise equivalent) integral involving $\bnabla \phi$ is not. 

Recalling that $\hat{\phi}$ is harmonic inside the body region, it must be analytic there. This means that its Taylor series about an arbitrary point $\bz_t \in \fB_t$ converges at least in some neighborhood of $\bz_t$. If that series converges throughout the body, it may be substituted into \eqref{CMevolveNewt2} and integrated term by term. The integral expression for the force is then equivalent to
\begin{eqnarray}
    \dot{p}_i(t) = - m \partial_i \hat{\phi}(\bz_t,t) 
    - \sum_{n=1}^\infty \frac{1}{n!} m^{j_1 \cdots j_n} (\bz_t, t) \partial_i \partial_{j_1} \cdots \partial_{j_n} \hat{\phi} (\bz_t,t) ,
    \label{MultipoleNewt}
\end{eqnarray}
where $m^{j_1 \cdots j_n}(\bz_t, t)$ denotes the body's $2^n$-pole mass moment about $\bz_t$:
\begin{equation}
    m^{j_1 \cdots j_n}(\bz_t,t) := \int_{\fB_t} \!\! (\bx - \bz_t)^{j_1} \cdots ( \bx - \bz_t)^{j_n} \rho(\bx,t) \rmd^3 \bx.
    \label{massMultipole}
\end{equation}
The series \eqref{MultipoleNewt} is referred to as a multipole expansion for the force. A similar series also exists for the angular momentum. If $\bz_t$ is chosen to coincide with the center of mass $\bm{\gamma}_t$, the dipole moment $m^i (\bm{\gamma}_t, t)$ vanishes by \eqref{CMNewt}. The $n=1$ term in \eqref{MultipoleNewt} therefore vanishes as well, so
\begin{eqnarray}
    \ddot{\gamma}^i_t = - \partial^i \hat{\phi}(\gamma_t,t) 
    - \frac{1}{m} \sum_{n=2}^\infty \frac{1}{n!} m^{j_1 \cdots j_n} (\gamma_t, t) \partial^i \partial_{j_1} \cdots \partial_{j_n} \hat{\phi} (\gamma_t,t) .
    \label{CMNewtAccel}
\end{eqnarray}
The utility of this expression is that there are many cases of interest where the multipole series can be truncated at low order without significant loss of accuracy. The simplest such truncation recovers \eqref{AccelNewtFull}. More generally, there are correction to this equation which involve a body's quadrupole and higher multipole moments.

Lastly, note that the moments \eqref{massMultipole} are somewhat different from the ones which are found in textbooks. The harmonicity of $\hat{\phi}$ implies that arbitrary traces may be added to the $m^{j_1 \cdots j_n}$ without affecting the force. The $m^{j_1 \cdots j_n}$ appearing in \eqref{MultipoleNewt} may therefore be replaced by different moments $\tilde{m}^{j_1 \cdots j_n}$ which are trace-free in all pairs of indices. It is these trace-free moments which are typically used in practical calculations. Besides the elimination of irrelevant components, the trace-free moments are also useful in that they may be determined purely using external measurements of an object's gravitational field.

\subsection{Reformulating Newtonian celestial mechanics}
\label{Sect:NewtonianNew}

The discussion which has just been presented relies heavily on the geometric peculiarities of Euclidean space. This is not essential, however. The only characteristic of (three-dimensional) Euclidean space which is truly important is that it is maximally symmetric: There exist a total of six linearly independent Killing vector fields. The Newtonian laws of motion are now rederived using methods which make this manifest.

As a consequence, certain aspects of the Newtonian problem are significantly clarified. The geometrical nature of the linear and angular momenta is made precise, for example. These are shown to be two aspects of a more fundamental vector which lives not in the physical space, but in a space which is dual to the space of Killing vector fields. The approach introduced in this section also emphasizes the importance of symmetries, and is fundamental to understanding motion in the more complicated theories discussed below. 

Another advantage of the reformulation discussed in this section is that certain generalizations of Newtonian gravity may be understood essentially ``for free.'' Noting that spherical and hyperbolic spaces are both maximally-symmetric, there are no new complications if the usual Euclidean background of Newtonian gravity is replaced by a space of constant curvature. It is also trivial to change the number of spatial dimensions, or to add, e.g., a mass term to the field equation. For concreteness, we restrict to three spatial dimensions and keep the gravitational field equation as-is. We do, however, allow the background space to be curved. This has interesting consequences which reappear in the more complicated relativistic theories considered in later sections.

\subsubsection{Geometric preliminaries}

The locations of Newtonian events may be viewed as points in a four-dimensional manifold $\mathcal{M}$. While a relativistic spacetime is defined using only a manifold and a non-degenerate metric, Newtonian spacetimes require more structure \cite{MarsdenHughes,TruesdellNoll,DixonSR}. One such structure is a preferred notion of time. This takes the form of an equivalence class\footnote{For any single time function $T: \mathcal{M} \rightarrow \mathbb{R}$ and any $c, d \in \mathbb{R}$ such that $c >0$, the map $c T+ d$ is also an acceptable time function.} of functions which associate each event in spacetime with ``the time'' at which it occurs. Associated with this is a preferred foliation of $\mathcal{M}$ into a one-parameter family of hypersurfaces $\{ \mathcal{S}_t \}$, the spaces of constant time. 

Newtonian spacetimes are difficult to work with directly. They simplify considerably in the presence of a frame, a structure that identifies events at different times as being at ``the same'' spatial point. It is assumed here that a frame has been fixed in such a way that all spaces $\mathcal{S}_t$ are mapped into a single space consisting of a three-dimensional manifold $\mathcal{S}$ together with a (fixed) Riemannian metric $g_{ab}$. This process also fixes a particular time function. It permits all physical quantities in spacetime to be viewed as time-dependent quantities on $\mathcal{S}$. We allow the spatial metric to be curved, but assume that its curvature is everywhere constant. Letting $\nabla_a$ and $R_{abc}{}^{d}$ denote the covariant derivative and Riemann tensor associated with $g_{ab}$, it follows that $\nabla_a R_{bcd}{}^{f} = 0$, and that $(\mathcal{S}, g_{ab})$ is maximally symmetric. 

Consider the motion of a material object instantaneously confined to a submanifold $\fB_t \subset \mathcal{S}$ which contains no other matter and has finite volume. Denote this body's mass density at time $t$ by $\rho(\cdot,t)$ and its velocity field at time $t$ by $v^a(\cdot,t)$. Local conservation of mass and momentum continue to hold in this context, so \eqref{MassCons} and \eqref{CauchyMomentum} carry over essentially without change:
\begin{align}
    &\frac{\partial}{\partial t} \rho + \nabla_a (\rho v^a ) = 0,
    \label{MassCons2}
    \\
    &\frac{\partial}{\partial t} (\rho v^a ) + \nabla_b ( \rho v^a v^b - \tau^{ab}) = - \rho \nabla^a \phi.
    \label{CauchyMomentum2}
\end{align}
The gravitational potential $\phi$ which appears here satisfies the obvious generalization of Poisson's equation:
\begin{equation}
    \nabla^a \nabla_a \phi = g^{ab} \nabla_a \nabla_b \phi = 4\pi \rho.
    \label{Poisson2}
\end{equation}

\subsubsection{Generalized momentum}
\label{Sect:GenPNewt}

Our first significant departure from the elementary discussion of Newtonian motion found in Section \ref{Sect:NewtonianOld} arises in the definitions for a body's linear and angular momenta. The usual integrals \eqref{pDef} make sense only when evaluated in a Cartesian coordinate system. Alternatively, they require a canonical identification of tangent spaces associated with different points in the spatial manifold. While this is easily accomplished in Euclidean space, it is not obvious what to do more generally. Our first task is therefore to define momenta which do not make reference to a specific coordinate system. Accomplishing this provides a notion of momentum which is easily generalized to curved Newtonian backgrounds, and even to completely generic relativistic spacetimes. It is a basic building block for all results discussed in this review. 

One problem with elementary definitions of mechanical momentum is that they attempt to represent this concept as a spatial vector or covector. This is physically unnatural, however (except for point particles or momentum densities); momenta are generically associated with extended regions, not individual points. There is no natural tangent or cotangent space in which to place, for example, the momentum of an extended region $\mathfrak{R} \subset \fB_t$. The simplest mathematical structure with which to represent a quasi-local quantity like a momentum must itself be quasi-local, and spatial tensors are not examples of such structures. 

Besides being quasi-local, momenta must also be extensive. For any two disjoint regions $\mathfrak{R}_1, \mathfrak{R}_2 \subset \fB_t$ which are ``physically independent,'' there must be a sense in which
\begin{equation}
    (\mbox{momentum in $\mathfrak{R}_1$}) + (\mbox{momentum in $\mathfrak{R}_2$}) = (\mbox{momentum in $\mathfrak{R}_1 \cup \mathfrak{R}_2$})
\end{equation}
for some binary operation ``$+$'' which is both associative and commutative. If $\mathfrak{R}_1$ and $\mathfrak{R}_2$ are identically prepared, it is also natural to suppose that
\begin{equation}
    (\mbox{momentum in $\mathfrak{R}_1$}) + (\mbox{momentum in $\mathfrak{R}_2$}) = 2 (\mbox{momentum in $\mathfrak{R}_1$}),
\end{equation}
thus motivating a notion of scalar multiplication. 

Together, these considerations and others suggest that momenta should be elements of a vector space. The most natural vector space is not, however, the space of tensors at any particular spatial point. A better choice may be motivated by recalling that conserved linear momenta arise naturally in theories which are derived from translation-invariant Lagrangians. Similarly, conserved angular momenta arise from Lagrangians which are invariant with respect to rotations. This suggests that both types of momenta can be associated explicitly with a collection of continuous symmetries. Consider, in particular, those symmetries --- the continuous isometries --- which preserve the spatial metric. While these are not necessarily symmetries for all physically-interesting quantities, they are extremely useful.

The continuous isometries of a Riemannian space $(\mathcal{S},g_{ab})$ are generated by its Killing vector fields. By definition, 
\begin{equation}
    \Lie_\xi g_{ab} = 0 
    \label{Killing}
\end{equation}
for every Killing vector $\xi^a$, where $\Lie_\xi$ denotes the Lie derivative with respect to $\xi^a$. We use $K$ to denote the vector space consisting of all Killing vector fields together with obvious notions of addition and scalar multiplication. It is well-known that the dimension of this vector space is finite. More specifically, if the dimension of the physical space is $\dim \mathcal{S} = \dim \fB_t = N$, it may be shown that (see, e.g., Appendix C of \cite{Wald})
\begin{equation}
    \dim K \leq \frac{1}{2} N (N+1).
    \label{MaxDim}
\end{equation}
This section restricts attention to maximally-symmetric spaces where $\dim K = \frac{1}{2} N(N+1)$. When $N=3$, Euclidean, spherical, and hyperbolic spaces are all maximally-symmetric, admitting six linearly-independent Killing fields. Given a preferred point, three Killing fields may be interpreted as translations and three as rotations. This split makes sense only near the given point, however, and is best avoided at this stage. Doing so implies that the linear and angular momenta should be treated as elements of a single object ``conjugate to'' the space of all Killing vector fields.

More specifically, consider a representation for a body's momentum as a vector in the space $K^*$ which is dual to $K$. An element of $K^*$ is, by definition, a linear map from $K$ to $\mathbb{R}$. The specific linear map which has the desired properties is 
\begin{equation}
    P_t[\mathfrak{R}](\xi) := \int_{\mathfrak{R}} \rho(x,t) v_a(x,t) \xi^a(x) \rmd V,
    \label{Pdef}
\end{equation}
where $\xi^a \in K$ and the volume element is the natural one associated with $g_{ab}$. \textit{We call this the generalized momentum contained in $\mathfrak{R} \subseteq \fB_t$ at time $t$}. It is often convenient to omit the dependence on $\mathfrak{R}$, in which case it is to be understood that $P_t = P_t[\fB_t]$.

The dimension of $K^*$ is equal to the dimension of $K$, so this momentum has six components in three spatial dimensions. These components correspond to the usual three components of linear momentum and three components of angular momentum. Such a split can be made explicit by introducing additional structure, namely a preferred point $z_t \in \fB_t$, and for any such point, $P_t[\mathfrak{R}]$ can be re-expressed in terms of spatial tensors $p_a$, $S^a$ at $z_t$. This decomposition is explained in Section \ref{Sect:PtopS}. For now, it suffices to consider $P_t[\mathfrak{R}]$ on its own. While the introduction of a preferred point allows this map to be replaced by spatial tensors, avoiding such representations whenever possible provides considerable calculational and conceptual simplifications.


In relativistic contexts where there exists a maximally-symmetric background geometry, the generalized momentum remains essentially unchanged. The infinitesimal momentum $\rho v_a \rmd V$ is merely replaced by $T_{a}{}^{b} \rmd S_b$, where $T_{a}{}^{b}$ is an appropriate stress-energy tensor and $\rmd S_b$ is the natural volume element on a three-dimensional hypersurface. If a spacetime is not maximally-symmetric, one also replaces $K$ by another vector space which has the correct dimensionality. The ``generalized Killing fields'' used for this purpose are discussed in Section \ref{Sect:GKF}.

\subsubsection{Generalized force}

How does Newtonian gravity affect the time evolution of the generalized momentum? Using local momentum conservation \eqref{CauchyMomentum2} and assuming that the boundary $\partial \mathfrak{R}$ is independent of time (or that there is no matter there),
\begin{align}
    \frac{\rmd}{\rmd t} P_t[\mathfrak{R}](\xi) &= \int_{\mathfrak{R}} \big[ - \rho \Lie_\xi \phi + \frac{1}{2} ( \rho v^a v^b - \tau^{ab}) \Lie_\xi g_{ab} \big] \rmd V
    = - \int_{\mathfrak{R}} \! \rho \Lie_\xi \phi \rmd V.
    \label{GenForceNewt}
\end{align}
The second equality here follows from Killing's equation \eqref{Killing}. If $\Lie_\psi \phi = 0$ for some specific Killing field $\psi^a$, it is clear that the associated momentum $P_t[\mathfrak{R}](\psi)$ is conserved. This means that if $\phi$ is constant along a translational Killing field, there can be no force in that direction. Similarly, a field which is invariant about rotations around a given axis exerts no torque about that axis. Both of these statements are physically obvious. They are also of limited value. Once the field equation \eqref{Poisson2} is taken into account, $\Lie_\psi \phi = 0$ implies that $\Lie_\psi \rho \propto \Lie_\psi \nabla^a \nabla_a \phi = \nabla^a \nabla_a \Lie_\psi \phi = 0$ as well. This is clearly impossible for any compact body if $\psi^a$ is a pure translation. Rotational symmetries fare somewhat better, although they are still a rather special case.

Transforming \eqref{GenForceNewt} into a surface integral results in a more interesting conservation law. Using the field equation and integrating by parts shows that
\begin{equation}
    \frac{\rmd}{ \rmd t }  P_t[\mathfrak{R}](\xi) = - \oint_{\partial \mathfrak{R}} T^{a}{}_{b} \xi^b \rmd S_a,
\end{equation}
where
\begin{equation}
    T_{ab} := \frac{1}{4\pi} ( \nabla_a \phi \nabla_b \phi - \frac{1}{2} g_{ab} \nabla^c \phi \nabla_c \phi ) 
\end{equation}
is the stress tensor associated with $\phi$. At least in flat space, one might imagine extending $\partial \mathfrak{R}$ (and perhaps $\partial \fB_t$) far outside of all matter of interest. If $\phi$ falls off sufficiently fast in this region, the surface integral can be seen to vanish. The generalized momentum is therefore conserved in such cases. Of course, the momentum associated with a single object in a larger system is not conserved. Understanding its dynamics requires a different argument.

\subsubsection{The self-field}

The generalized force \eqref{GenForceNewt} involves the physical field $\phi$. As discussed in Section \ref{Sect:NewtonianOld}, this is too complicated to work with directly. We therefore isolate its most complicated part --- the self-field --- and compute what it does directly. Once this is accomplished, the remaining undetermined portion of the force is relatively simple to understand.

The self-field in this context is defined in Section \ref{Sect:NewtonianOld} in terms of a certain two-point function $G$. More specifically, $G$ is a Green function. If $G(x,x') = G(x',x)$, the two constraints \eqref{GradG}, \eqref{GradG2} which implied a notion of Newton's third law in Euclidean space generalize to the statement that
\begin{equation}
    \Lie_\xi G(x,x') = \big[ \xi^a(x) \nabla_a + \xi^{a'}(x') \nabla_{a'} \big] G(x,x') = 0
    \label{LieG}
\end{equation}
for all $\xi^a \in K$. In the Euclidean case, translational invariance alone implies the weak form of Newton's third law. Further imposing rotational invariance recovers the strong form of Newton's third law. In general, though, symmetries of $G$ imply only ``portions of'' Newton's third law.

It is always possible to find Green functions which satisfy \eqref{LieG} in maximally-symmetric backgrounds. Indeed, these Green functions depend only on the geodesic distance between their arguments. Introducing Synge's function (also known as the world function) \cite{PoissonRev,DeWittBrehme, Synge}
\begin{equation}
    \sigma(x,x') := \frac{1}{2} (\mbox{squared geodesic distance between $x$ and $x'$}),
    \label{Synge}
\end{equation}
the Euclidean  Green function \eqref{GreenNewt} can be written as $G = - 1/\sqrt{2 \sigma}$. Green functions associated with spherical and hyperbolic spaces are merely more complicated functions of $\sigma$ \cite{Diacu}. In any of these cases, $\Lie_\xi G \propto \Lie_\xi \sigma = 0$.

Using the symmetric Green function which satisfies \eqref{LieG} to define the self-field, let
\begin{equation}
    \phiS (x,t) := \int_{\fB_t} \rho(x',t) G(x,x') \rmd V'.
    \label{SelfFieldNewt2}
\end{equation}
Substituting this into \eqref{GenForceNewt} then shows that
\begin{align}
    \frac{\rmd}{\rmd t} P_t =& - \int_{\fB_t} \!\! \rho(x,t) \Lie_\xi \hat{\phi}(x,t) \rmd V 
    \nonumber
    \\
    &~ - \frac{1}{2} \int_{\fB_t} \!\! \rmd V \int_{\fB_t} \!\! \rmd V' \rho(x,t) \rho(x',t) \Lie_\xi G(x,x') 
    \nonumber
    \\
     =& - \int_{\fB_t} \rho(x,t) \Lie_\xi \hat{\phi}(x,t) \rmd V ,
    \label{GenForceNewt2}
\end{align}
where $\hat{\phi} = \phi - \phiS$ and $\mathfrak{R}$ has been replaced by the entire body region $\fB_t$. It is clear from this that the self-force and self-torque both vanish as an immediate consequence of \eqref{LieG}. All forces and torques may therefore be computed using $\hat{\phi}$ instead of $\phi$. Furthermore, the effective field satisfies the vacuum equation
\begin{equation}
  \nabla^a \nabla_a \hat{\phi} = 0,
  \label{phiHat}
\end{equation}
and can clearly be computed by subtracting the self-field from the physical field. Alternatively, Stokes' theorem may be used together with \eqref{phiHat} to write $\hat{\phi}$ as a kind of average of $\phi$ over a closed surface which surrounds the body of interest. 

It has already been mentioned that $P_t(\psi)$ is conserved if $\Lie_\psi \phi =0$. Equation \eqref{GenForceNewt2} shows that this also true if $\Lie_\psi \hat{\phi} =0$, a much weaker condition. For a closed system, one typically has $\phi = \phiS$ and hence $\hat{\phi} = 0$. All components of the generalized momentum are therefore conserved in such cases.

Equation \eqref{GenForceNewt2} has been established by showing that the generalized force exerted by $\phiS$ always vanishes. This force involves an integral over $\fB_t \times \fB_t$, and may therefore be interpreted as a two-point interaction. It can sometimes be interesting to also consider interactions between three or more points. Let
\begin{align}
    \tilde{\phi}_\mathrm{S}(x,t) := \sum_{n=1}^{ n_\mathrm{max}  } c_n \int_{\fB_t} \!\! \rmd V_1 \cdots \int_{\fB_t} \!\! \rmd V_n \rho(y_1,t) \cdots \rho(y_n,t) G_{n} (x,y_1,\ldots,y_n),
    \label{PhiSGen}
\end{align}
where the $c_n$ are arbitrary constants and the $(n+1)$-point propagators $G_{n}$ are symmetric in their arguments and satisfy $\Lie_\xi G_{n}$ for all $\xi^a \in K$. It is straightforward to show that the generalized force exerted by any such field vanishes. Given the two-point $G$ used to define $\phiS$, an appropriate three-point interaction may be chosen using, e.g.,
\begin{equation}
    G_3 (x,y,z) = G(x,y) G(y, z) G(z,x).
\end{equation}
Other choices are also possible, of course. Higher-order propagators typically lead to fields $\tilde{\phi}_\mathrm{S}$ which are not really Newtonian self-fields in the sense  that $\nabla^a \nabla_a \tilde{\phi}_\mathrm{S} \neq 4 \pi \rho$. Series like \eqref{PhiSGen} can nevertheless be useful for understanding different theories where matter couples to nonlinear fields. In those cases, the sum in $\tilde{\phi}_\mathrm{S}$ might be compared to a kind of Dyson series for an object's self-field. Regardless of the field equation, however, the existence of a Killing field $\psi^a$ which satisfies $\Lie_\psi (\phi - \tilde{\phi}_\mathrm{S}) = 0$ for \textit{some} $\tilde{\phi}_\mathrm{S}$ always implies that $P_t(\psi)$ is conserved. Although this conservation law might be manifest only for a particular choice of $\tilde{\phi}_\mathrm{S}$, the value of $P_t(\psi)$ does not depend on that choice.

\subsubsection{Multipole expansions}
\label{Sect:multipoleNewt}

Returning to the main development, note that \eqref{GenForceNewt} and \eqref{GenForceNewt2} differ only by the replacement $\phi \rightarrow \hat{\phi}$. Although both of these integrals are numerically equivalent, the latter is often simpler to evaluate. This is because $\Lie_\xi \hat{\phi}$ can be readily approximated throughout $\fB_t$ in many more physically-interesting situations than can $\Lie_\xi \phi$. Such approximations are based on a Taylor expansion of $\hat{\phi}$. While this has an obvious meaning in Euclidean space, a technical diversion is needed to explain what is meant by Taylor expansions more generally. 

Given an origin $z_t \in \fB_t$ about which a particular Taylor expansion is to be performed, the most natural Cartesian-like coordinate systems are the Riemann normal coordinates with origin $z_t$. These are unique up to rotations, and may be used to perform Taylor expansions in the usual way. 

To be more precise, recall that the exponential map $\exp_{x} X^a = x'$ takes as input a point $x$ and a vector $X^a$ at that point. The point $x'$ which is returned is found by considering an affinely-parametrized geodesic $y_u$ satisfying $y_0 = x$ and $\dot{y}_0^a = X^a$. The point $x'$ is then equal to $y_1$. An equivalent statement may be expressed using Synge's function \eqref{Synge}. Letting $\sigma_a (x',x)$ denote $\nabla_a \sigma(x',x)$,
\begin{equation}
    \exp_{x} [-\sigma^a (x',x) ] = x'.
\end{equation}
First derivatives of Synge's function therefore generalize the concept of a ``separation vector.'' The $\bx'-\bx$ of a conventional Taylor series in Cartesian coordinates naturally turns into $-\sigma^a(x',x)$ in more general contexts. If a scalar field $\lambda(x)$ is to be expanded in a Taylor series about some $x$, it is convenient to first rewrite this as a function on the tangent bundle by defining
\begin{equation}
    \Lambda (x,X^a) := \lambda (\exp_{x} X^a).
\end{equation}
Now let the $n$th tensor extension of $\lambda$ at $x$ be
\begin{equation}
    \lambda_{,a_1\cdots a_n}(x) := \left[ \frac{\partial^n \Lambda (x,X^b) }{ \partial X^{a_1} \cdots \partial X^{a_n} } \right]_{X^b = 0}.
    \label{ExtensionDef}
\end{equation}
This is the unique tensor field which reduces to $n$ partial derivatives of $\lambda$ in a Riemann normal normal coordinate system with origin $x$. In flat space, $\lambda_{,a_1\cdots a_n} = \nabla_{a_1} \cdots \nabla_{a_n} \lambda$. More generally, the curvature can appear. Further discussion of tensor extensions may be found in \cite{HarteBrokenSyms, Dix74}.

Combining all of these concepts, a natural Taylor series for $\hat{\phi}$ which applies regardless of the background geometry is 
\begin{equation}
    \hat{\phi} (x',t) = \sum_{n=0}^\infty \frac{(-1)^n}{n!} \sigma^{a_1} (x',z_t) \cdots \sigma^{a_n} (x',z_t) \hat{\phi}_{,a_1 \cdots a_n}(z_t,t).
    \label{MultipoleNewt2}
\end{equation}
All distances are assumed to be sufficiently small that $\sigma$ remains single-valued and its derivative well-defined. Furthermore, a Taylor series like this is --- even if it does not converge everywhere of interest --- assumed to be at least a useful asymptotic approximation throughout $\fB_s$. Substituting \eqref{MultipoleNewt2} into \eqref{GenForceNewt2} and integrating term-by-term then results in a multipole expansion for the generalized force. Noting that
\begin{equation}
    \Lie_\xi \sigma^a = \Lie_\xi (g^{ab} \nabla_b \sigma) = g^{ab} \nabla_b \Lie_\xi \sigma = 0
    \label{LieSigma}
\end{equation}
for any Killing field $\xi^a$, the multipole expansion for the generalized force is
\begin{equation}
    \frac{\rmd}{\rmd t} P_t(\xi) = - \sum_{n=0}^\infty \frac{1}{n!} m^{a_1 \cdots a_n} (z_t,t) \Lie_\xi \hat{\phi}_{, a_1 \cdots a_n} (z_t,t),
    \label{MultipoleForceNewt}
\end{equation}
where the mass moments depend on $\rho$ via
\begin{equation}
    m^{a_1 \cdots a_n}(z_t, t) := (-1)^n \int_{\fB_t} \!\! \sigma^{a_1}(x',z_t) \cdots \sigma^{a_n} (x',z_t) \rho(x',t) \rmd V'.
    \label{MultipoleDef}
\end{equation}
It follows from \eqref{MassCons2} that the zeroth moment, the mass, is independent of time. Conservation laws do not, however, fix the evolution of the higher moments. These depend on the type of matter under consideration.

If $\Lie_\psi \hat{\phi} = 0$ for some Killing field $\psi^a$, it follows that $\Lie_\psi \hat{\phi}_{,a_1 \cdots a_n} = 0$ for any $n$. The conservation  of $P_t(\psi)$ in such a case is therefore preserved by any approximation which truncates the multipole series at finite $n$. This is an important property which contributes to the accuracy of these approximations over long times.

\subsubsection{Linear and angular momenta}
\label{Sect:PtopS}

Thus far, $P_t = P_t[\fB_t]$ has been loosely described as being equivalent to a body's linear and angular momenta at time $t$. Similarly, time derivatives of the generalized momentum have been interpreted as ``forces and torques.'' These identifications are now made precise.

Recall that the generalized momentum is a vector in $K^*$, the vector space dual to the Killing fields $K$. While it is productive to view $P_t$ simply as a linear map from $K$ to $\mathbb{R}$, it can also be useful to find its components with respect to a particular basis. It is in this context that the linear and angular momenta arise in their more familiar form.

A basis for $K$ may be found by recalling that knowledge of a Killing field and its first derivative at any one point fixes it everywhere \cite{Wald}. Choosing an arbitrary point $x$, the space of Killing vectors is in one-to-one correspondence with the space of all 1- and 2-forms at $x$. There exist two-point tensor fields $\Xi^{a'a}(x',x)$, $\Xi^{a' a b}(x',x)$ such that
\begin{equation}
    \xi^{a'} (x') = \Xi^{a' a} (x',x) A_{a} + \Xi^{a' ab}(x',x) B_{ab}
    \label{KillingPropagator}
\end{equation}
is an element of $K$ for any $A_{a}$ and any $B_{ab} = B_{[ab]}$, and also
\begin{equation}
    A_{a} = \xi_{a} (x), \qquad B_{ab} = B_{[ab]} =  \nabla_{a} \xi_{b} (x).
    \label{KillingData}
\end{equation}
In a physical space of dimension $N$, there exist $N$ linearly independent 1-forms and $N(N-1)/2$ linearly independent 2-forms. Together, these generate the requisite $N(N+1)/2$ linearly independent Killing vectors. In Euclidean space and in a Cartesian coordinate system,
\begin{equation}
    \Xi^{i' i} (x',x) = \delta^{i i'}, \qquad \Xi^{i' i j}(x',x) = (\bx'-\bx)^{[i} \delta^{j] i' }.
\end{equation}
More generally, $\Xi^{a'a}$ and $\Xi^{a'ab}$ are related to the geodesic deviation equation and form a basis for $K$. They can be computed using the first two derivatives of Synge's function \cite{Dix70a}. Defining $\sigma_{ab} := \nabla_b \sigma_a = \nabla_b \nabla_a \sigma$, $\sigma_{aa'} := \nabla_{a'} \sigma_a$, and $H^{a'}{}_{a} := [ -\sigma^{a}_{a'} ]^{-1}$,
\begin{equation}
    \Xi^{a'a} = H^{a'}{}_{b} \sigma^{ba} , \qquad \Xi^{a'ab} = H^{a'[a} \sigma^{b]}.
    \label{JacobiPropagators}
\end{equation}

Substituting \eqref{KillingPropagator} into \eqref{Pdef} shows that $P_t(\xi)$ can be written as a linear combination of $\xi_{a}(x)$ and $\nabla_a \xi_b(x)$. The coefficients in this combination are identified with the linear momentum $p^{a}(x,t)$ and the angular momentum bivector $S^{ab} = S^{[ab]}(x,t)$:
\begin{equation}
    P_t(\xi) = p^{a}(x,t) \xi_{a}(x) + \frac{1}{2} S^{ab}(x,t) \nabla_a \xi_b (x).
    \label{Pps}
\end{equation}
This is an implicit definition. Varying amongst all possible $\xi_{a}$ and $\nabla_a \xi_b$ recovers the explicit formulae
\begin{align}
    p^{a}(x,t) = \int_{\fB_t} \!\! \rho(x',t) v_{a'}(x',t) H^{a'}{}_{b}(x',x) \sigma^{ba} (x',x) \rmd V',
    \\
    S^{ab}(x,t) = 2 \int_{\fB_t} \!\! \rho(x',t) v_{a'}(x',t) H^{a'[a}(x',x) \sigma^{b]}(x',x) \rmd V'.
\end{align}
In three spatial dimensions, the angular momentum bivector is dual to an angular momentum 1-form $S_{a}$ via
\begin{equation}
    S_{a} = \frac{1}{2} \epsilon_{abc} S^{bc}.
    \label{SvectNewt}
\end{equation}
Introducing Cartesian coordinates in a flat background, it is easily verified that the $p^i$ and $S_i$ derived from $P_t$ in this way reproduce the elementary definitions \eqref{pDef}. Explicit coordinate expressions are more difficult to obtain in curved backgrounds, but these are rarely necessary.

Thus far, the spatial curvature has played no explicit role in any of our discussion. It does appear, however, in the evolution equations for $p^{a}$ and $S^{ab}$. First note the general identity \cite{Wald}
\begin{equation}
    \nabla_a \nabla_b \xi_c = -R_{bca}{}^{d} \xi_d,
    \label{d2K}
\end{equation}
which holds for any Killing field $\xi^a$. Time derivatives of the Killing data $(A_a, B_{ab})$ along a path $z_t$ therefore satisfy
\begin{equation}
    \frac{\rmD}{\rmd t} A_a = \dot{z}^b_t B_{ba}, \qquad \frac{\rmD}{\rmd t} B_{ab} = - R_{abc}{}^{d} \dot{z}^c_t A_d. 
    \label{KillingDataDot}
\end{equation}
These are known as the Killing transport equations \cite{Wald, Geroch}, and are ordinary differential equations which can be used to relate Killing data at one point to Killing data at another point. 

Consider linear and angular momenta defined about some $z_t$, so, e.g., $p^a = p^a ( z_t, t)$. Substituting \eqref{KillingData} and \eqref{KillingDataDot} into \eqref{Pps} then shows that 
\begin{equation}
     \left( \frac{\rmD p^a}{\rmd t} - \frac{1}{2} R_{bcd}{}^{a} S^{bc} \dot{z}^{d}_t \right) \xi_{a} + \frac{1}{2} \left( \frac{\rmD S^{ab}}{\rmd t} - 2 p^{[a} \dot{z}^{b]}_t  \right)  \nabla_a \xi_b = \frac{\rmd}{\rmd t} P_t (\xi)
    \label{PdotPapapetrou}
\end{equation}
for all $\xi^a \in K$. Varying over all Killing vector fields finally recovers the individual evolution equations
\begin{align}
    \dot{p}^a = \frac{1}{2} R_{bcd}{}^{a} S^{bc} \dot{z}^d_t + F^a, \qquad \dot{S}^{ab}= 2 p^{[a} \dot{z}^{b]}_t + N^{ab},
    \label{pDot}
\end{align}
where the force $F^a$ and torque $N^{ab} = N^{[ab]}$ are determined by matching appropriate coefficients in $\rmd P_t/\rmd t$. \textit{Exact} expressions for these quantities follow from \eqref{GenForceNewt2}, \eqref{KillingPropagator}, and \eqref{JacobiPropagators}: 
\begin{align}
    F_a = - \int_{\fB_t} \!\! \rho H^{a'b} \sigma_{ab} \nabla_{a'} \hat{\phi} \rmd V',
    \label{ForceMultipoleNewtExact}
    \\
    N_{ab} = - 2 \int_{\fB_t} \!\! \rho H^{a'[a} \sigma^{b]} \nabla_{a'} \hat{\phi} \rmd V'.
    \label{TorqueMultipoleNewtExact}
\end{align}
Multipole expansions for the force and torque could be obtained directly from these integrals, although it is simpler to instead start from \eqref{MultipoleForceNewt}. Regardless,
\begin{align}
    F_a = - \sum_{n=0}^\infty \frac{1}{n!} m^{b_1 \cdots b_n} \nabla_a \hat{\phi}_{, b_1 \cdots b_n} ,
    \label{ForceMultipoleNewt}
    \\
    N^{ab} = \sum_{n=0}^\infty \frac{2}{n!} g^{c[a} m^{b] d_1 \cdots d_n} \hat{\phi}_{,c d_1 \cdots d_n}.
    \label{TorqueMultipoleNewt}
\end{align}

Note that the velocity $\dot{z}^a_t$ of the (arbitrarily-chosen) origin does not appear in $F_a$ or $N^{ab}$. Those portions of \eqref{pDot} which do involve the velocity are spatial analogs of the Mathisson-Papapetrou terms typically used to describe the motion of spinning particles in general relativity. It is apparent here that similar terms arise even in non-relativistic theories. Their origin is essentially kinematic, being related  to the decomposition of $K$ into pure translations and pure rotations. It follows from \eqref{Pps} that $p_a(z_t,t)$ is associated with Killing vectors which appear translational at $z_t$ in the sense that $\nabla_a \xi_b (z_t) = 0$. Similarly, $S_{ab}(z_t,t)$ is associated with Killing fields which are purely rotational in the sense that $\xi_a (z_t) = 0$. The Mathisson-Papapetrou terms arise in the laws of motion because, e.g., a Killing vector which is purely translational at $z_t$ is not necessarily purely translational at a neighboring point $z_{t+dt}$. A given Killing field may have different proportions of ``translation'' and ``rotation'' at different points, and this inevitably mixes the evolution equations for $p^a$ and $S^{ab}$. A simple version of this effect occurs even in flat space, where a pure rotation about one origin is not necessarily a pure rotation about another origin. This explains the $p^{[a} \dot{z}^{b]}_t$ term in \eqref{pDot} and the $-\dot{\bz}_t \times \bm{p}$ term in \eqref{sDotNewtOld}.

Essentially the same explanation for the Mathisson-Papapetrou terms applies in general relativity. In that case, the spacetime may not admit any Killing vectors at all. Regardless, there still exists a ten-dimensional space of ``generalized Killing fields'' as described in Section \ref{Sect:GKF}. Given a particular event, these are naturally decomposed into a four-dimensional space of translations and a six-dimensional space of rotations and boosts. Whether or not a particular generalized Killing field is, e.g., purely translational varies from point to point just as it does for ordinary Killing fields. The evolution equations for relativistic momenta therefore acquire velocity-dependent terms which are closely analogous to those which appear in the generalized Newtonian theory discussed here.

Confining attention only to the generalized momentum whenever possible avoids the complications associated with the Mathisson-Papapetrou terms. It also simplifies the discussion of conservation laws. Recall that the presence of a particular spatial Killing field $\psi^a$ which satisfies $\Lie_\psi \hat{\phi} = 0$ implies that $P_t(\psi)$ must be conserved. It follows from \eqref{Pps} that a particular  linear combination of $p_a$ and $S_{ab}$ must be conserved as well:
\begin{equation}
    p^a(z_t,t) \psi_a(z_t) + \frac{1}{2} S^{ab}(z_t,t) \nabla_a \psi_b(z_t) = (\mbox{constant}).
\end{equation}
This constant is independent of $z_t$. Its existence implies that a particular combination of forces and torques must vanish. Specifically, comparison with \eqref{PdotPapapetrou} shows that
\begin{equation}
    F^a \psi_a + \frac{1}{2} N^{ab} \nabla_a \psi_b = 0.
\end{equation}
Although these results could be deduced directly from \eqref{pDot}-\eqref{TorqueMultipoleNewtExact}, they are considerably more clear from the perspective of the generalized momentum and its evolution equation \eqref{GenForceNewt2}.

\subsubsection{Center of mass}

The laws of motion for $p^a$ and $S^{ab}$ have left $z_t$ undetermined. One convenient choice is to set $z_t = \gamma_t$, where $\gamma_t$ denotes the body's center of mass at time $t$.

This is straightforward when the background space is flat, in which case it is standard to define the center of mass to be the origin about which the mass dipole moment vanishes: $m^a( \gamma_t, t) = 0$. Enforcing this while differentiating \eqref{MultipoleDef} recovers the standard relation $p^a = m \dot{\gamma}^a_t$ between an object's velocity and its linear momentum. Using \eqref{pDot} and \eqref{ForceMultipoleNewt} then shows that
\begin{equation}
    \ddot{\gamma}^a_t  = -\nabla^a \hat{\phi}(\gamma_t,t) - \frac{1}{m} \sum_{n=2}^\infty \frac{1}{n!} m^{b_1 \cdots b_n}(\gamma_t,t) \nabla^a \nabla_{b_1} \cdots \nabla_{b_n} \hat{\phi} (\gamma_t,t),
    \label{CMnewt}
\end{equation}
which is equivalent to \eqref{CMNewtAccel}.

Similar results do not appear to hold when the background space is curved. It is still possible to demand that the dipole moment vanish [which, among other benefits, eliminates the $n=1$ term in \eqref{ForceMultipoleNewt} and the $n=0$ term in \eqref{TorqueMultipoleNewt}], but then the velocity of such a trajectory may be shown to satisfy
\begin{equation}
    \dot{\gamma}^{b}_t \int_{\fB_t} \!\! \rho(x',t) \sigma^{a}{}_{b}(x',\gamma_t) \rmd V'  = -\int_{\fB_t} \!\! \rho(x',t) v^{a'}(x',t) \sigma^{a}{}_{a'} (x',\gamma_t) \rmd V'.
    \label{CMProb}
\end{equation}
The integral on the left-hand side of this equation can (typically) be inverted to yield an explicit expression for $\dot{\gamma}^{a}_t$. Unfortunately, the result does not depend on $p^{a}$ in any simple way. Simplifications are possible when a body's dimensions are much smaller than the curvature scale, in which case $\sigma^{a}{}_{b}$ and $\sigma^{a}{}_{a'}$ can be expanded in Taylor series about $\gamma_t$, yielding the ordinary momentum-velocity relation at lowest order. More generally, though, \eqref{CMProb} is problematic. Higher-order corrections require more information about the body than is required for the evolution equations of the momenta alone. Moments of a body's momentum distribution --- its ``current moments'' --- appear to be needed together with its mass moments.

It is only in this very last step where a celestial mechanics of ``curved Newtonian gravity'' appears to be problematic. Similar complications do not arise in relativistic systems. Among other differences, the presence of boost-type Killing fields (or their generalizations) on spacetime provide additional constraints which imply well-behaved momentum-velocity relations. 

\subsubsection{Equations of motion}

Results such as \eqref{pDot} are properly described as laws of motion, not equations of motion \cite{HavasGoldberg, EhlersReview}. They are incomplete in the sense that even if $z_t$ is chosen as a body's center of mass, we still have not described how to compute $\hat{\phi}$ or the higher multipole moments.

The traditional approach is to introduce smallness assumptions. Consider, for simplicity, the $n$-body problem in flat space. If a particular body in such a system has characteristic size $\ell$ and mass $m$, its $2^n$-pole moments must be smaller\footnote{Large astrophysically-relevant objects like planets tend to be very nearly spherical due to the limited shear stresses which can be supported. The trace-free components of the moments, which are all that couple to the motion, are then much smaller than $m \ell^n$. These tend to be induced mainly by rotation and external tidal fields, and are typically modeled using Love numbers.} than approximately $m \ell^n$. Letting $r$ denote a minimum distance between bodies and assuming that all masses are comparable, the $n$th term in \eqref{CMnewt} is at most of order $\frac{1}{n!} (m/r)^2 (\ell/r)^n$. Considerable simplifications therefore result if $\ell \ll r$. At lowest order, only the monopole term is retained in the law of motion. Each $\hat{\phi}$ in such an approximation may also be computed by assuming that all other masses are pure monopoles, thus recovering the typical Newtonian $n$-body equations of motion. More details may be found in, e.g., \cite{DamourReview, Dix79}.

\section{An introduction to relativistic motion}
\label{Sect:SR}

Despite being considerably more abstract than the traditional presentation of Newtonian gravity, the formalism which has just been described is very powerful. It does not rely on any particular coordinate systems, and the majority of the discussion doesn't even require that the metric be flat. Indeed, most of the results well-known in ordinary Newtonian gravity continue to hold in generalizations of this theory which employ spherical or hyperboloidal geometries. It is also trivial to change the number of spatial dimensions, or even to amend the field equation in certain ways. It is physically more interesting, however, to consider motion in relativistic theories such as electromagnetism or general relativity.

This section describes how the formalism of Section \ref{Sect:NewtonianNew} generalizes for objects coupled to relativistic fields. For simplicity, we consider the motion of an extended mass coupled to a scalar field $\phi$ which satisfies the Klein-Gordon equation
\begin{equation}
    (\nabla^a \nabla_a - \mu^2) \phi = 4 \pi \rho.
    \label{phiField}
\end{equation}
$\mu$ represents a (constant) field mass and $\rho$ the body's charge density. Following the Newtonian problem as closely as possible, the four-dimensional background spacetime $(\mathcal{M},g_{ab})$ is assumed to be maximally-symmetric.
Understanding motion in more general curved spacetimes requires eliminating our reliance on a maximal set of Killing vector fields. This is indeed possible, but is somewhat technical. Its discussion is therefore delayed to Section \ref{Sect:Curved} below. Motion in electromagnetic fields is discussed there as well.

Scalar charges in maximally-symmetric spacetimes provide a simple example with which to introduce the relativistic theory of motion. They differ from their Newtonian counterparts in only one important respect: Self-forces no longer vanish. Still, self-forces are ``almost ignorable'' in the sense that they effectively renormalize a body's momentum, but do nothing else\footnote{The notion of self-force used here is consistent with the usual Newtonian definition, but is unconventional in relativistic contexts. Its precise meaning is made clear below.}. This is a finite renormalization, meaning that self-forces conspire to, e.g., make the mass computed by integrating over a body's stress-energy tensor differ from the mass inferred by watching how that body accelerates in response to external fields.

Physically, renormalization arises because as a charge accelerates, its field must be accelerated as well. Although portions of that field may break away as radiation or otherwise change, there is a sense in which charges and their fields remain inseparably coupled. The energy contained in a body's self-field implies that it must resist acceleration and contribute to that body's inertia. 

Now, self-forces vanish in Newtonian theory because of Newton's third law. The self-field is sourced by a body's instantaneous mass distribution and exerts forces on that same mass distribution. Interactions are no longer instantaneous, however, in theories which involve hyperbolic field equations. Fields are sourced by charge in a four-dimensional region of spacetime, but act only on configurations in three-dimensional hypersurfaces. It is impossible to maintain an exact concept of ``action-reaction pairs'' in this context, and the imbalance which results turns out to exert forces and torques which precisely mimic changes to a body's linear and angular momenta. This type of effect is generic for any coupling to long-range fields which satisfies hyperbolic field equations (or otherwise depends on a system's history).

\subsection{Relativistic continuum mechanics}
\label{Sect:ScalarSR}

The simplest relativistic modification of Newtonian gravity\footnote{This is to be considered as a model problem. If interpreted as a theory of gravity, the type of scalar field theory described here is not compatible with observations. Of course, it is not necessary to interpret $\phi$ as a gravitational potential (so $\rho$ needn't be a ``mass density'' in any sense).} involves objects with scalar charge density $\rho$ interacting via a field $\phi$ which satisfies the wave equation \eqref{phiField}. Suppose that the body of interest resides inside a worldtube $\mathfrak{W} \subset \mathcal{M}$ containing no other matter, and also that $\mathfrak{W}$ is spatially bounded in the sense that its spacelike slices have finite volume.

Our description for the motion of a compact object is based on its stress-energy tensor $T^{ab}_\mathrm{body}$. This encodes many of a body's mechanical properties, and is analogous to the $(\rho, v^a, \tau^{ab})$ triplet used to analyze Newtonian objects in Section \ref{Sect:Newton}. Like those variables, the stress-energy satisfies differential equations which are independent of the specific type of material under consideration. Although these laws do not determine $T^{ab}_\mathrm{body}$ completely, they do provide significant constraints.

If $\phi$ vanishes everywhere and there are no other long-range fields, $\nabla_b T^{ab}_\mathrm{body} = 0$. More generally, scalar fields contribute to a system's total stress-energy. It is only this total $T_\mathrm{tot}^{ab} = T_\mathrm{tot}^{(ab)}$ which is necessarily\footnote{In a Lagrangian formalism, the total stress-energy tensor considered here is derived from a functional derivative of the action with respect to the metric. It is conserved whenever the action is diffeomorphism-invariant \cite{Wald}.} conserved:
\begin{equation}
    \nabla_b T^{ab}_\mathrm{tot} = 0.
    \label{StressCons}
\end{equation}
Consider splitting this total into ``body'' and ``field'' components:
\begin{equation}
    T^{ab}_\mathrm{tot} = T^{ab}_\mathrm{field} + T^{ab}_\mathrm{body}.
\end{equation}
Away from any matter, it is clear that $T^{ab}_\mathrm{body} = 0$ and
\begin{equation}
    T_\mathrm{field}^{ab} = \frac{1}{4\pi} \left[ \nabla^a \phi \nabla^b \phi - \frac{1}{2} g^{ab} ( \nabla_c \phi \nabla^c \phi + \mu^2 \phi^2) \right] .
    \label{Tfield}
\end{equation}
Elsewhere, local interactions between the matter and the field make it physically difficult to motivate any particular split.

One possible way forward is to work only with $T^{ab}_\mathrm{tot}$. Unfortunately, the momentum obtained from this stress-energy tensor might be very different if computed first in a slice of $\mathfrak{W}$, and then in a slightly larger hypersurface. There is no natural boundary where integrations can be stopped. Although momentum integrals might settle down when performed over very large volumes, it is unclear how useful this is. The relevant distance scale could be so large that the only ``total momenta'' which are interesting encompass the entire (modeled) universe, thus precluding any ability to learn about the dynamics of individual masses. Results based on $T^{ab}_\mathrm{tot}$ alone are known to be useful in certain approximations involving the motion of very small bodies \cite{GrallaHarteWaldEM}, but this is insufficiently general for our purposes.

The approach adopted here is mathematically the simplest. Let $T^{ab}_\mathrm{field}$ be given by \eqref{Tfield} throughout $\mathfrak{W}$. The remaining stress-energy tensor is then defined to be the body's: $T^{ab}_\mathrm{body} = T^{(ab)}_\mathrm{body} := T^{ab}_\mathrm{tot} - T^{ab}_\mathrm{field}$. Equations \eqref{phiField}-\eqref{Tfield} imply that this satisfies
\begin{equation}
    \nabla_b T^{ab}_\mathrm{body} = - \rho \nabla^a \phi,
    \label{StressConsScalar}
\end{equation}
which generalizes the Newtonian conservation laws \eqref{MassCons2} and \eqref{CauchyMomentum2}.

\subsection{Generalized momentum}

Recall the generalized momentum \eqref{Pdef} defined for Newtonian mass distributions. This requires very little modification for use in relativistic systems. The one complication which does arise is that there is no longer any preferred notion of time. A time parameter must be supplied as an additional ingredient, which is accomplished by foliating $\mathfrak{W}$ with a 1-parameter family of hypersurfaces $\{ \fB_s \}$. Each $\fB_s$ may be viewed as the body region at time $s$, and is assumed to have finite volume. The precise nature of these body regions may be considered arbitrary for now, and can be spacelike or even null\footnote{Consider, e.g., the past-directed light cones associated with a timelike worldline.}.

Supposing that a particular foliation has been fixed, the generalized momentum contained in any three-dimensional region $\mathfrak{R} \subseteq \fB_s$ is most obviously defined as
\begin{equation}
    P_s [\mathfrak{R}] (\xi) = \int_{\mathfrak{R}} T^{ab}_\mathrm{body} \xi_a \rmd S_b,
    \label{PdefRel}
\end{equation}
where $\xi^a$ is any Killing vector field. $P_s [\mathfrak{R}] (\cdot)$ defines a linear map on the space of $K$ of Killing vector fields. It is therefore a vector in the dual space $K^*$. For the maximally-symmetric four-dimensional spacetimes considered here, $\dim K = \dim K^* = 10$. Given a particular event, four of these dimensions correspond to translations and six to rotations and boosts. As in the Newtonian case, such decompositions allow the generalized momentum to be expressed in a basis which recovers linear and angular momenta associated with a preferred event. The details of this correspondence are described more precisely in Section \ref{Sect:PtopsScalar}.

\subsection{Generalized force}

Forces and torques are determined by $s$-derivatives of the generalized momentum. Considering only the momenta in $\fB_s$, it is convenient to simplify the notation by defining $P_s = P_s [\fB_s]$. Using \eqref{StressConsScalar} together with Killing's equation then shows that
\begin{equation}
    \frac{\rmd}{\rmd s} P_s (\xi) = - \int_{\fB_s} \!\! \rho \Lie_\xi \phi \rmd S
    \label{ForceScalar}
\end{equation}
for all $\xi^a \in K$, where $\rmd S := t^a \rmd S_a$ and $t^a$ denotes a time evolution vector field for the foliation $\{ \fB_s \}$. The relativistic generalized force \eqref{ForceScalar} is essentially identical to its Newtonian analog \eqref{GenForceNewt}. As in that context, the force can be immediately put into a practical form only if the object of interest does not significantly contribute to $\phi$. More generally, the self-field introduces considerable complications. Progress is made by finding a precise definition for the self-field, computing its effects analytically, and then subtracting it out. The ``effective field'' which remains after this process is typically much simpler to analyze than the physical one. 

\subsection{The self-field}
\label{Sect:SelfFieldSR}

At least in part, the Newtonian self-field \eqref{SelfFieldNewt} can be generalized essentially as-is. Let the relativistic self-field $\phiS$ be obtained by convolving an object's charge density with a particular two-point\footnote{It is also possible to introduce $(n+1)$-point self-fields similar to \eqref{PhiSGen}. This is not considered any further here.} scalar $G$. This must be a Green function, so
\begin{equation}
    (\nabla^a \nabla_a - \mu^2) G (x,x') = 4 \pi \delta(x,x'). 
    \label{GDefScalar}
\end{equation}
Still more constraints are necessary to fix $G$ uniquely. One of these follows from requiring that the self-field\footnote{The term self-field is used in several different ways in the literature. The definition adopted here is uncommon, and is sometimes described as the ``Coulomb-like'' component of the self-field.} depend only quasi-locally on a body's ``instantaneous'' configuration. It should not, for example, involve distantly-imposed boundary conditions, the behavior of other objects, or a body's history in the distant past. Such conditions can be ensured by demanding that $G(x,x') = 0$ whenever $x$ and $x'$ are timelike-separated. Lastly, suppose that $G(x,x') = G(x',x)$. Such objects exist (at least in finite regions), are unique, and are referred to as S-type or ``singular'' Detweiler-Whiting Green functions \cite{PoissonRev, DetWhiting}. In the maximally-symmetric backgrounds considered here, $G$ satisfies $\Lie_\xi G = 0$ for all $\xi^a \in K$, implying a relativistic form of Newton's third law. For massless fields in Minkowski spacetime, $G = \frac{1}{2} (G_+ + G_-)$ where $G_\pm$ are the advanced and retarded Green functions. More generally, 
\begin{equation}
    G = \frac{1}{2} ( G_+ + G_- - V)
    \label{Gexpand}
\end{equation}
for a certain symmetric biscalar $V(x,x')$ which satisfies the homogeneous field equation. $G$ can also be expressed in terms of Synge's function $\sigma$. Using $\Delta$ to denote the van Vleck determinant \cite{PoissonRev} (which depends on second derivatives of $\sigma$), $\delta$ the Dirac distribution, and $\Theta$ the Heaviside distribution,
\begin{equation}
    G = \frac{1}{2} [ \Delta^{1/2} \delta(\sigma) - V \Theta(\sigma)].
    \label{GHadamard}
\end{equation}
This shows that $G(x, \cdot)$ can have support on and \textit{outside} the light cones of $x$.

The self-field ``due to'' charge contained in a given spacetime volume $\mathfrak{R} \subseteq \mathfrak{W}$ can now be expressed in terms of the S-type Detweiler-Whiting Green function:
\begin{equation}
    \phiS[\mathfrak{R}](x) = \int_\mathfrak{R} \! G(x,x') \rho(x') \rmd V'.
    \label{SelfFieldScalar}
\end{equation}
If the argument $\mathfrak{R}$ is omitted, the integral is understood to be carried out over an object's entire worldtube $\mathfrak{W}$.

\begin{figure}
\begin{center}
\includegraphics[height=3.5cm]{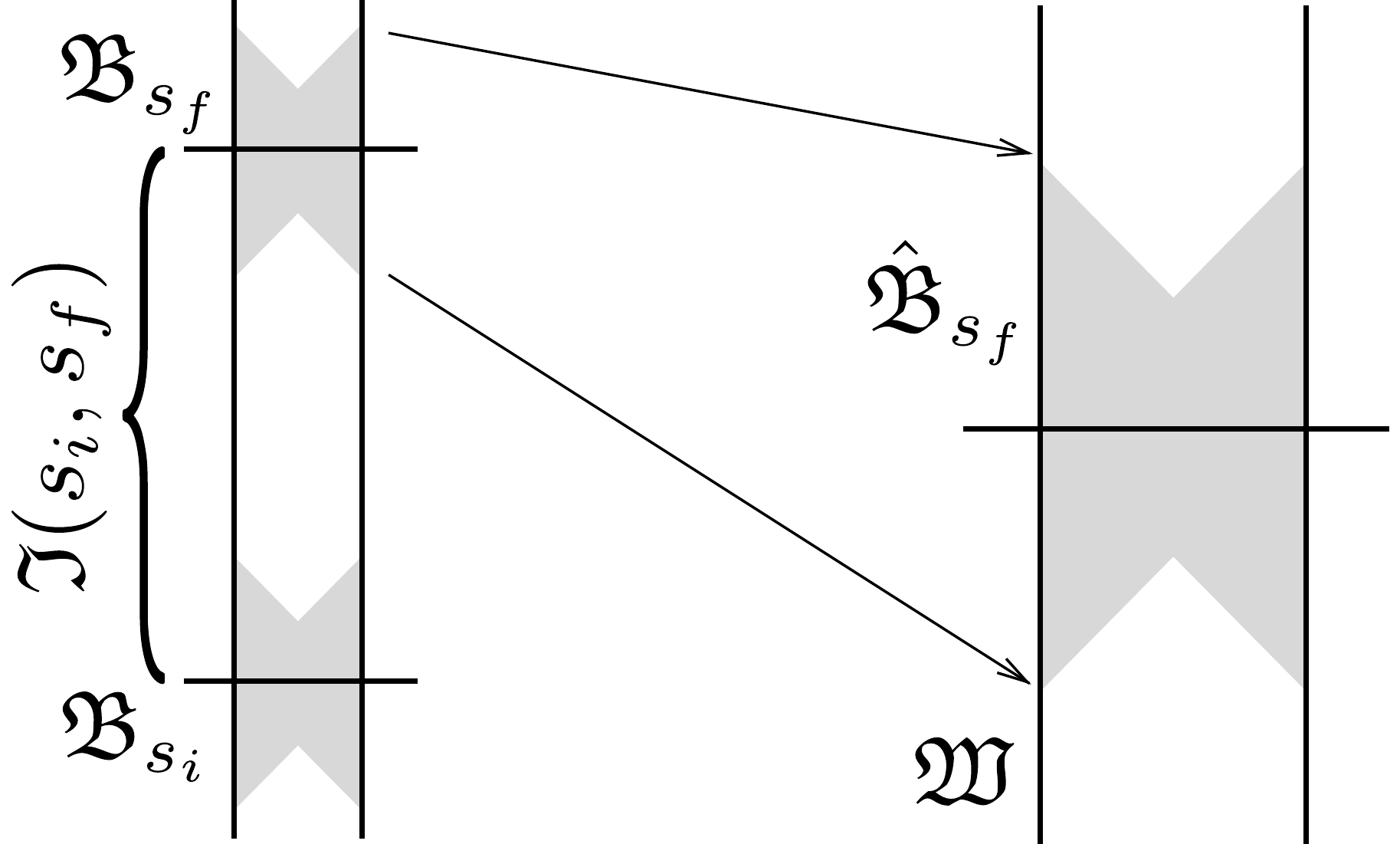}
\end{center}
\caption{Schematic illustrations of a body's worldtube $\fW$ together with hypersurfaces $\fB_{s_i}$ and $\fB_{s_f}$ (drawn spacelike). The region $\mathcal{I}(s_i, s_f) \subset \fW$ bounded by these hypersurfaces and appearing in \eqref{Pdiff} is indicated. The shaded 4-volumes $\hat{\fB}_{s_i}$ and $\hat{\fB}_{s_f}$ [see \eqref{BHatDef}] denote the domains of dependence associated with the self-momenta $E_{s_i}$ and $E_{s_f}$ defined by \eqref{Edef}. Although $P_{s_f}(\xi) - P_{s_i}(\xi)$ depends on $\rho \Lie_\xi \hat{\phi}$ throughout $\mathfrak{I}(s_i, s_f)$, it depends on more complicated aspects of a body's internal structure only in the shaded regions. These contributions are always confined to within approximately one light-crossing time of the bounding hypersurfaces, and are therefore ``quasi-local.''}
\label{Fig:Renorm}
\end{figure}

We now compute the self-force. It simpler not to consider this directly, but rather its integral over a finite interval of time. Letting $s_\mathrm{f} > s_\mathrm{i}$, it is clear from \eqref{ForceScalar} that
\begin{equation}
    P_{s_\mathrm{f}} (\xi) - P_{s_\mathrm{i}} (\xi) = \int_{ s_\mathrm{i} }^{ s_\mathrm{f} } \frac{\rmd}{\rmd s} P_s(\xi) \rmd s  = -\int_{ \mathfrak{I} } \rho \Lie_\xi \phi \rmd V
    \label{Pdiff}
\end{equation}
for any $\xi^a \in K$. The 4-volume $\mathfrak{I} = \mathfrak{I}(s_\mathrm{i}, s_\mathrm{f}) \subset \mathfrak{W}$ which appears here represents that part of an object's worldtube which lies in between the initial and final hypersurfaces $\fB_{s_\mathrm{i}}$, $\fB_{s_\mathrm{f}}$. See Figure \ref{Fig:Renorm}. Substitution of the self-field definition \eqref{SelfFieldScalar} into \eqref{Pdiff} shows that the total change in momentum due to $\phiS$ alone is
\begin{equation}
    \int_\mathfrak{I} \rmd V \int_\mathfrak{W} \rmd V' f(x,x'),
    \label{calFdef}
\end{equation}
where
\begin{equation}
    f(x,x') = -\rho(x) \rho(x') \xi^a(x) \nabla_a G(x,x')
    \label{Fscalar}
\end{equation}
may be interpreted as the density of generalized force exerted at $x$ by $x'$. Independently of any specific form for $f$, double integrals with the form \eqref{calFdef} can be rewritten as
\begin{equation}
    \frac{1}{2} \int_\mathfrak{I} \rmd V \left( \int_\fW \rmd V' [f(x,x') + f(x',x)] + \! \! \int_{\fW \setminus \mathfrak{I}} \! \! \rmd V'  [f(x,x') - f(x',x)] \right) 
    \label{ForceAv}
\end{equation}
whenever the relevant integrals commute. This identity is very general, and is central to understanding motion in every relativistic theory we discuss. It is therefore worthwhile to examine it in detail.

The first term in \eqref{ForceAv} can be interpreted as an average of ``action-reaction pairs'' in the sense of Newton's third law. It is very similar to the types of identities used to simplify the motion of objects coupled to elliptic fields in Section \ref{Sect:Newton}. Recalling that discussion, the reciprocity relation $G(x,x') = G(x',x)$ implies that
\begin{equation}
    \int_\mathfrak{I} \rmd V \! \int_\fW \! \rmd V' [f(x,x') + f(x',x)] = - \int_\mathfrak{I} \rmd V \! \int_\fW \! \rmd V' \rho(x) \rho(x') \Lie_\xi G(x,x') .
    \label{force1}
\end{equation}
Lie derivatives of $G$ are again associated with sums over action-reaction pairs, and as in the Newtonian case, these sums vanish. The relativistic scalar self-force is therefore determined only by the second group of terms in \eqref{ForceAv}. Those terms do not vanish in general, but are instead connected to the finite speed of propagation associated with $\phi$. They are responsible for renormalizing a body's momentum.

\subsection{Renormalization}
\label{Sect:Renormalize}

The generalized force exerted by $\phiS$ is entirely determined by the final part of \eqref{ForceAv}. To understand this, first let $\fB^+_s$ (resp. $\fB^-_s$) denote the four-dimensional future (past) of $\fB_s$ inside the body's worldtube:
\begin{equation}
    \fB^\pm_s (s) := \bigcup_{ \pm (\tau -s ) > 0 } \fB_\tau .
\end{equation}
Also define
\begin{equation}
    E_s(\xi) := \frac{1}{2} \left( \int_{\fB^+_s} \! \! \rho \Lie_\xi \phiS[\fB^-_s] \rmd V - \int_{\fB^-_s} \! \! \rho \Lie_\xi \phiS[\fB^+_s] \rmd V  \right) .
    \label{Edef}
\end{equation}
Like $P_s$, this represents an $s$-dependent vector in $K^*$. Using it, the second term in the expansion \eqref{ForceAv} for the self-force is simply
\begin{equation}
  \frac{1}{2} \int_\mathfrak{I} \rmd V \! \! \int_{\fW \setminus \mathfrak{I}} \! \! \rmd V'  [f(x,x') - f(x',x)] = - [ E_{s_\mathrm{f}} ( \xi) - E_{s_\mathrm{i}} (\xi)]  . 
  \label{force2}
\end{equation}
Taking the limit $s_\mathrm{f} \rightarrow s_\mathrm{i}$ while combining \eqref{Pdiff}, \eqref{ForceAv}, \eqref{force1}, and \eqref{force2} finally shows that the generalized force can be written as
\begin{equation}
    \frac{\rmd}{\rmd s} P_s(\xi) = - \int_{\fB_s} \!\! \rho \Lie_\xi \hat{\phi} \rmd S - \frac{\rmd}{\rmd s} E_s(\xi).
    \label{ScalarForceSR}
\end{equation}
Replacing the physical field $\phi$ with $\hat{\phi} = \phi - \phiS$ can therefore be accomplished only at the cost of the counterterm $-\rmd E_s/\rmd s$. That this is a total derivative suggests the introduction of an ``effective generalized momentum'' $\hat{P}_s$ satisfying
\begin{equation}
    \hat{P}_s := P_s + E_s.
    \label{Phat}
\end{equation}
For any finite scalar charge in a maximally-symmetric spacetime,
\begin{equation}
    \frac{\rmd}{\rmd s} \hat{P}_s(\xi) = - \int_{\fB_s} \!\! \rho \Lie_\xi \hat{\phi} \rmd S .
    \label{EOMScalar}
\end{equation}
There is a sense in which $E_s$ renormalizes the (bare) momentum $P_s$. The sum of $P_s$ and $E_s$ behaves instantaneously as though it were the momentum of a test charge placed in the effective field $\hat{\phi}$. Furthermore, 
\begin{equation}
  (\nabla^a \nabla_a - \mu^2) \hat{\phi} = 0.
\end{equation}

Physically, it is not sufficient to motivate the renormalization $P_s \rightarrow \hat{P}_s$ merely by fact that the self-force is a total derivative. Essentially \textit{any} function of one variable can be written as the total derivative of its integral. Indeed, one might introduce a constant $s_0$ and define
\begin{equation}
    \tilde{P}_s(\xi) := P_s(\xi) + E_s(\xi) + \int^s_{s_0} \rmd \tau \int_{\fB_\tau} \! \! \rmd S \rho \Lie_\xi \hat{\phi} .
\end{equation}
This does not vary at all with $s$. While it may be useful for some purposes, $\tilde{P}_s$ is not a physically acceptable momentum. This is because it depends in an essential way on the configuration of the system for all times between $s_0$ and $s$. While $\tilde{P}_s$ would be approximately local for $s \approx s_0$, it otherwise depends on a system's history in a complicated way. 

The renormalized momentum $\hat{P}_s$ defined by \eqref{Phat} does not share this deficiency. Like $P_s$, it depends only on the body's configuration in regions ``near'' $\fB_s$. The relevant region is, however, somewhat larger for $\hat{P}_s$ than it is for $P_s$. Definitions \eqref{SelfFieldScalar} and \eqref{Edef} imply that $E_s(\xi)$ depends on a neighborhood $\hat{\fB}_s \supset \fB_s$ defined to be the set of all points in $\fW$ which are null- or spacelike-separated to at least one point in $\fB_s$. In terms of Synge's function,
\begin{equation}
  \hat{\fB}_s = \{ x \in \fW \mid \mbox{$\sigma(x,y) \geq 0$ for some $y \in \fB_s$}  \} .
  \label{BHatDef}
\end{equation}
$\hat{\fB}_s$ is a (finite) four-dimensional region of spacetime. It extends into the past and future of $\fB_s$ by roughly the body's light-crossing time. See Figure \ref{Fig:Renorm}.

One might have guessed that a self-momentum at time $s$ could be defined by integrating the stress-energy tensor associated with\footnote{Recalling \eqref{Tfield}, $T^{ab}_\mathrm{field}$ is quadratic in $\phi$. The stress-energy tensor ``associated with $\phiS$'' is taken to mean that portion of $T^{ab}_\mathrm{field}$ which is quadratic in $\phiS$. Terms linear in $\phiS$ are not included.} $\phiS$ over a large hypersurface which contains $\fB_s$. Unfortunately, such integrals depend on gradients of $\phiS$ far outside of $\fB_s$. These, in turn, depend on the body's state in the distant past and future. Such a definition is physically unacceptable in general. Nevertheless, it does make sense in the stationary limit, and may be shown to coincide with $E_s$ in that case \cite{HarteScalar}. In more dynamical cases, the $E_s$ defined here appears to be the only well-motivated possibility.

\subsection{Multipole expansions}
\label{Sect:MultipoleScalarSR}

Forces and torques exerted on relativistic scalar charges may be expanded exactly as in the Newtonian theory. Assuming that $\hat{\phi}$ can be accurately approximated using a Taylor series about some origin $z_s \in \fB_s$, the techniques of Section \ref{Sect:multipoleNewt} may be used to show that \eqref{EOMScalar} admits the multipole expansion
\begin{equation}
    \frac{\rmd}{\rmd s} \hat{P}_s (\xi) = - \sum_{n=0}^\infty \frac{1}{n!} q^{a_1 \cdots a_n}(z_s, s) \Lie_\xi \hat{\phi}_{,a_1 \cdots a_n}(z_s).
    \label{EOMscalarMultipole}
\end{equation}
The $2^n$-pole moment of $\rho$ which appears here is defined by
\begin{equation}
    q^{a_1 \cdots a_n}(s) := (-1)^n \int_{\fB_s} \! \! \sigma^{a_1}(x,z_s) \cdots \sigma^{a_n}(x,z_s) \rho(x) \rmd S,
    \label{ScalarMultipoleRel}
\end{equation}
and $\hat{\phi}_{,a_1 \cdots a_n}$ denotes the $n$th tensor extension of $\hat{\phi}$. Equation \eqref{EOMscalarMultipole} may be compared with the Newtonian generalized force \eqref{MultipoleForceNewt}. Unlike its Newtonian counterpart, however, the relativistic scalar monopole moment $q$ may depend on time; the total charge is not necessarily conserved. Also note that the relativistic multipole expansion is intended only to be asymptotic. It may require truncation at large $n$ (see, e.g., \cite{Dix67}). 

\subsection{Linear and angular momenta}
\label{Sect:PtopsScalar}

Like $P_s$, the effective generalized momentum $\hat{P}_s$ is an element of $K^*$. Expanding this in an appropriate basis recovers objects which may be interpreted as a body's linear and angular momenta. The appropriate arguments are almost identical to those described in Section \ref{Sect:PtopS}.

Choosing a point $z_s \in \fB_s$, every Killing field may be written as a linear combination of 1- and 2-forms at $z_s$ [cf. \eqref{KillingPropagator}]. $P_s(\xi)$ and $\hat{P}_s(\xi)$ are clearly linear in $\xi^a$, so they too may be expanded in linear combinations of 1- and 2-forms at $z_s$. Recalling \eqref{Pps}, the coefficients in this combination may be identified as a body's linear and angular momentum:
\begin{align}
    \hat{P}_s(\xi)  = \hat{p}^a(z_s,s) \xi_a(z_s) + \frac{1}{2} \hat{S}^{ab}(z_s,s) \nabla_a \xi_b(z_s).
    \label{PtopSscalar}
\end{align}
Hats have been placed on $\hat{p}^a$ and $\hat{S}^{ab}$ to emphasize that these are renormalized momenta. Bare quantities defined in terms of $P_s$ may be introduced as well and shown to coincide with the momenta introduced by Dixon \cite{Dix74,Dix79,Dix70a} for objects without electromagnetic charge-currents. The bare momenta obey more complicated laws of motion, and are not considered any further.

Differentiating \eqref{PtopSscalar} while using \eqref{d2K} leads to an implicit evolution equation for $\hat{p}^a$ and $\hat{S}^{ab}$:
\begin{equation}
     \bigg( \frac{\rmD \hat{p}^a}{\rmd s} - \frac{1}{2} R_{bcd}{}^{a} \hat{S}^{bc} \dot{z}^{d}_s \bigg) \xi_{a} + \frac{1}{2} \bigg( \frac{\rmD \hat{S}^{ab}}{\rmd s} - 2 \hat{p}^{[a} \dot{z}^{b]}_s  \bigg)  \nabla_a \xi_b = \frac{\rmd}{\rmd s} \hat{P}_s (\xi).
     \label{PdotPapapetrouScalar}
\end{equation}
Varying over all $\xi_a$ and all $\nabla_a \xi_b = \nabla_{[a} \xi_{b]}$ recovers the explicit equations
\begin{align}
    \frac{\rmD \hat{p}^a}{\rmd s} = \frac{1}{2} R_{bcd}{}^{a} \hat{S}^{bc} \dot{z}^d + \hat{F}^a, \qquad
    \frac{\rmD \hat{S}^{ab}}{\rmd s} = 2 \hat{p}^{[a} \dot{z}^{b]} + \hat{N}^{ab}. 
    \label{pDotScalar}
\end{align}
The force $\hat{F}^a$ and torque $\hat{N}^{ab} = \hat{N}^{[ab]}$ appearing here may be found in integral form by comparing \eqref{EOMScalar} and \eqref{PdotPapapetrouScalar}. Using the multipole expansion \eqref{EOMscalarMultipole} instead,
\begin{align}
    \hat{F}_a = - \sum_{n=0}^\infty \frac{1}{n!} q^{b_1 \cdots b_n} \nabla_a \hat{\phi}_{, b_1 \cdots b_n} ,
    \label{ForceMultipoleScalar}
    \\
    \hat{N}^{ab} = \sum_{n=0}^\infty \frac{2}{n!} g^{c[a} q^{b] d_1 \cdots d_n} \hat{\phi}_{,c d_1 \cdots d_n}.
    \label{TorqueMultipoleScalar}
\end{align}
The laws of motion \eqref{pDotScalar}-\eqref{TorqueMultipoleScalar} describe bulk features of essentially arbitrary self-interacting scalar charge distributions in maximally-symmetric backgrounds. As in the Newtonian case, all explicit dependence on $\dot{z}^a_s$ is contained in the Mathisson-Papapetrou terms. These terms are kinematic, and may again be traced to the changing character of Killing fields evaluated about different origins.

All conservation laws discussed in the Newtonian context generalize immediately. If $\Lie_\psi \hat{\phi} = 0$ for some particular Killing field $\psi^a$,
\begin{equation}
    \hat{P}_s(\psi) = \hat{p}^a \psi_a + \frac{1}{2} \hat{S}^{ab} \nabla_a \psi_b = (\mbox{constant}).
\end{equation}
Similarly, 
\begin{equation}
    \hat{F}^a \psi_a + \frac{1}{2} \hat{N}^{ab} \nabla_a \psi_b = 0.
    \label{ForceSymmetry}
\end{equation}
These results are exact. They hold independently of any choices made for $z_s$, and also apply to approximate momenta evolved via any consistent truncation of the multipole series. 

Although the relativistic multipole expansions are structurally almost identical to their Newtonian counterparts, it is important to emphasize that the effective field is far more difficult to compute in relativistic contexts. In Newtonian gravity, $\hat{\phi}$ is simply the external potential and is easily computed given the instantaneous external mass distribution of the universe. The relativistic effective potential can, however, depend in complicated ways on boundary conditions, initial data, and past history. Using retarded boundary conditions, the relativistic $\hat{\phi}$ typically depends on $\rho$, and is therefore not interpretable as a purely external field. 

Another property of the relativistic theory is that the angular momentum has six independent components rather than three. This can be interpreted by introducing a local frame at $z_s$ in the form of a unit timelike vector $u^a$. Such a vector allows $\hat{S}^{ab}$ to be decomposed into two components $\hat{S}^a$ and $\hat{m}^a$ satisfying
\begin{equation}
    \hat{S}_{ab} = \epsilon_{abcd} u^c \hat{S}^d - 2 u_{[a} \hat{m}_{b]}
    \label{Sdecomp}
\end{equation}
and $u_a \hat{S}^a = u_a \hat{m}^a = 0$. Writing out $\hat{S}^{ab}$ explicitly in flat spacetime in the limit of negligible self-interaction suggests that $\hat{S}^a$ represents a body's ``ordinary'' angular momentum about $z_s$. Similarly, $\hat{m}^a$ may be interpreted as the dipole moment of a body's energy distribution. Relativistically, these are two aspects of the same physical structure. The split $\hat{S}^{ab} \rightarrow (\hat{S}^a , \hat{m}^a)$ is closely analogous to the decomposition $F_{ab} \rightarrow (E_a, B_a)$ of an electromagnetic field into its electric and magnetic components. 

\subsection{Center of mass}
\label{Sect:CMSR}

Thus far, the foliation $\{ \fB_s \}$ of $\fW$ used to define the generalized momentum has been left unspecified. The collection of events $\{ z_s \}$ used to perform the multipole expansion \eqref{EOMscalarMultipole} has been arbitrary as well. This constitutes a considerable amount of freedom. 

One simplifying strategy is to first associate a hypersurface with each possible point in $\mathfrak{W}$. This could be accomplished by, e.g, defining $\fB_s$ to be the past- (or future-) null cone with vertex $z_s$. A timelike worldline parametrized by $\{z_s\}$ then results in a null foliation of $\mathfrak{W}$. Alternatively, a spacelike foliation may be chosen as described in \cite{Dix79, Dix70a, EhlersRudolph}.

Regardless, defining each $\fB_s$ in terms of $z_s$ reduces all freedom in the law of motion to the choice of a single worldline (and its parametrization). Recall that in Newtonian gravity, a body's center of mass is the location about which its mass dipole moment vanishes. Relativistically, the dipole moment of a body's stress-energy tensor is proportional to $\hat{S}^{ab}(z_s,s)$. In general, there is no choice of $z_s$ which can be used to make this vanish entirely. It is, however, possible to use translations to set $\hat{m}^a = 0$ as defined in \eqref{Sdecomp}. This requires the introduction of a frame with which to choose an appropriate dipole moment. Consider the zero-momentum frame $u^a$ where 
\begin{equation}
    \hat{p}^a = \hat{m} u^a.
    \label{uDef}
\end{equation}
$u^a$ is defined to be a unit vector, so the rest mass $\hat{m}$ must satisfy
\begin{equation}
  \hat{m} := \sqrt{ - \hat{p}^a \hat{p}_a}.
  \label{mDef}
\end{equation}
A center of mass $\gamma_s$ may now be defined by demanding that
\begin{equation}
    \hat{S}^{ab}(\gamma_s,s) \hat{p}_a (\gamma_s,s) = 0.
    \label{CM}
\end{equation}
This can be interpreted as requiring that the dipole moment of a body's energy distribution vanish as seen by a zero-momentum observer at $\gamma_s$. It is a highly implicit definition. Good existence and uniqueness results are known for the closely-related Dixon momenta \cite{Schattner1, Schattner2}, but not more generally. We nevertheless assume that a unique worldline (and associated foliation) can be found found in this way. Other choices are also possible, however.

A general relation between the center of mass 4-velocity and the linear momentum may be found by differentiating \eqref{CM}. The result of this differentiation is solved explicitly for $\dot{\gamma}_s^a$ in \cite{Dix79, EhlersRudolph}, resulting in\footnote{References \cite{Dix79, EhlersRudolph} derive the momentum-velocity relation using Dixon's momenta without a scalar field, but in a spacetime which is not maximally-symmetric. Here, Dixon's momenta are modified by $E_s$, there is a scalar field, and the spacetime is maximally-symmetric. Despite these differences, the relevant tensor manipulations are identical.}
\begin{equation}
    \hat{m} \dot{\gamma}^a_s = \hat{p}^a - \hat{N}^{a}{}_{b} u^b - \frac{ \hat{S}^{ab} [ \hat{m} \hat{F}_b - \frac{1}{2} (\hat{p}^c - \hat{N}^{c}{}_{h} u^h ) \hat{S}^{df} R_{bcdf} ] }{ \hat{m}^2 + \frac{1}{4} \hat{S}^{pq} \hat{S}^{rs} R_{pqrs} }.
    \label{pCM}
\end{equation}
This assumes that the parameter $s$ has been chosen such that $\dot{\gamma}_s^a \hat{p}_a = -\hat{m}$, and also that all instances of $z_s$ have been replaced with $\gamma_s$. In principle, it is possible for the denominator $\hat{m}^2 + \frac{1}{4} \hat{S}^{pq} \hat{S}^{rs} R_{pqrs}$ here to vanish, indicating a breakdown of the center of mass condition. This can occur only if the curvature scale is comparable to a body's own size, in which case it is unlikely that any simple description of an extended body in terms of its center of mass is likely to be useful. In more typical cases, it is straightforward to obtain multipole approximations of \eqref{pCM} by substituting appropriately-truncated versions of \eqref{ForceMultipoleScalar} and \eqref{TorqueMultipoleScalar}. 

Also note that $\dot{\gamma}^a_s$ is not necessarily collinear with $\hat{p}^a$. The difference $\hat{p}^a - \hat{m} \dot{\gamma}^a_s$ may be interpreted as a ``hidden mechanical momentum.'' Simple examples of hidden momentum are commonly discussed in electromagnetic problems (see, e.g., \cite{Jackson, Hidden1, Hidden2}), but occur much more generally. Some consequences of the hidden momentum are discussed in \cite{Bobbing, CostaHidden}.

The center of mass condition provides more than just a relation between the momentum and the velocity. It also implies that $\hat{S}^{ab}(\gamma_s,s)$ can be written entirely in terms of the spin vector $\hat{S}^a(\gamma_s, s)$. Inverting \eqref{Sdecomp} while using \eqref{CM},
\begin{equation}
    \hat{S}_a (\gamma_s,s) = - \frac{1}{2} \epsilon_{abcd} u^b \hat{S}^{cd}.
\end{equation}
Differentiating this and applying \eqref{pDotScalar}, the spin vector evaluated about the center of mass is seen to satisfy
\begin{equation}
    \frac{ \rmD \hat{S}_a }{ \rmd s} = - \frac{1}{2} \epsilon_{abcd} u^b \hat{N}^{cd} + u_a \left( \hat{S}_b \frac{ \rmD u^b }{ \rmd s} \right) .
    \label{spinEvolve}
\end{equation}
The first term here represents a torque in the ordinary sense. The second term is responsible for the Thomas precession, and may be made more explicit by substituting \eqref{pDotScalar} and \eqref{uDef}.

An evolution equation may also be be derived for the mass $\hat{m}$, which is not necessarily constant. Variations in $\hat{m}$ are not an exotic effect; masses change even for monopole test bodies coupled to relativistic scalar fields. In general, use of \eqref{pDotScalar} and \eqref{mDef} shows that the mass evaluated using $z_s = \gamma_s$ satisfies \cite{Dix70a}
\begin{equation}
    \frac{ \rmd \hat{m}}{\rmd s} = - \dot{\gamma}^a_s \hat{F}_a + \hat{N}_{ab} u^a \frac{\rmD u^b}{\rmd s}.
    \label{mEvolve}
\end{equation}
The final term here may be made more explicit by using \eqref{pDotScalar} and \eqref{uDef} to eliminate $\rmD u^b/\rmd s$.

\subsection{Monopole approximation}
\label{Sect:MonopoleSR}

The equations derived here are quite complicated in general. Some intuition for them may be gained by truncating the multipole series at monopole order. Inspection of \eqref{ForceMultipoleScalar} and \eqref{TorqueMultipoleScalar} then shows that
\begin{equation}
    \hat{F}_a =-  q \nabla_a \hat{\phi} , \qquad \hat{N}_{ab} = 0.
    \label{ForceMono}
\end{equation}
Further restricting to cases where $q = (\mbox{constant})$, it follows from \eqref{mEvolve} that
\begin{equation}
    \hat{m} - q \hat{\phi} = (\mbox{constant}).
    \label{MassMono}
\end{equation}
$\hat{m} - q \hat{\phi}$ may therefore be viewed as a conserved energy for the system. Note that it is the effective field $\hat{\phi}$ which occurs here, not the physical field $\phi$.

Contracting \eqref{spinEvolve} with $\hat{S}^a$ also shows that $\hat{S}^2 := \hat{S}^a \hat{S}_a = (\mbox{constant})$, meaning that the spin vector can only precess in the monopole approximation. The rate at which this occurs may be simplified by first recalling that in any maximally-symmetric spacetime, there exists a constant $\kappa$ such that 
\begin{equation}
    R_{abcd} = \kappa g_{a [c } g_{d] b}.
\end{equation}
Of course, $\kappa = 0$ in the flat background of special relativity. Regardless, the spin evolution is independent of $\kappa$:
\begin{equation}
    \frac{ \rmD \hat{S}_a }{ \rmd s } = - (q/\hat{m}) u_a \hat{S}^b \nabla_b \hat{\phi}.
    \label{SpinEvolveMono}
\end{equation}
It experiences a purely Thomas-like precession. The momentum-velocity relation \eqref{pCM} does, by contrast, retain explicit evidence of the curvature, reducing to
\begin{equation}
    \hat{m} \dot{\gamma}_s^a = \hat{p}^a + \left( \frac{ q/\hat{m}  }{ 1 + \frac{1}{2} \kappa (\hat{S}/\hat{m})^2  } 
    \right) \epsilon^{abcd} u_b \hat{S}_c \nabla_d \hat{\phi} .
\end{equation}
The linear momentum is also affected by $\kappa$:
\begin{align}
    \frac{\rmD \hat{p}_a}{\rmd s} &= - q \nabla_a \hat{\phi} + q \left( \frac{ \frac{1}{2} \kappa ( \hat{S} /\hat{m} )^2 }{ 1 + \frac{1}{2} \kappa (\hat{S}/\hat{m})^2 } \right) \left( \delta^b_a + u_a u^b - \hat{S}_a \hat{S}^b /\hat{S}^2 \right) \nabla_b \hat{\phi}.        \label{pEvolveMono}
\end{align}
The overall force is therefore a particular linear transformation of $-q \nabla_a \hat{\phi}$. Up to an overall factor, the second term here extracts that component of $q \nabla_a \hat{\phi}$ which is orthogonal to both $\hat{p}^a$ and $\hat{S}^a$. 

Together, \eqref{MassMono} and \eqref{SpinEvolveMono}-\eqref{pEvolveMono} determine the evolution of a scalar charge in the monopole approximation. Despite the approximations which have already been made, these equations remain rather formidable. They may be simplified further by demanding that $\hat{S} = 0$ at some initial time. The monopole approximation implies that an initially non-spinning particle remains non-spinning, so it is consistent to set $\hat{S} = 0$ for all time. It also follows that $\hat{p}^a = \hat{m} \dot{\gamma}^a_s$ and 
\begin{equation}
    \ddot{\gamma}^a_s = - (q/\hat{m}) ( g^{ab} + \dot{\gamma}^a_s \dot{\gamma}^b_s ) \nabla_b \hat{\phi} .
    \label{CMscalarMonopole}
\end{equation}
In the test body limit where $\hat{\phi} \approx \phi$, this is the typical equation adopted for the motion of a point particle with scalar charge $q$. A point particle limit of this equation which still allows for self-interaction is equivalent to what is known as the Detweiler-Whiting regularization \cite{PoissonRev, DetWhiting}. This regularization --- which originally arose via heuristic arguments associated with the singularity structure of point particle fields --- is a special case of the much more general results derived here (and which first appeared in \cite{HarteScalar}). Its origin  is unrelated to any point particle limits or to the singularities associated with them.

It is shown in Section \ref{Sect: ChargeCurved} below that standard self-force results follow easily from \eqref{CMscalarMonopole}. That section also generalizes all laws of motion derived here to apply to scalar charges moving in arbitrarily-curved spacetimes.

\section{Motion in curved spacetimes}
\label{Sect:Curved}

The discussion up to this point has made extensive use of Killing vector fields. This is familiar and simple, but not necessary. The first step towards understanding motion in generic spacetimes is to find a suitable replacement for the space of Killing vector fields. Once this is accomplished, the problem of motion for extended charges coupled to scalar fields is considered once again. This example is used to illustrate a new type of renormalization which occurs in spacetimes without symmetries. The techniques used to solve the scalar problem are then adapted to discuss motion in electromagnetic fields. Lastly, we consider motion in general relativity, where the objects of interest dynamically modify the geometry itself.

\subsection{Generalized Killing fields}
\label{Sect:GKF}

Recalling \eqref{Pdef} or its relativistic analog \eqref{PdefRel}, the generalized momenta used in maximally-symmetric spacetimes are defined as linear operators over the space $K$ of Killing vector fields. There is, however, no obstacle to replacing $K$ by some other vector space $K_G$. This is the approach we take to defining momenta in generic spacetimes. Although several notions of generalized or approximate Killing fields exist in the literature \cite{MatznerSyms, CookWhitingSyms, BeetleSyms}, only one of these \cite{HarteSyms} appears to be suitable for our purposes. We describe it now.

The space of generalized Killing fields used here can be motivated axiomatically. First note that avoiding significant modifications to the formalism developed thus far requires that all elements of $K_G$ be vector fields on spacetime (or space in non-relativistic problems). It is also reasonable to require that:
\begin{enumerate}
    \item All genuine Killing vectors which might exist are also generalized Killing vectors: $K \subseteq K_G$.
    
    \item $K = K_G$ in maximally-symmetric spacetimes.
    
    \item The dimensionality of $K_G$ can depend only on the number of spacetime dimensions.
\end{enumerate}
The first of these conditions is clearly necessary for any $K_G$ which may be said to generalize $K$. The second condition ensures that there are not ``too many'' generalized Killing fields in simple cases. The final condition is more subtle. It guarantees that the generalized momentum contains the ``same amount'' of information regardless of the metric. More specifically, a generalized momentum defined using $K_G$ must be decomposable in terms of linear and angular momenta with the correct number of components. Recalling \eqref{MaxDim},
\begin{equation}
    \dim K_G = \frac{1}{2} N (N+1) \geq \dim K
\end{equation}
in any $N$-dimensional space.

The given conditions restrict $K_G$, but do not define it. An additional constraint is needed which describes how those elements of $K_G$ which are not also elements of $K$ preserve an appropriate geometric structure. It is not, of course, possible to demand that they preserve the metric. Symmetries of the connection or curvature are unsuitable as well. The only reasonable possibilities are nonlocal. 

\subsubsection{Symmetries about a point}

First consider the Riemannian case\footnote{The same geometric conditions can also be imposed in Lorentzian geometries. Physically, however, the vector fields discussed here are most useful in non-relativistic contexts. A more complicated structure described in Section \ref{Sect:worldlineGKF} is better-suited to Lorentzian physics.}. Recalling \eqref{LieSigma}, any Killing vector field used to drag two points $x$ and $x'$ preserves the separation vector $-\sigma^a (x,x')$ between those points. A slightly weaker condition can be used to define generalized Killing vectors even when no genuine Killing vectors exist. Suppose that a particular point $x$ has been fixed and demand that $A_G (x)$ be defined as the set of all vector fields $\xi^a$ satisfying
\begin{equation}
    \Lie_\xi \sigma^a( x' ,x) = 0.
    \label{LieSigma2}
\end{equation}
The result clearly forms a vector space. Unfortunately $A_G(x)$ is too large. In flat space, it becomes independent of $x$ and coincides with the space of affine collineations: vector fields satisfying $\nabla_a \Lie_\xi g_{bc} = 0$. Geometrically, affine collineations represent symmetries which preserve the connection. In generic spaces, $A_G(x)$ may be described as a space of generalized affine collineations with respect to $x$. 

The space of Killing vector fields is known to be a vector subspace of the affine collineations. Similarly, generalized Killing fields $K_G(x)$ may be obtained as an appropriate subspace of $A_G(x)$. It is sufficient to demand only that the appropriate vector fields be exactly Killing at $x$:
\begin{equation}  
    \Lie_\xi g_{ab} ( x ) = 0.
    \label{GKF1}
\end{equation}
The set of all vector fields satisfying this and \eqref{LieSigma2} are denoted by $K_G(x)$. They are the generalized Killing fields with respect to $x$. Any genuine Killing fields which may exist are also elements of $K_G(x)$.

Many geometric structures are preserved by generalized Killing fields. Equation \eqref{GKF1}, can, for example, be shown to generalize to
\begin{equation}
        \Lie_\xi g_{ab} ( x ) = \nabla_a \Lie_\xi g_{bc} ( x ) = 0,
        \label{AlmostKillingPoint}
\end{equation}
but only at $x$. More generally, the ``projected Killing equations''
\begin{equation}
    \sigma^{a'}(x,x') \Lie_\xi g_{a'b'}(x')  = \sigma^{a'}(x,x') \sigma^{b'}(x,x') \nabla_{a'} \Lie_\xi g_{b'c'}(x') = 0
\end{equation}
are valid wherever the generalized Killing fields are defined \cite{HarteSyms}. Elements of $K_G(x)$ also preserve all distances from $x$ in the sense that $\Lie_\xi \sigma (x,x') = 0$. In general, $\xi^{a'} (x')$ is a solution to the geodesic deviation (or Jacobi) equation along the geodesic connecting $x$ and $x'$.

The generalized Killing fields may be shown to admit an expansion in terms of 1- and 2-forms at $x$. Expanding \eqref{LieSigma2} in terms of covariant derivatives shows that for every $\xi^a \in K_G(x)$,
\begin{equation}
    \xi^{a'} = \Xi^{a'a} \xi_{a} + \Xi^{a'ab} \nabla_{a} \xi_{b}.
    \label{Jacobi}
\end{equation}
The bitensors $\Xi^{a'a}$ and $\Xi^{a'ab}$ which appear here --- known as Jacobi propagators --- are explicitly given by \eqref{JacobiPropagators}. The expansion \eqref{KillingPropagator} of Killing vector fields is therefore identical to the expansion \eqref{Jacobi} of generalized Killing vector fields. Varying $\xi_a$ and $\nabla_a \xi_b$ arbitrarily, it is clear that $\dim K_G(x) = N + \frac{1}{2} N(N-1) = \frac{1}{2} N(N+1)$ in $N$ dimensions. 

The generalized Killing fields defined by \eqref{LieSigma2} and \eqref{GKF1} provide a notion of symmetry with respect to a point. They may be used to analyze non-relativistic motion in geometries which do not admit any exact symmetries. This is not pursued here. We instead focus on relativistic motion, in which case it is more appropriate to consider a different kind of generalized Killing field which provides a notion of symmetry near a worldline instead of a point. 

\subsubsection{Symmetries about a worldline}
\label{Sect:worldlineGKF}

In relativistic contexts, it is useful to define a $K_G$ which takes as arguments a timelike worldline and a foliation instead than a single point. Given a worldtube $\mathfrak{W} = \{ \fB_s \mid s\in \mathbb{R} \}$ in a Lorentzian spacetime $(\mathcal{M},g_{ab})$ of dimension $N$, consider a foliation $\{\fB_s \}$. Also consider a timelike worldline $\mathcal{Z}$ parametrized by $z_s := \mathcal{Z} \cap \fB_s$. The definition of $K_G(\mathcal{Z}, \{ \fB_s \})$ in this context is as follows: First, \eqref{AlmostKillingPoint} is enforced for all $z_s$. This provides a sense in which the elements of $K_G(\mathcal{Z}, \{ \fB_s \})$ generalize Poincar\'{e} symmetries ``near'' $\mathcal{Z}$. It implies that the generalized Killing fields and their first derivatives satisfy the Killing transport equations on $\mathcal{Z}$. Moreover,
\begin{equation}
    \left. \nabla_a \nabla_b \xi_c \right|_{\mathcal{Z}} = -R_{bca}{}^{d} \xi_d,
\end{equation}
which generalizes \eqref{d2K}. This describes, e.g., how generalized Killing fields which might appear purely rotational or boost-like at one point acquire translational components at nearby points. When applied to problems of motion, it leads to the Mathisson-Papapetrou spin-curvature force. 

Demanding that \eqref{AlmostKillingPoint} hold on $\mathcal{Z}$ describes the generalized Killing fields only on that worldline. They may be extended outwards into $\fW$ by demanding that
\begin{equation}
    \Lie_\xi \sigma^{a'} (x',z_s) = 0
\end{equation}
for each $x' \in \fB_s$. This is merely a restriction of \eqref{LieSigma2}. It implies that \eqref{Jacobi} holds whenever there exists some $s$ such that $x = z_s$ and $x' \in \fB_s$. Elements of $K_G(\mathcal{Z}, \{ \fB_s \})$ may therefore be specified using an arbitrary pair of 1- and 2-forms at any point on $\mathcal{Z}$. As required, the generalized Killing fields form a vector space with dimension $\frac{1}{2} N (N+1)$. Additionally, $\Lie_\xi \sigma(x',z_s) = 0$ and
\begin{equation}
    \sigma^{a'} (x',z_s) \sigma^{b'} (x',z_s) \Lie_\xi g_{a'b'} (x',z_s) = 0
\end{equation}
whenever $x' \in \fB_s$ \cite{HarteSyms}. The elements of $K_G( \mathcal{Z}, \{ \fB_s \} )$ are the generalized Killing fields used in the remainder of this work.

\subsection{Scalar charges in curved spacetimes}
\label{Sect: ChargeCurved}

We now return to the motion of scalar charges as discussed in Section \ref{Sect:SR}, but no longer require that the background spacetime admit any symmetries. Consider a body coupled to a Klein-Gordon field $\phi$ in a four-dimensional spacetime $(\mathcal{M},g_{ab})$. This body is assumed to be contained inside a worldtube $\fW \subset \mathcal{M}$ with finite spatial extent and no other matter. Its stress-energy tensor $T^{ab}_\mathrm{body}$ is assumed to satisfy \eqref{StressConsScalar}.

A generalized momentum is easily defined by reusing \eqref{PdefRel}, but with the space $K$ employed there replaced by an appropriate space $K_G (\mathcal{Z}, \{ \fB_s \})$ of generalized Killing fields as described in Section \ref{Sect:GKF}. This requires the introduction of a timelike worldline $\mathcal{Z}$ and a foliation $\{ \fB_s \}$ of $\fW$. Supposing that these structures have been chosen --- perhaps using center of mass conditions --- let
\begin{equation}
    P_s (\xi) := \int_{\fB_s} T^{ab}_\mathrm{body} \xi_a \rmd S_b
    \label{PdefGen}
\end{equation}
for all $\xi^a \in K_G (\mathcal{Z}, \{ \fB_s \} )$. For each $s$, this is a vector in the ten-dimensional space $K^*_G (\mathcal{Z}, \{ \fB_s \})$. The associated linear and angular momenta $p_a$ and $S_{ab}$ coincide with those introduced by Dixon \cite{Dix74, Dix79, Dix70a} for matter which does not couple to an electromagnetic field\footnote{Dixon's papers never considered matter coupled to scalar fields. The momenta associated with \eqref{PdefGen} are those which arise naturally for objects falling freely in curved spacetimes.}.

Using stress-energy conservation to differentiate the generalized momentum with respect to the time parameter $s$,
\begin{equation}
    \frac{\rmd}{\rmd s} P_s(\xi) = \int_{\fB_s} \left( \frac{1}{2} T^{ab}_\mathrm{body} \Lie_\xi g_{ab} - \rho \Lie_\xi \phi \right) \rmd S.
    \label{ForceScalarCurved}
\end{equation}
The first term on the right-hand side of this expression is not present in its maximally-symmetric counterpart \eqref{ForceScalar}; extra forces arise when the $\xi^a$ are not Killing. These may be interpreted as gravitational effects. While sufficiently small test bodies fall along geodesics in curved spacetimes, the same is not true for more extended masses. Gravity generically exerts nonzero 4-forces and 4-torques which are described by the $\Lie_\xi g_{ab}$ term in \eqref{ForceScalarCurved}. If expanded in a multipole series, \eqref{AlmostKillingPoint} implies that such effects first appear at quadrupole order. This is described more fully in Section \ref{Sect:ChargeCurvedMultipole}. 

The effect of the scalar field on the second term on the right-hand side of \eqref{ForceScalarCurved} may be understood by repeating the arguments of Sections \ref{Sect:SelfFieldSR} and \ref{Sect:Renormalize}. This results in Equation \eqref{EOMScalar} being replaced by
\begin{align}
    \frac{\rmd}{\rmd s} \hat{P}_s(\xi) = \int_{\fB_s} \left[ \frac{1}{2} T^{ab}_\mathrm{body}(x) \Lie_\xi g_{ab}(x) - \rho(x) \Lie_\xi \hat{\phi}(x) \right.
    \nonumber
    \\
    \left. ~ - \frac{1}{2} \int_\fW \! \rmd V' \rho(x) \rho(x') \Lie_\xi G (x,x') \right] \rmd S .
    \label{ScalarForceCurved}
\end{align}
The effective field which appears here is defined by $\hat{\phi} := \phi - \phiS$, where $\phiS$ satisfies \eqref{SelfFieldScalar} and $G$ is the Detweiler-Whiting S-type Green function described in Section \ref{Sect:SelfFieldSR}. The generalized momentum $P_s$ has also been replaced by the renormalized momentum $\hat{P}_s := P_s + E_s$, where the self-momentum $E_s$ is given by \eqref{Edef}.

An absence of spacetime symmetries explicitly affects self-interaction only via the second line of \eqref{ScalarForceCurved}. Unlike in maximally-symmetric backgrounds, the Detweiler-Whiting Green function does not satisfy $\Lie_\xi G = 0$ in general. Indeed, no Green function can be constructed with this property, a result which could be viewed as implying that it is impossible to define an analog of Newton's third law in generic spacetimes. It follows that the self-force cannot be entirely absorbed into a redefinition of the momentum. Its remainder can, however, be understood as equivalent to another type of renormalization which affects the quadrupole and higher multipole moments of $T^{ab}_\mathrm{body}$.

\subsubsection{Breakdown of Newton's third law}
\label{Sect:RenormalizeCurved}

The self-force which remains after renormalizing $P_s$ depends on $\Lie_\xi G$. It may be understood physically by recalling that there is a sense in which $G$ is constructed purely from the spacetime metric. It therefore follows that for any vector field $\xi^a$, whether it is a generalized Killing field or not, $\Lie_\xi G$ must depend linearly on $\Lie_\xi g_{ab}$. In this sense, the first and third terms on the right-hand side of \eqref{ScalarForceCurved} are both linear in $\Lie_\xi g_{ab}$. They are physically very similar, both representing different aspects of the gravitational force \cite{HarteBrokenSyms}.

As remarked in Section \ref{Sect:SR}, a body's inertia depends on both its own stress-energy tensor and the properties of its self-field. The inertia due to $T^{ab}_\mathrm{body}$ is described by $P_s$ and the inertia due to $\phiS$ by $E_s$. A body's passive gravitational mass experiences a similar split. The gravitational force exerted on a body due to its stress-energy tensor is 
\begin{equation}
    \frac{1}{2} \int_{\fB_s} T^{ab}_\mathrm{body} \Lie_\xi g_{ab} \rmd S,
    \label{GForce}
\end{equation}
while the gravitational force exerted on a body's self-field is instead described by
\begin{equation}
    - \frac{1}{2} \int_{\fB_s} \!\! \rmd S \int_\fW \!\! \rmd V' \rho\rho \Lie_\xi G .
    \label{CurvedSelfForce}
\end{equation}
Although it is difficult to do so explicitly, this latter expression can be transformed into a linear operator on $\Lie_\xi g_{ab}$. It effectively adds to a body's gravitational mass distribution as inferred by observing responses to different spacetime curvatures. \textit{In a multipole expansion, the force \eqref{CurvedSelfForce} may be viewed as renormalizing the quadrupole and higher multipole moments of a body's stress-energy tensor.} 

The failure of Newton's third law, or equivalently $\Lie_\xi G \neq 0$, therefore provides a second mechanism by which self-fields lead to renormalizations. It is distinct --- both in its origin and in the quantities it affects --- from the mechanism described in Section \ref{Sect:Renormalize}. The renormalization of a body's momentum was shown to be closely connected to the hyperbolicity of the underlying field equation. The geometry-induced\footnote{This type of renormalization fundamentally arises from the connection between $\Lie_\xi G$ and $\Lie_\xi g_{ab}$ which occurs for Green functions associated with the Klein-Gordon equation. In different theories, Lie derivatives of $G$ can depend on fields other than the metric. Self-forces then renormalize whichever moments are coupled to these fields.} failure of Newton's third law can instead arise even for matter coupled to elliptic field equations. It affects only the quadrupole and higher moments of a body's stress-energy tensor. Combined, the two types of renormalization affect all multipole moments of $T^{ab}_\mathrm{body}$. In this sense, one is led to the concept of an effective stress-energy tensor. This is defined quasi-locally, and can be identified with $T^{ab}_\mathrm{tot}$ only in special cases.

A fully explicit demonstration of this effect is not known. It may nevertheless be motivated more directly in terms of a wave equation satisfied by $\Lie_\xi G$. Noting that
\begin{equation}
    \Lie_\xi \delta(x,x') = - \frac{1}{2} \delta(x,x') g^{ab}(x) \Lie_\xi g_{ab}(x),
\end{equation}
a Lie derivative of \eqref{GDefScalar} yields
\begin{align}
    (\nabla^a \nabla_a - \mu^2) \Lie_\xi G = \nabla_a \big[ \big(g^{ac} g^{bd} - \frac{1}{2} g^{ab} g^{cd} \big) \nabla_b G \Lie_\xi g_{cd} \big] 
    \nonumber
    \\
    ~ + \frac{\mu^2}{2} (g^{ab} \Lie_\xi g_{ab}) G.
\end{align}
Viewing $\Lie_\xi G$ on the left-hand side of this equation as ``independent'' of the $G$ appearing on the right-hand side suggests that $\Lie_\xi G$ is a solution to a wave equation sourced by $\Lie_\xi g_{ab}$ and its first derivative. A source which is independent of $G$ may be found by applying the Klein-Gordon operator to both sides:
\begin{align}
    (\nabla^{a'} \nabla_{a'} - \mu^2) ( \nabla^a \nabla_a - \mu^2) \Lie_\xi G = 4 \pi \nabla_a \big[ \big(g^{ac} g^{bd} - \frac{1}{2} g^{ab} g^{cd} \big) (\Lie_\xi g_{cd}) \nabla_b \delta \big] 
    \nonumber
    \\
    ~ - 2 \pi \mu^2 ( g^{ab} \Lie_\xi g_{ab}) \delta.
\end{align}
This describes $\Lie_\xi G$ as the solution to a fourth-order distributional differential equation sourced by $\Lie_\xi g_{ab}$ and $\nabla_a \Lie_\xi g_{bc}$.

Results like these may be used together with the field equation to integrate \eqref{CurvedSelfForce} by parts. Let
\begin{equation}
    T^{ab}_\mathrm{field,S} := \frac{1}{4\pi} \left[ \nabla^a \phiS \nabla^b \phiS - \frac{1}{2} g^{ab} ( \nabla_c \phiS \nabla^c \phiS + \mu^2 \phiS^2) \right] 
\end{equation}
be the stress-energy tensor associated with $\phiS$ and define
\begin{align}
    \mathcal{I}^a := &\frac{1}{8\pi} (g^{ac} g^{bd} - \frac{1}{2} g^{ab} g^{cd} ) \phiS \nabla_b \phiS \Lie_\xi g_{cd} 
    \nonumber
    \\
    &~ + \frac{1}{8\pi } \int_\fW \rho' ( \nabla^a \phiS \Lie_\xi G - \phiS \nabla^a \Lie_\xi G) \rmd V' .
\end{align}
The law of motion \eqref{ScalarForceCurved} then reduces to
\begin{align}
    \frac{\rmd}{\rmd s} \hat{P}_s = \int_{\fB_s} \left[ \frac{1}{2} (T^{ab}_\mathrm{body} + T^{ab}_\mathrm{field,S} ) \Lie_\xi g_{ab} - \rho \Lie_\xi \hat{\phi} \right] \rmd S 
    \nonumber
    \\
    ~ - \frac{\rmd}{\rmd s} \int_{\fB_s} \!\! \mathcal{I}^a \rmd S_a - \oint_{\partial \fB_s} \!\! \mathcal{I}^a t^b \rmd S_{ab}. 
\end{align}
The gravitational force in this expression clearly acts on the combined stress-energy tensor $T^{ab}_\mathrm{body} + T^{ab}_\mathrm{field,S}$. Unfortunately, the stress-energy tensor associated with the self-field does not have compact spatial support. The finite integration volume $\fB_s$ must therefore be compensated by the two boundary terms involving $\mathcal{I}^a$. If these can be ignored, it is evident that gravitational forces are determined only by $T^{ab}_\mathrm{body} + T^{ab}_\mathrm{field,S}$. This cannot, however, be expected to hold generically. In general, there does not appear to be any reason to neglect $\mathcal{I}^a$. 

Approximations may instead be introduced which allow the renormalized multipole moments to be computed essentially using local Taylor series \cite{HarteBrokenSyms}. Another approach, described here for the first time, is to use Hadamard series. Recalling the Hadamard form \eqref{GHadamard} for $G$,
\begin{equation}
    \Lie_\xi G = \frac{1}{2} \left[ \Delta^{1/2} \delta'(\sigma) \Lie_\xi \sigma + (\Lie_\xi \Delta^{1/2} - H \Lie_\xi \sigma) \delta(\sigma) - \Lie_\xi V \Theta(\sigma) \right].
    \label{LieGExpl}
\end{equation}
Our task is now to convert this into an expression where all Lie derivatives act on the metric.

This is easily accomplished for the Lie derivatives of $\sigma$ and $\Delta$. Differentiating the well-known identity $\sigma^{a'} \sigma_{a'} = 2\sigma$ shows that
\begin{align}
    \sigma^{a'} \nabla_{a'} \Lie_\xi \sigma - \Lie_\xi \sigma  = \frac{1}{2} \sigma^{a'} \sigma^{b'} \Lie_\xi g_{a'b'}.
    \label{SigmaTransport}
\end{align}
The differential operator $\sigma^{a'} (x,x') \nabla_{a'}$ appearing here is a covariant derivative along the geodesic connecting $x$ and $x'$, so \eqref{SigmaTransport} may be viewed as an ordinary differential equation for $\Lie_\xi \sigma$. Letting $y_\tau$ denote a geodesic which is affinely parametrized by $\tau$ while satisfying $y_0 = x$ and $y_1 = x'$, it follows that
\begin{equation}
    \Lie_\xi \sigma(x,x') = \frac{1}{2} \int_0^1 \dot{y}^a_\tau \dot{y}^b_\tau \Lie_\xi g_{ab}(y_\tau ) \rmd \tau.
    \label{LieSigmaInt}
\end{equation}
Moreover, an argument found in \cite{HarteBrokenSyms} shows that Lie derivatives of the van Vleck determinant $\Delta$ depend on $\Lie_\xi \sigma$:
\begin{equation}
    \Lie_\xi \ln \Delta^{1/2} = - \frac{1}{4} \left[ g^{ab} \Lie_\xi g_{ab} + g^{a'b'} \Lie_\xi g_{a'b'} + 2 H^{aa'} \nabla_a \nabla_{a'} \Lie_\xi \sigma \right].
    \label{LieDelta}
\end{equation}
Both $\Lie_\xi \sigma$ and $\Lie_\xi \Delta^{1/2}$ may therefore be written as line integrals --- solutions to transport equations --- which are linear in $\Lie_\xi g_{ab}$. Substituting these expressions into \eqref{LieGExpl} goes much of the way towards expressing $\Lie_\xi G$ in terms of $\Lie_\xi g_{ab}$.

All that remains is to consider $\Lie_\xi V$. This is more difficult. Even $V$ itself is complicated to compute. It is a solution to the homogeneous field equation which is symmetric in its arguments and satisfies
\begin{equation}
    V(x,x') = G_+ (x,x') + G_- (x,x') 
\end{equation}
whenever $x'$ is timelike-separated from $x$. Alternatively, $V$ can be computed using a transport equation along null geodesics \cite{PoissonRev}. For each $x'$, this transport equation may be used as boundary data with which to obtain $V(\cdot,x')$. Extending this data outside of the null cones is essentially an ``exterior characteristic problem:'' One seeks a solution to a hyperbolic differential equation in the exterior of a null cone given values of the solution on that cone. Unlike interior characteristic problems, the general mathematical status of such problems is unclear. 

One way to proceed is to construct a Hadamard series. This is an ansatz which supposes that $V$ can be expanded via \cite{Friedlander}
\begin{equation}
    V(x,x') = \sum_{n=0}^\infty V_n(x,x') \sigma^n(x,x').
    \label{HadamardAnsatz}
\end{equation}
The $V_n$ here are determined by demanding that each explicit power of $\sigma$ vanish separately when this series is inserted into $(\nabla^a \nabla_a - \mu^2) V = 0$. The result is an infinite tower of ordinary differential equations. The $n = 0$ case is governed by
\begin{align}
    \Delta^{1/2} \sigma^a \nabla_a (\Delta^{-1/2} V_0) + V_0 = \frac{1}{2} \nabla^a \nabla_a \Delta^{1/2}.
    \label{H0}
\end{align}
For each $x'$, this determines $V(\cdot,x')$ on the light cones of $x'$. Higher-order terms are needed in the exteriors of these light cones. For all $n \geq 1$,
\begin{align}
    \Delta^{1/2} \sigma^a \nabla_a (\Delta^{-1/2} V_n)  + (n+1) V_n = - \frac{1}{2n} \nabla^a \nabla_a V_{n-1}.
    \label{Hn}
\end{align}
It should be emphasized that the Hadamard series is not a Taylor expansion. The $V_n$ are two-point scalar fields, not constants. Furthermore, the Hadamard series is known to converge only if the metric is analytic (and even then, it might converge only in a finite region) \cite{Friedlander}. Although analyticity is quite a strong assumption, there may be other interesting cases where a finite Hadamard series can be used to approximate $V$ up to  a well-controlled remainder.

Assuming that \eqref{HadamardAnsatz} is valid, it is easily used to compute Lie derivatives of $V$. This results in
\begin{equation}
    \Lie_\xi V = \sum_{n=0}^\infty \big[ \Lie_\xi V_n + (n+1) V_{n+1} \Lie_\xi \sigma \big] \sigma^n.
\end{equation}
Lie derivatives of $\sigma$ may already be transformed into Lie derivatives of $g_{ab}$ via \eqref{LieSigmaInt}. Lie derivatives of the $V_n$ may instead be found by differentiating \eqref{H0} and \eqref{Hn}. The $n=0$ case satisfies
\begin{align}
    \Delta^{1/2} \sigma^a \nabla_a ( \Delta^{-1/2} \Lie_\xi V_0)  + \Lie_\xi V_0 = \frac{1}{2} \big( \Lie_\xi \nabla^a \nabla_a \Delta^{1/2} - V_0 \Lie_\xi \sigma^{a}{}_{a} \big) 
    \nonumber
    \\
    ~ + (g^{ab} \sigma^c \Lie_\xi g_{bc} - \nabla^a \Lie_\xi \sigma )\nabla_a V_0,
\end{align}
for example. The left-hand side of this equation may be interpreted as an ordinary differential operator along the geodesic which connects the two arguments of $V_0$.  All Lie derivatives in the source on the right-hand of this equation may, with the help of \eqref{LieSigmaInt} and \eqref{LieDelta}, be rewritten in terms of $\Lie_\xi g_{ab}$. It follows that $\Lie_\xi V_0$ can be expressed as a line integral involving $\Lie_\xi g_{ab}$. Similar results also hold for $\Lie_\xi V_n$ with $n>0$. The detailed forms of these integrals are complicated and are not displayed here. The important point, however, is that all parts of $\Lie_\xi G$ may be expressed as line integrals involving $\Lie_\xi g_{ab}$. Changing variables appropriately and using the spatially-compact support of $\rho$ then shows that the self-force \eqref{CurvedSelfForce} does indeed renormalize the gravitational force \eqref{GForce}. 

\subsubsection{Multipole expansions}
\label{Sect:ChargeCurvedMultipole}

It is useful to expand the scalar force in a multipole series when $\hat{\phi}$ varies slowly inside the body. Similarly, the gravitational force may be expanded in its own multipole series when there is an appropriate sense in which $g_{ab}$ does not vary too rapidly\footnote{More precisely, the coordinate components $g_{ij}$ must not vary too rapidly when expressed in a Riemann normal coordinate system with origin $z_s$. Physically, this is a significant restriction. It would be far better to perform multipole expansions only using effective metrics where gravitational self-fields have been appropriately removed. It is not known how to do to this for the full Einstein-Klein-Gordon system.} inside each $\fB_s$. Such expansions can be obtained using the techniques of Sections \ref{Sect:multipoleNewt} and \ref{Sect:MultipoleScalarSR}. Recalling \eqref{EOMScalar}, \eqref{EOMscalarMultipole}, and \eqref{ScalarForceCurved}, first note that
\begin{align}
    \frac{\rmd}{\rmd s} \hat{P}_s(\xi) = - \sum_{n=0}^\infty \frac{1}{n!} q^{a_1 \cdots a_n} \Lie_\xi \hat{\phi}_{,a_1 \cdots a_n} + 
    \frac{1}{2} \int_{\fB_s} \!\! T^{ab}_\mathrm{body} \Lie_\xi g_{ab} \rmd S
    \nonumber
    \\
    ~ - \frac{1}{2} \int_{\fB_s} \!\! \rmd S \int_\fW \! \rmd V' \rho \rho' \Lie_\xi G .
    \label{CurvedSelfForce2}
\end{align}
The two integrals which remain here are intrinsically gravitational. 

Given \eqref{AlmostKillingPoint}, it is evident that multipole expansion for the generalized force must begin at quadrupole order. More specifically, it may be shown that \cite{HarteBrokenSyms, Dix74, Dix79}
\begin{equation}
    \frac{1}{2} \int_{\fB_s} T^{ab}_\mathrm{body} \Lie_\xi g_{ab} \rmd S = \frac{1}{2} \sum_{n=2}^\infty \frac{1}{n!} I^{c_1 \cdots c_n ab} \Lie_\xi g_{ab,c_1 \cdots c_n},
    \label{TMultipoles}
\end{equation}
where $g_{ab,c_1 \cdots c_n}$ represents the $n$th tensor extension of $g_{ab}$ and $I^{c_1 \cdots c_n ab}$ is Dixon's \cite{Dix74} $2^n$-pole moment of $T^{ab}_\mathrm{body}$. Tensor extensions in this context are somewhat more complicated than for the scalar case discussed in Section \ref{Sect:multipoleNewt}. While they are still defined as those tensors which reduce to $n$ partial derivatives when evaluated at the origin of a Riemann normal coordinate system, equations like \eqref{ExtensionDef} must be generalized for objects with nonzero tensorial rank (see, e.g., \cite{HarteBrokenSyms}). For this reason, explicit integrals relating $I^{c_1 \cdots c_n ab}$ to $T^{ab}_\mathrm{body}$ are significantly more complicated than their scalar analogs. These are not needed here, and may be found in \cite{HarteBrokenSyms, Dix74}. Additionally, note that the first few tensor extensions of the metric are $g_{ab,c} = 0$ and
\begin{align}
    g_{ab,c_1 c_2} = \frac{2}{3} R_{a (c_1 c_2) b}, \qquad g_{ab,c_1 c_2 c_3} = \nabla_{(c_1} R_{|a|c_2 c_3) b},
    \label{MetricExtensions}
    \\
    g_{ab,c_1 c_2 c_3 c_4} = \frac{6}{5} \nabla_{(c_1 c_2} R_{|a| c_3 c_4) b} + \frac{16}{15} R_{a(c_1 c_2}{}^{d} R_{|b|c_3 c_4) d}.
\end{align}

Expanding both integrals in \eqref{CurvedSelfForce2} while identifying coefficients in front of $\Lie_\xi g_{ab,c_1 \cdots c_n}$ results in a series structurally identical to \eqref{TMultipoles}, but with all bare multipole moments $I^{c_1 \cdots c_n ab}$ replaced by their renormalized counterparts $\hat{I}^{c_1 \cdots c_n ab}$. The final multipole expansion for the generalized force acting on a self-interacting scalar charge distribution is therefore
\begin{align}
    \frac{\rmd}{\rmd s} \hat{P}_s(\xi) = \frac{1}{2} \sum_{n=2}^\infty \frac{1}{n!} \hat{I}^{c_1 \cdots c_n ab} \Lie_\xi g_{ab,c_1 \cdots c_n} - \sum_{n=0}^\infty \frac{1}{n!} q^{a_1 \cdots a_n} \Lie_\xi \hat{\phi}_{,a_1 \cdots a_n}.
    \label{ScalarGenForce}
\end{align}
Gravitational terms first arise at quadrupole order, while scalar terms appear even at monopole order.

\subsubsection{Forces and torques}

As in Sections \ref{Sect:PtopS} and \ref{Sect:PtopsScalar}, the generalized momentum can be decomposed into linear and angular components $\hat{p}^a$, $\hat{S}^{ab}$. These obey the law of motion \eqref{pDotScalar}, where the force and torque are now supplemented by gravitational terms at quadrupole and higher orders:
\begin{align}
    \hat{F}_a = \frac{1}{2} \sum_{n=2}^\infty \frac{1}{n!} \hat{I}^{d_1 \cdots d_n bc} \nabla_a g_{bc, d_1 \cdots d_n} - \sum_{n=0}^\infty \frac{1}{n!} q^{b_1 \cdots b_n} \nabla_a \hat{\phi}_{, b_1 \cdots b_n} ,
    \\
    \hat{N}^{ab} = \sum_{n=2}^\infty \frac{2}{n!} g^{f[b} \big( \hat{I}^{|c_1 \cdots c_n|a]d} g_{df,c_1 \cdots c_n} + \frac{n}{2} \hat{I}^{a] c_1 \cdots c_{n-1} dh} g_{dh,c_1 \cdots c_{n-1} f}  \big) 
    \nonumber
    \\
    ~ + \sum_{n=0}^\infty \frac{2}{n!} g^{c[a} q^{b] d_1 \cdots d_n} \hat{\phi}_{,c d_1 \cdots d_n}.
\end{align}

A center of mass may be defined exactly as described in Section \ref{Sect:CMSR}. Applying \eqref{CM}, the momentum-velocity relation remains \eqref{pCM}.

\subsubsection{Monopole approximation}
\label{Sect:ScalarMonopole}

Suppose that $\mathcal{Z}$ is identified with the center of mass worldline $\{ \gamma_s \}$. It is then interesting to consider the laws of motion truncated at monopole order. As explained in Section \ref{Sect:MonopoleSR}, the spin magnitude is conserved in such cases. It is therefore consistent to consider cases where $\hat{S}_a$ is negligible. Assuming this, an object's mass $\hat{m}$ and center of mass $\gamma_s$ satisfy
\begin{equation}
    \hat{m} - q \hat{\phi} = (\mbox{constant}), \qquad 
    \hat{m} \ddot{\gamma}^a_s = - q ( g^{ab} + \dot{\gamma}^a_s \dot{\gamma}^b_s ) \nabla_b \hat{\phi} 
    \label{MonopoleScalarGR}
\end{equation}
whenever $q = (\mbox{constant})$. These are formally equivalent to the standard equations of motion adopted for scalar test charges, but with the physical field replaced everywhere by the effective field; laws of motion applicable to self-interacting charges involve $\hat{\phi}$, not $\phi$.

Classical results on the scalar self-force are easily derived from \eqref{MonopoleScalarGR}. In the absence of any external charges (which does not necessarily imply trivial motion in curved spacetimes), it is natural to suppose that initial data for $\phi$ has been prescribed in the distant past. If all details of that data have decayed sufficiently, the only remaining field is the retarded solution associated with the body's own charge distribution. Generalizing this somewhat to also allow for a prescribed ``external field'' $\phi_\mathrm{ext}$, 
\begin{equation}
    \phi(x) = \phi_\mathrm{ext}(x) + \int_\fW \!\rho(x') G_-( x,x') \rmd V'.
\end{equation}
While the retarded Green function $G_-$ can be difficult to compute in nontrivial spacetimes, we assume that this problem has been solved. Using \eqref{Gexpand} and \eqref{SelfFieldScalar} then results in the effective potential\footnote{The two-point scalar $V$ is to be understood here as equivalent to $G_\pm$ when its arguments are timelike-separated. This defines it even in the presence of caustics and other potential complications.}
\begin{equation}
    \hat{\phi} = \phi_\mathrm{ext} + \frac{1}{2} \int_\fW \! \rho' ( G_- - G_+ + V) \rmd V'.
    \label{phiHatRet}
\end{equation}
Although the ``self-field'' $\phiS$ has been removed from $\phi$, it is clear from this that remnants of the body's charge distribution do remain. These are responsible for self-forces as they are commonly defined in the literature, and give rise to physical phenomena such as radiation reaction.

The ``traditional'' self-force problem involves point particle limits. Such limits consist of appropriate one-parameter families of charge distributions ${}_{\lambda} \rho$ and stress-energy tensors ${}_\lambda T^{ab}_\mathrm{body}$ whose supports collapse to a single timelike worldline as $\lambda \rightarrow 0$ (see, e.g., \cite{GrallaHarteWaldEM}). Suppose that these families are chosen such that a body's physical size and mass asymptotically scale like $\lambda^1$ as $\lambda \rightarrow 0$, suggesting that ${}_\lambda q^{a_1 \cdots a_n}$ scales like $\lambda^{n+1}$ for all $n \geq 0$ and that ${}_\lambda \hat{I}^{c_1 \cdots c_n ab}$ scales like $\lambda^{n+1}$ for all $n \geq 2$. These conditions guarantee that the multipole series can be truncated at low order [as has already been assumed in \eqref{MonopoleScalarGR}].

It is also important to demand that the time dependence of ${}_\lambda \rho$ remain smooth as $\lambda \rightarrow 0$. This guarantees that a body's internal dynamics remain slow compared to its light-crossing timescale, which is required to ensure that a charge does not self-accelerate in the absence of any external influence. Self-acceleration could occur if, e.g., internal oscillations conspired to generate strongly-collimated beams of radiation. Such cases are physically possible, but are typically excluded from self-force discussions.


The $\lambda \rightarrow 0$ limit of the approximation which has just been sketched results in a timelike worldline satisfying the usual equations for a monopole test charge accelerated by $\phi_\mathrm{ext}$. Effects typically described as self-forces occur at the following order. In principle, interactions between dipole moments and $\phi_{\mathrm{ext}}$ also occur at this order. These are dropped here for simplicity, leaving only \eqref{MonopoleScalarGR}. Without entering into technical details, the appropriate approximation for the effective field in this limit may be shown to be given by \eqref{phiHatRet} with a point particle charge density. Dropping all labels involving $\lambda$, 
\begin{equation}
    \hat{\phi}(x) = \phi_\mathrm{ext}(x) + \frac{1}{2} q \int [ G_-(x,\gamma_{s'}) - G_+(x, \gamma_{s'}) + V(x,\gamma_{s'})] \rmd s'
\end{equation}
plus terms of order $\lambda^2$. Unlike the $\lambda \rightarrow 0$ limit of $\phi$, the limit of $\hat{\phi}$ is well-defined even at the particle's location. The same is also true of its gradient. All necessary regularizations have automatically been taken into account by first deriving the correct laws of motion for nonsingular extended bodies.

The computations needed to evaluate $\nabla_a \hat{\phi}$ at $\gamma_s$ are rather tedious, so we merely cite\footnote{The point particle field derived in \cite{PoissonRev} includes a derivative of the particle's acceleration. A careful treatment of the perturbation theory shows, however, that such terms refer only to accelerations at lower order \cite{GrallaHarteWaldEM}. The self-consistent discussion which is implicit here therefore requires that accelerations be simplified using the zeroth order equation of motion. This is taken into account in \eqref{ScalarPtParticle}. \label{Foot:OrderReduction} } the result \cite{PoissonRev}. Defining the projection operator $h^{ab} := g^{ab} + \dot{\gamma}^a_s \dot{\gamma}^b_s$ and assuming that $q = (\mbox{constant})$,
\begin{align}
    \nabla^a \hat{\phi} = \nabla^a \phi_\mathrm{ext} + \frac{1}{3} (q^2/\hat{m}) h^{ab} \dot{\gamma}^c_s \big[ \nabla_b \nabla_c \phi_\mathrm{ext} - 2 (q/\hat{m}) \nabla_b \phi_\mathrm{ext} \nabla_c \phi_\mathrm{ext} \big] 
    \nonumber
    \\
    ~ + \frac{1}{12} q ( R \dot{\gamma}^a_s - 2 h^{ab} R_{bc} \do{\gamma}^c_s ) + q \lim_{\epsilon \rightarrow 0^+} \int_{-\infty}^{s-\epsilon} \!\! \nabla^a G_- (\gamma_s,\gamma_{s'} ) \rmd s'.
    \label{ScalarPtParticle}
\end{align}
Substituting this into \eqref{MonopoleScalarGR}, the final equations of motion are
\begin{align}
    \ddot{\gamma}^a_s = - (q/\hat{m}) h^{ab} \bigg( \nabla_b \phi_\mathrm{ext} + \frac{1}{3} (q^2/\hat{m}) \dot{\gamma}^c_s \big[ \nabla_b \nabla_c \phi_\mathrm{ext} - 2 (q/\hat{m}) \nabla_b \phi_\mathrm{ext} \nabla_c \phi_\mathrm{ext} \big] 
    \nonumber
    \\
    ~ - \frac{1}{6} q R_{bc} \dot{\gamma}^c_s + q \lim_{\epsilon \rightarrow 0^+} \int_{-\infty}^{s-\epsilon} \nabla_b G_- (\gamma_s,\gamma_{s'}) \rmd s' \bigg)
    \label{ScalarEOM}
\end{align}
and
\begin{equation}
    \hat{m} - q \phi_\mathrm{ext} + q^2 \left( \frac{1}{12} R - \lim_{\epsilon \rightarrow 0^+} \int_{-\infty}^{s-\epsilon} \!\! G_- (\gamma_s, \gamma_{s'}) \rmd s' \right) = (\mbox{constant})
\end{equation}
through first order in $\lambda$. Essentially\footnote{Some sign conventions in \cite{HarteScalar} and \cite{Quinn} are different from those adopted here.} the same results were obtained by Quinn using an axiomatic argument \cite{Quinn}, and later derived from first principles in \cite{HarteScalar}. Note that the integrals over the tail of $G_-$ which appear here indicate that a charge's motion can depend on its past history. This is related to the failure of Huygens' principle, implying that the field a body sources in the past can scatter back towards it at later times. This effect disappears for matter coupled to massless fields in flat spacetime, but is almost always present otherwise.

\subsection{Electromagnetic charges}
\label{Sect:emSR}

The problem of electromagnetic self-force has inspired considerable discussion over the past century. Here, we show how it can be understood using a straightforward application of the formalism just described for scalar charges. The body of interest is assumed to be smooth and to be confined to a finite worldtube $\fW \subset \mathcal{M}$ in a fixed four-dimensional spacetime $(\mathcal{M},g_{ab})$. It couples to an electromagnetic field $F_{ab} = F_{[ab]}$ satisfying Maxwell's equations
\begin{equation}
    \nabla_{[a} F_{bc]} = 0, \qquad \nabla^b F_{ab} = 4\pi J_a,
\end{equation}
from which it follows as an integrability condition that the body's 4-current $J^a$ must be divergence-free. Its total charge
\begin{equation}
    q := \int_{\fB_s} \!\! J^a \rmd S_a
\end{equation}
is therefore a constant independent of $\fB_s$  (as long as this hypersurface completely contains the body of interest). Furthermore, the stress-energy tensor associated with $F_{ab}$ can be defined throughout $\fW$ via
\begin{equation}
    T^{ab}_\mathrm{field} = \frac{1}{4 \pi} (F_{a}{}^{c} F_{bc} - \frac{1}{4} g_{ab} F^{cd} F_{cd} ) .
\end{equation}
Identifying all remaining stress-energy as $T^{ab}_\mathrm{body}$, stress-energy conservation implies that
\begin{equation}
        \nabla_b T^{ab}_\mathrm{body} = F^{ab} J_b.
        \label{StressConsEM}
\end{equation}
The right-hand side of this equation is the Lorentz force density.

\subsubsection{An electromagnetic momentum}

A generalized momentum may again be defined using \eqref{PdefGen}. This requires a choice of worldline $\mathcal{Z}$ and a foliation $\{ \fB_s \}$, leading to a vector which resides in $K^*_G( \mathcal{Z}, \{ \fB_s \})$. Applying \eqref{StressConsEM} shows that forces and torques follow from
\begin{equation}
    \frac{\rmd}{\rmd s} P_s(\xi) = \int_{\fB_s} \left( \frac{1}{2} T^{ab}_\mathrm{body} \Lie_\xi g_{ab} + F_{ab} \xi^a J^b \right) \rmd S.
    \label{ForceEM} 
\end{equation}
This is the simplest approach. It is not, however, the only reasonable possibility. $P_s$ has the unfortunate property that symmetries do not necessarily imply conservations laws. Even if there exists a vector field $\psi^a$ satisfying  $\Lie_\psi F_{ab} = \Lie_\psi g_{ab} = 0$, the associated momentum $P_s(\psi)$ is not necessarily conserved.

Dixon \cite{Dix74, Dix70a, BaileyIsrael} has proposed a different set of linear and angular momenta in this context which do form simple conservation laws in the presence of symmetries (among other desirable properties more generally). Translating his definitions into a generalized momentum $P^{\mathrm{D}}_s$ results in
\begin{align}
    P^\mathrm{D}_s (\xi) := P_s (\xi) 
     + \int_{\fB_s} \!\! \rmd S_a J^a(x) \int_0^1 \rmd u u^{-1} \sigma^{b'} (y_u',z_s) \xi^{c'}(y_u') F_{b'c'}(y_u').
    \label{PDixon}
\end{align}
The curve $\{y_u' \mid u \in [0,1] \}$ describes an affinely-parametrized geodesic satisfying $y_0 = z_s$ and $y_1 = x$. The correction to $P_s$ which is used here represents a type of interaction between the electromagnetic field and its source. It can be motivated using symmetry considerations \cite{Dix70a}, the theory of multipole moments \cite{Dix74}, or Lagrangian methods \cite{BaileyIsrael}. Some intuition for this interaction may be gained by introducing a vector potential $A_a$ so that $F_{ab} = 2 \nabla_{[a} A_{b]}$. Then, 
\begin{equation}
    P^\mathrm{D}_s + \left. q (A_a \xi^a) \right|_{z_s}= P_s + \int_{\fB_s} \!\! \rmd S_a J^a \left( A_b \xi^b - \int_0^1 \rmd u u^{-1} \sigma^{b'} \Lie_\xi A_{b'} \right).
    \label{PDixon2}
\end{equation}
Although this has been written in terms of a gauge-dependent vector potential, it is manifest from \eqref{PDixon} that $P_s$ and $P^\mathrm{D}_s$ are both gauge-invariant. 

Combining \eqref{PDixon2} with \eqref{ForceEM} shows that the generalized force associated with Dixon's momentum is
\begin{align}
    \frac{\rmd}{\rmd s} ( P_s^\mathrm{D} + q A_a \xi^a ) =& \int_{\fB_s} \!\! \rmd S \left( \frac{1}{2} T^{ab}_\mathrm{body} \Lie_\xi g_{ab} + J^a \Lie_\xi A_a \right)
    \nonumber
    \\
    &~ - \frac{\rmd}{\rmd s} \int_{\fB_s} \rmd S_a J^a \int_0^1 \rmd u u^{-1} \sigma^{a'} \Lie_\xi A_{a'}.
\end{align}
This is awkward to write more explicitly without introducing additional technical tools. Even so, it is simple to temporarily consider special cases where there exists a Killing vector field $\psi^a$ satisfying $\Lie_\psi F_{ab} = \Lie_\psi g_{ab} = 0$. It is then possible to find a vector potential $A^{(\psi)}_a$ such that $\Lie_\psi A^{(\psi)}_a = 0$. Using this,  the component of momentum conjugate to $\psi^a$ is seen to be conserved in the sense that
\begin{equation}
    P^\mathrm{D}_s (\psi) + q A^{(\psi)}_a \xi^a = (\mbox{constant}),
    \label{EMConsLaw}
\end{equation}
where $A_a^{(\psi)} \xi^a$ is understood to be evaluated at $z_s$. Although this is reminiscent of the canonical momentum associated with a pointlike test charge, it is valid for essentially arbitrary extended charge distributions.

\subsubsection{The self-field and self-force}

Electromagnetic self-forces may be defined and removed from either $P_s$ or $P^\mathrm{D}_s$. In both cases, it is convenient to work in the Lorenz gauge:
\begin{equation}
    \nabla^a A_a = 0.
\end{equation}
Maxwell's equations then reduce to the single hyperbolic equation
\begin{equation}
    \nabla^b \nabla_b A_a - R_{a}{}^{b} A_b = -4 \pi J_a.
    \label{BoxA}
\end{equation}
Introducing the parallel propagator $g^{a}{}_{a'} (x,x')$ \cite{PoissonRev}, consider a Green function $G_{aa'}$ satisfying
\begin{enumerate}
    \item $\Box G_{aa'} - R_{a}{}^{b} G_{ba'} = - 4 \pi g_{aa'} \delta(x,x')$,
    \item $G_{aa'}(x,x') = G_{a'a}(x',x)$,
    \item $G_{aa'}(x,x') = 0$ when $x$, $x'$ are timelike-separated.
\end{enumerate}
These are the closest possible analogs to the constraints used to define $G$ in Section \ref{Sect:SelfFieldSR}. They characterize the S-type Detweiler-Whiting Green function for the reduced Maxwell equation \eqref{BoxA}. In terms of the advanced and retarded Green functions $G_{aa'}^{\pm}$, there exists a homogeneous solution $V_{aa'}$ such that
\begin{equation}
    G_{aa'} = \frac{1}{2} ( G^+_{aa'} + G^-_{aa'} - V_{aa'} ) = \frac{1}{2} [ g_{aa'} \Delta^{1/2} \delta(\sigma) - V_{aa'} \Theta(\sigma)].
    \label{GexpandEM}
\end{equation}
Although it can be difficult to find $G_{aa'}$ explicitly in a particular spacetime, we assume that it is known.

A self-field $A_a^\mathrm{S}$ may be defined by convolving $J^a$ with $G_{aa'}$. Considering a spacetime volume $\mathfrak{R} \subseteq \fW$, let
\begin{equation}
    A_a^\mathrm{S}[\mathfrak{R}] := \int_{\mathfrak{R}} G_{aa'} J^{a'}\rmd V'.
\end{equation}
In the scalar case, the analog of this expression represents the ``self-field'' sourced in the region $\mathfrak{R}$. The interpretation here is somewhat more obscure, as the restriction of $J^a$ to arbitrary regions is not necessarily conserved and is therefore unphysical. This definition is nevertheless useful. Without specification of any particular region, it is to be understood that $\mathfrak{R} = \fW$ so $A_a^\mathrm{S} = A_a^\mathrm{S} [\fW]$. This (full) self-field is sourced by a conserved current, implying that $F_{ab}^\mathrm{S} := 2 \nabla_{[a} A_{b]}^\mathrm{S}$ satisfies the complete Maxwell equations
\begin{equation}
    \nabla_{[a} F^\mathrm{S}_{bc]} = 0, \qquad \nabla^b F^\mathrm{S}_{ab} = 4\pi J_a.
\end{equation}
The same cannot necessarily be said for fields $2 \nabla_{[a} A_{b]}^\mathrm{S} [ \mathfrak{R} ]$ where $\mathfrak{R} \neq \fW$.

Now consider changes in $P_s$ due to $A_a^\mathrm{S}$. Applying the arguments of Sections \ref{Sect:SelfFieldSR} and \ref{Sect:Renormalize} to \eqref{ForceEM} shows that the self-force has the form \eqref{calFdef} with
\begin{equation}
    f(x,x') = 2 \xi^a J^{b} J^{b'} \nabla_{[a} G_{b]b'}.
\end{equation}
The average $f(x,x') + f(x',x)$ of action-reaction pairs reduces in this case to
\begin{equation}
    J^a J^{a'} \Lie_\xi G_{aa'} - \nabla_a ( J^a J^{a'} \xi^b G_{ba'} ) - \nabla_{a'} ( J^a J^{a'} \xi^{b'} G_{ab'} ).
\end{equation}
The divergences which appear here are new features of the electromagnetic problem. Only their integrals matter, however, so they are easily dealt with. Defining the homogeneous effective field
\begin{equation}
    \hat{F}_{ab} := F_{ab} - F_{ab}^\mathrm{S},
    \label{FhatDefEM}
\end{equation}    
the final result is that 
\begin{equation}
    \frac{\rmd}{\rmd s} \hat{P}_s = \int_{\fB_s} \!\! \rmd S \left( \frac{1}{2} T^{ab}_\mathrm{body} \Lie_\xi g_{ab} + \xi^a J^b \hat{F}_{ab} + \frac{1}{2} \int_{\fW} \! \! \rmd V' J^a J^{a'} \Lie_\xi G_{aa'} \right),
    \label{forceEM1}
\end{equation}
where $\hat{P}_s := P_s+E_s$ and 
\begin{align}
    E_s := \frac{1}{2} \left( \int_{\fB_s^+} \!\! \rmd V J^a \Lie_\xi A_a^\mathrm{S} [\fB^-_s] - \int_{\fB_s^-} \!\! \rmd V J^a \Lie_\xi A_a^\mathrm{S} [\fB_s^+] \right) 
    \nonumber
    \\
    ~ + \int_{\fB_s} \! \! (\xi^a A_a^\mathrm{S}) J^b \rmd S_b.
\end{align}
The renormalized momentum $\hat{P}_s$ therefore evolves via $\hat{F}_{ab}$ rather than $F_{ab}$. As explained in Section \ref{Sect:RenormalizeCurved}, the forces involving $\Lie_\xi g_{ab}$ and $\Lie_\xi G_{aa'}$ in \eqref{forceEM1} combine in a natural way to form an effective gravitational force. Furthermore, $E_s$ is known to reduce in simple cases to the expected expression involving the stress-energy tensor associated with $F_{ab}^\mathrm{S}$ \cite{HarteEM}. More generally, it should be thought of as a quasi-local functional of $J^a$.

A very similar result may be derived using the Dixon momentum $P^\mathrm{D}_s$. This is most easily accomplished if a new self-momentum is introduced which satisfies
\begin{align}
    E^\mathrm{D}_s & := E_s - \int_{\fB_s} \rmd S_a J^a \int_0^1 \rmd u u^{-1} \sigma^{b'} \xi^{c'} F^\mathrm{S}_{b'c'}.
\end{align}
Defining $\hat{P}^\mathrm{D}_s := P_s^\mathrm{D} + E^\mathrm{D}_s$, it follows that
\begin{align}
    \frac{\rmd}{\rmd s} \hat{P}_s^\mathrm{D} = \int_{\fB_s} \!\! \rmd S \left( \frac{1}{2} T^{ab}_\mathrm{body} \Lie_\xi g_{ab} + \xi^a J^b \hat{F}_{ab} + \frac{1}{2} \int_{\fW} \! \! \rmd V' J^a J^{a'} \Lie_\xi G_{aa'} \right)
    \nonumber
    \\
    ~ + \frac{\rmd}{\rmd s} \int_{\fB_s} \rmd S_a J^a \int_0^1 \rmd u u^{-1} \sigma^{b'} \xi^{c'} \hat{F}_{b'c'},
\end{align}
or equivalently
\begin{align}
    \frac{\rmd}{\rmd s} ( \hat{P}_s^\mathrm{D} + q \hat{A}_a \xi^a) = \int_{\fB_s} \!\! \rmd S \left( \frac{1}{2} T^{ab}_\mathrm{body} \Lie_\xi g_{ab} + J^a \Lie_\xi \hat{A}_a  + \frac{1}{2} \int_{\fW} \! \! \rmd V' J^a J^{a'} \Lie_\xi G_{aa'} \right)
    \nonumber
    \\
    ~ - \frac{\rmd}{\rmd s} \int_{\fB_s} \rmd S_a J^a \int_0^1 \rmd u u^{-1} \sigma^{a'} \Lie_\xi \hat{A}_{a'}
\end{align}
for any vector potential $\hat{A}_a$ satisfying $\hat{F}_{ab} = 2 \nabla_{[a} \hat{A}_{b]}$. 

If a Killing vector field $\psi^a$ exists which satisfies $\Lie_\psi \hat{F}_{ab} = \Lie_\psi g_{ab} = 0$, it is possible to choose a vector potential for $\hat{F}_{ab}$ with the property that $\Lie_\psi \hat{A}^{(\psi)}_a = 0$. It then follows immediately that $\hat{P}_s^\mathrm{D} + q \hat{A}^{(\psi)}_a \xi^a$ is conserved. No similarly-simple conservation law is associated with $\hat{P}_s$.

\subsubsection{Multipole expansions}

Integral expressions for the generalized force are not particularly useful on their own. It is instead more interesting to consider their multipole expansions.  Unlike in the scalar theories discussed earlier, more than one ``reasonable'' force may be considered in the electromagnetic case. As a matter of computation, it is simplest to expand the force associated with $\hat{P}_s$. Dixon's momentum is otherwise more attractive, however. Multipole series for the associated forces and torques have already been derived in a test body approximation \cite{Dix74}. The mechanics of the calculation are exactly the same here, and result in
\begin{align}
    \frac{ \rmd }{ \rmd s } \hat{P}^\mathrm{D}_s = \frac{1}{2} \sum_{n=2}^\infty \frac{1}{n!} \hat{I}^{c_1 \cdots c_n ab} \Lie_\xi g_{ab,c_1 \cdots c_n} + q \hat{F}_{ab} \xi^a \dot{z}^b_s 
    \nonumber
    \\
    ~ + \sum_{n=1}^\infty \frac{n}{(n+1)!} q^{b_1 \cdots b_n a} \Lie_\xi \hat{F}_{a b_1, b_2 \cdots b_n}.
    \label{EMmultipole}
\end{align}
The coefficients $q^{b_1 \cdots b_n a}$ represent $2^n$-pole moments of $J^a$ as defined (using the notation $m^{b_1 \cdots b_n a}$) by Dixon \cite{Dix74, Dix70b}. They are not to be confused with the scalar charge moments satisfying \eqref{ScalarMultipoleRel}. The $\hat{I}^{c_1 \cdots c_n ab}$ appearing in \eqref{EMmultipole} represent $2^n$-pole moments of $T^{ab}_\mathrm{body}$ as renormalized by Lie derivatives of $G_{aa'}$. Again, these are different from the renormalized stress-energy moments which appear in the scalar law of motion \eqref{ScalarGenForce}. In the limit of negligible self-interaction, however, both definitions reduce to Dixon's stress-energy moments \cite{Dix74}. For reference, the first tensor extensions of $\hat{F}_{ab}$ are explicitly
\begin{equation}
    \hat{F}_{ab,c} = \nabla_c \hat{F}_{ab}, \qquad \hat{F}_{ab,c d} = \nabla_{(c} \nabla_{d)} \hat{F}_{ab} - \frac{2}{3} \hat{F}_{f[a} R_{b](cd)}{}^{f}.
\end{equation}

If there exists a Killing vector $\psi^a$ which satisfies $\Lie_\psi \hat{F}_{ab} = \Lie_\psi \hat{A}^{(\psi)}_a = 0$, it has already been stated that $\hat{P}^\mathrm{D}_s(\psi) + q \hat{A}^{(\psi)}_a \psi^a$ is conserved exactly. It is evident from \eqref{EMmultipole} that this quantity is also conserved in any consistent truncation of the multipole series. If a particular $\psi^a$ is Killing but does not necessarily preserve $\hat{F}_{ab}$, all gravitational terms vanish from \eqref{EMmultipole} and
\begin{equation}
        \frac{ \rmd }{ \rmd s } \hat{P}^\mathrm{D}_s(\psi) = q \hat{F}_{ab} \psi^a \dot{z}^b_s + \sum_{n=1}^\infty \frac{n}{(n+1)!} q^{b_1 \cdots b_n a} \Lie_\psi \hat{F}_{a b_1, b_2 \cdots b_n}.
\end{equation}
This holds for all $\psi^a \in K \subseteq K_G$. It is all that arises for charges moving in flat or de Sitter spacetimes.

\subsubsection{Linear and angular momentum}

Linear and angular momenta may be extracted from $\hat{P}^\mathrm{D}_s$ using the methods described in Sections \ref{Sect:PtopS} and \ref{Sect:PtopsScalar}. Let $\hat{p}^a$ and $\hat{S}^{ab}$ be defined by
\begin{equation}
    \hat{P}^\mathrm{D}_s = \hat{p}^a \xi_a + \frac{1}{2} \hat{S}^{ab} \nabla_a \xi_b.
\end{equation}
Differentiating this and varying over all generalized Killing fields shows that
\begin{align}
    \frac{\rmD \hat{p}^a}{\rmd s} = q \hat{F}^{a}{}_{b} \dot{z}^b_s + \frac{1}{2} R_{bcd}{}^{a} \hat{S}^{bc} \dot{z}^d_s + \hat{F}^a, \qquad     \frac{\rmD \hat{S}^{ab}}{\rmd s} = 2 \hat{p}^{[a} \dot{z}_s^{b]} + \hat{N}^{ab},
    \label{EOMem}
\end{align}
where
\begin{align}
    \hat{F}_a = \frac{1}{2} \sum_{n=2}^\infty \frac{1}{n!} \hat{I}^{d_1 \cdots d_n bc} \nabla_a g_{bc, d_1 \cdots d_n} + \sum_{n=1}^\infty \frac{n}{(n+1)!} q^{c_1 \cdots c_n b} \nabla_a \hat{F}_{b c_1, c_2 \cdots c_n},
    \label{ForceEMmultipole}
    \\
    \hat{N}^{ab} =  \sum_{n=2}^\infty \frac{2}{n!} g^{f[b} \big( \hat{I}^{|c_1 \cdots c_n|a]d} g_{df,c_1 \cdots c_n} + \frac{n}{2} \hat{I}^{a] c_1 \cdots c_{n-1} dh} g_{dh,c_1 \cdots c_{n-1} f}  \big)
    \nonumber
    \\
    ~ + \sum_{n=0}^\infty \frac{2}{n!} g^{f[b} q^{a] c_1 \cdots c_n d} \hat{F}_{df,c_1 \cdots c_n}.
    \label{TorqueEMmultipole}
\end{align}
Electromagnetic forces and torques defined in this way first arise at dipole order. The (monopole) Lorentz force depends --- unlike any other electromagnetic terms  --- on $\dot{z}^a_s$. It has therefore been separated out explicitly in \eqref{EOMem}. There is no similarly velocity-dependent electromagnetic torque.

Defining the force to exclude the Lorentz component has the advantage that if $\psi^a$ is Killing and preserves $\hat{F}_{ab}$, the associated conservation law implies that
\begin{equation}
    \hat{F}_a \psi^a + \frac{1}{2} \hat{N}_{ab} \nabla^a \psi^b = 0.
\end{equation}
This does not involve $q$, and is directly analogous to the scalar result \eqref{ForceSymmetry}.

\subsubsection{Center of mass}

The electromagnetic laws of motion \eqref{EOMem}-\eqref{TorqueEMmultipole}  depend on both the foliation $\{ \fB_s \}$ and the worldline $\mathcal{Z}$. Center of mass conditions may be used to fix these structures as described in Section \ref{Sect:CMSR}. An evolution equation for the center of mass position $\gamma_s$ can then be obtained by differentiating $\hat{p}_a \hat{S}^{ab} = 0$. The result differs slightly from \eqref{pCM} due to the additional velocity-dependence associated with the Lorentz force. Adapting the methods of \cite{EhlersRudolph}, the electromagnetic momentum-velocity relation may be shown to be
\begin{equation}
    \hat{m} \dot{\gamma}^a_s = \hat{p}^a - \hat{N}^{a}{}_{b} u^b - \frac{ \hat{S}^{ab} [ \hat{m} \hat{F}_b + (\hat{p}^c - \hat{N}^{c}{}_{h} u^h ) ( q \hat{F}_{bc} - \frac{1}{2}  R_{bcdf} \hat{S}^{df} )] }{ \hat{m}^2 - \frac{1}{2} \hat{S}^{pq} (q \hat{F}_{pq} - \frac{1}{2} R_{pqrs} \hat{S}^{rs} ) }
    \label{pCMem}
\end{equation}
when $s$ has been chosen such that $\dot{\gamma}^a_s \hat{p}_a = - \hat{m}$ and $u^a$ is the unit timelike vector satisfying $\hat{p}^a = \hat{m} u^a$. 

Although a precise set of assumptions which imply this are not known, it is assumed here that the center of mass condition \eqref{pCM} admits a unique timelike solution in a broad class of physical systems. Inspection of \eqref{pCMem} shows a sufficient (but not necessary) condition for this to fail is
\begin{equation}
  \hat{m}^2 - \frac{1}{2} \hat{S}^{bc} (q \hat{F}_{bc} - \frac{1}{2} R_{bcdf} \hat{S}^{df} ) = 0.
\end{equation}
While the $q = 0$ case of this equation was dismissed in Section \ref{Sect:CMSR} as unlikely to be physically relevant, the charged case is potentially more interesting. Consider an electron\footnote{It is unclear that there is any sense in which an electron's behavior can be modeled using equations derived for classical extended charges. Nevertheless, the example appears to be suggestive.} in a magnetic field of order $B$. Setting $q = e$, $\hat{m} = m_e$, $R_{abc}{}^{d} = 0$, and $\hat{S} = \hbar/2$, the denominator in \eqref{pCMem} can diverge when $B \sim 2 m_e^2/e \hbar \sim 10^{14} \, \mathrm{Gauss}$. This may be viewed as the field strength at which an electron's dipole energy $\hbar B/2$ becomes comparable to its rest mass. Quantum mechanically, it is also the field strength at which the separation between (Landau) energy levels becomes comparable to the rest mass energy. Indeed, this is the scale at which quantum electrodynamics is expected to become dominant. Although systems with $10^{14} \, \mathrm{Gauss}$ magnetic fields are far from direct laboratory experience, such fields are believed to exist around some neutron stars \cite{NeutronStarB}. Even in somewhat smaller magnetic fields, hidden momentum effects predicted by the classical theory can become very large. Whether or not this has qualitative consequences for neutron star astrophysics is an open question.

If the center of mass can be defined and the classical laws of physics remain valid, \eqref{pCMem} may be combined with \eqref{EOMem}-\eqref{TorqueEMmultipole} to yield very general laws of motion. As in the scalar case, $\hat{S}^{ab}$ can also be reduced to a single spin vector $\hat{S}^a$ satisfying \eqref{spinEvolve}. Similarly, the mass varies according to \eqref{mEvolve}. Unlike in the scalar case, matter coupled to electromagnetic fields can change mass only at dipole and higher orders. Such effects are related to changes in a body's internal energy due to work performed by (or against) the ambient fields \cite{Dix70a}.

\subsubsection{Monopole approximation}

In simple cases, the laws of motion governing an extended charge distribution may be truncated at monopole order. Within this approximation, $\hat{F}_a = \hat{N}_{ab} = 0$ and \eqref{EOMem} reduces to
\begin{equation}
    \frac{ \rmD \hat{p}^a}{ \rmd s} = q \hat{F}^{a}{}_{b} \dot{\gamma}^b_s + \frac{1}{2} R_{bcd}{}^{a} \hat{S}^{bc} \dot{\gamma}^d_s, \qquad \frac{ \rmD \hat{S}_{ab} }{ \rmd s} = 2 \hat{p}^{[a} \dot{\gamma}^{b]}_s.
\end{equation}
Contracting the second of these equations with $\hat{S}_{ab}$ while using \eqref{CM} again shows that $\hat{S}^2 := \hat{S}^a \hat{S}_a = (\mbox{constant})$. An object which is initially not spinning therefore remains non-spinning in this approximation. Consider these cases for simplicity. The momentum-velocity relation then reduces to $\hat{p}^a = \hat{m} \dot{\gamma}^a_s$, the mass remains constant, and the body accelerates via the Lorentz force law
\begin{equation}
    \hat{m} \ddot{\gamma}^a_s  = q \hat{F}^{a}{}_{b} \dot{\gamma}^b_s 
    \label{Lorentz}
\end{equation}
in the effective electromagnetic field $\hat{F}_{ab}$. The effective field always satisfies the vacuum Maxwell equations, and is generically distinct from the physical field $F_{ab}$ which governs the acceleration of nearby test charges.

In many cases of interest, it is useful to model $F_{ab}$ as the sum of some external field $F_{ab}^\mathrm{ext}$ and the retarded field associated with a body's charge distribution:
\begin{equation}
    F_{ab} = F_{ab}^\mathrm{ext} + 2 \int_\fW \! \nabla_{[a} G^-_{b]b'} J^{b'} \rmd V'.
\end{equation}
In these cases, it follows from \eqref{GexpandEM} and \eqref{FhatDefEM} that the effective field must satisfy
\begin{equation}
    \hat{F}_{ab} = F_{ab}^\mathrm{ext} + \int_\fW \! \nabla_{[a} ( G^-_{b]b'} - G^+_{b]b'} + V_{b]b'}) J^{b'} \rmd V'.
    \label{FhatEM}
\end{equation}

Performing a point particle limit as discussed in Section \ref{Sect:ScalarMonopole} and \cite{HarteEM, GrallaHarteWaldEM}, the lowest-order self-interaction effects follow from \eqref{FhatEM} with a pointlike current. Through first order in the expansion parameter $\lambda$,
\begin{equation}
    \hat{F}_{ab} = F_{ab}^\mathrm{ext} + q \int \nabla_{[a} ( G^-_{b]b'} - G^+_{b]b'} + V_{b]b'}) \dot{\gamma}^{b'}_s \rmd s'.
    \label{PointField}
\end{equation}
Evaluating this on $\gamma_s$ and recalling the projection operator $h^{ab} = g^{ab} + \dot{\gamma}^a_s \dot{\gamma}^b_s$, it may be shown that \cite{PoissonRev}

\begin{align}
    \hat{F}_{ab} = F_{ab}^\mathrm{ext} + \frac{4}{3} q \dot{\gamma}^l_s g_{l[a} h_{b]}{}^{c} \dot{\gamma}^d_s \big[ (q/\hat{m}) \dot{\gamma}_s^f \nabla_f F^{\mathrm{ext}}_{cd} + (q/\hat{m})^2 g^{fh} F_{cf}^\mathrm{ext} F^\mathrm{ext}_{hd} 
    \nonumber
    \\
    ~ + \frac{1}{2} R_{cd} \big] + 2 q \lim_{\epsilon \rightarrow 0^+} \int_{-\infty}^{s-\epsilon} \!\! \nabla_{[a} G^-_{b]b'} (\gamma_s , \gamma_{s'})  \dot{\gamma}_{s'}^{b'} \rmd s'.
\end{align}
Substitution into \eqref{Lorentz} finally yields the equation of motion for a self-interacting ``pointlike'' electric charge:
\begin{align}
  \hat{m} \ddot{\gamma}_s^a = q g^{ab} F_{bc}^\mathrm{ext} \dot{\gamma}^c_s + \frac{2}{3} (q^3/\hat{m}) h^{ab} \dot{\gamma}_s^c \big[ \dot{\gamma}^d_s \nabla_d F_{bc}^\mathrm{ext} + (q/\hat{m}) g^{df} F_{bd}^\mathrm{ext} F_{fc}^\mathrm{ext} \big]
  \nonumber
  \\
  ~ + \frac{1}{3} q^2 h^{ab} R_{bc} \dot{\gamma}^c_s + 2 q^2 \lim_{\epsilon \rightarrow 0^+} \int_{-\infty}^{s-\epsilon} \!\! \nabla_{[a} G^-_{b]b'} \dot{\gamma}_s^b \dot{\gamma}_{s'}^{b'} \rmd s'.
  \label{LorentzDirac}
\end{align}
In curved spacetime, this is a reduced-order version --- see Footnote \ref{Foot:OrderReduction} --- of a result first obtained by DeWitt and Brehme \cite{DeWittBrehme} (with corrections due to Hobbs \cite{Hobbs}). The second line of \eqref{LorentzDirac} vanishes in flat spacetime, leaving only
\begin{align}
  \hat{m} \ddot{\gamma}_s^a = q g^{ab} F_{bc}^\mathrm{ext} \dot{\gamma}^c_s + \frac{2}{3} (q^3/\hat{m}) h^{ab} \dot{\gamma}_s^c \big[ \dot{\gamma}^d_s \nabla_d F_{bc}^\mathrm{ext} + (q/\hat{m}) g^{df} F_{bd}^\mathrm{ext} F_{fc}^\mathrm{ext} \big].
\end{align}
This is essentially the Abraham-Lorentz-Dirac equation \cite{Jackson}, but in a reduced-order form typically attributed to Landau and Lifshiftz \cite{Landau}. 

While instructive, the neglect of spin and electromagnetic dipole effects in \eqref{LorentzDirac} can be overly restrictive. This restriction is easily dropped by making use of the general multipole expansions derived above. The resulting changes are qualitatively significant \cite{HarteEM} (see also \cite{GrallaHarteWaldEM}): A hidden momentum appears, external fields may change an object's rest mass, and the spin magnitude may change due to torques exerted by $F_{ab}^\mathrm{ext}$.

For monopole point particles in flat spacetime, the use of $\hat{F}_{ab}$ instead of $F_{ab}$ in the laws of motion was first suggested by Dirac \cite{Dirac}. The appropriate generalization for monopole particles in curved spacetimes was obtained much more recently by Detweiler and Whiting \cite{DetWhiting}. Both of these proposals were essentially physically-motivated axioms intended to define the dynamics of point charges. The discussion here, which follows \cite{HarteEM}, shows that these regularization schemes are actually limits of laws of motion which hold rigorously for nonsingular extended charge distributions. Once the general laws of motion \eqref{EOMem}-\eqref{TorqueEMmultipole} and \eqref{pCMem} have been derived, more explicit results such as \eqref{LorentzDirac} follow very easily, as do their spin-dependent generalizations.

\subsection{General relativity}
\label{Sect: GR}

All results discussed up to this point assume that the spacetime metric is known beforehand and has been fixed. This assumption may be relaxed. Doing so allows the consideration of self-gravitating masses in general relativity. We take a minimal approach to this problem by adapting the techniques of the previous sections as closely as possible. Although there are some deficiencies to this strategy, considerable progress can still be made.

Let the spacetime be described by $(\mathcal{M}, g_{ab})$ and the body of interest be contained inside a spatially-compact worldtube $\fW \subset \mathcal{M}$. Only the purely gravitational problem is considered here, meaning that objects can interact with each other solely via their influence on the metric; electromagnetic and similar long-range interactions are excluded. It follows that $T^{ab}_\mathrm{body} = T^{ab}_\mathrm{tot}$ inside $\fW$. Letting $g_{ab}$ be a solution to Einstein's equation 
\begin{equation}
    R_{ab} - \frac{1}{2} g_{ab} R = 8\pi g_{ac} g_{bd} T^{cd}_\mathrm{body},
    \label{Einstein}
\end{equation}
it follows that
\begin{equation}
    \nabla_b T^{ab}_{\mathrm{body}} = 0.
    \label{StressConsGR}
\end{equation}
These two equations replace, e.g., \eqref{phiField} and \eqref{StressConsScalar} used to understand the motion of scalar charges. Unlike in the scalar or electromagnetic cases, the gravitational law of motion \eqref{StressConsGR} is a consequence of the field equation, not an independent assumption. 

Both $T^{ab}_{\mathrm{body}}$ and the metric inside $\fW$ are assumed to be smooth. This precludes the consideration of black holes. Although unfortunate, it appears difficult to remove this restriction in a non-perturbative theory which describes the motions of individual objects (and not only characteristics of the entire spacetime). While quantities such as momenta might be associated with a black hole horizon \cite{BHp1, BHp2}, adopting such definitions typically excludes the discussion of objects without horizons. Alternatively, one may consider an effective background and  compute momenta associated with, e.g., the Landau-Lifshitz pseudotensor \cite{KipBobbing1, KipBobbing2}. This is troublesome as well. Nevertheless, the motion of black holes can be sensibly discussed within certain approximation schemes \cite{PoissonRev, GrallaWald, PoundSF}. These consider only the metric outside of the body of interest, and apply versions of matched asymptotic expansions in appropriate ``buffer regions.'' We instead consider internal metrics as well as external ones, but do not require the existence of a buffer region. 

As in other theories, the laws of motion derived here involve an effective field which is distinct from the physical one. In general relativity, the relevant field is the metric. We therefore use two metrics, which makes index raising and lowering ambiguous. All factors of the appropriate metric are therefore displayed explicitly in this section.

\subsubsection{Generalized momentum}

The first step to understanding the motion of a self-gravitating mass is to write down a generalized momentum which describes an object's large-scale behavior. Introducing a foliation $\{ \fB_s \}$ of $\fW$ and a worldline $\mathcal{Z}$, the linear and angular momenta proposed by Dixon \cite{Dix74, Dix79, Dix70a} are conjugate to generalized Killing fields constructed using the physical metric $g_{ab}$. Adding an extra argument to $K_G$ to reflect this metric dependence, the appropriate generalized momentum which contains Dixon's definitions is
\begin{equation}
    P^\mathrm{D}_s (\xi) := \int_{\fB_s} \!\! T^{ab}_\mathrm{body} g_{bc} \xi^c \rmd S_a, \qquad \xi^a \in K_G( \mathcal{Z}, \{ \fB_s \}; g).
\end{equation}
For each $s$, this is an element of the ten-dimensional vector space $K^*_G ( \mathcal{Z}, \{ \fB_s \}; g)$. The associated linear and angular momenta have a number of useful properties \cite{Dix74, EhlersRudolph, SchattStreub1, SchattStreub2}. Using \eqref{StressConsGR}, their time evolution satisfies
\begin{equation}
    \frac{\rmd}{\rmd s} P^\mathrm{D}_s(\xi) = \frac{1}{2} \int_{\fB_s} \!\! T^{ab}_\mathrm{body} \Lie_\xi g_{ab}  \rmd S.
    \label{pDotGR}
\end{equation}
If $\Lie_\xi g_{ab}$ varies slowly throughout $\fB_s$, forces and torques can be expanded in multipole series as described in \cite{HarteBrokenSyms, Dix74} and in Section \ref{Sect:ChargeCurvedMultipole}.

While such assumptions can be useful for test bodies, they are too strong for self-gravitating masses. Moving beyond the test body regime first requires the introduction of an effective metric $\hat{g}_{ab}$ such that --- after appropriate renormalizations --- the $\Lie_\xi g_{ab}$ appearing in \eqref{pDotGR} can be replaced by $\Lie_\xi \hat{g}_{ab}$. If $\Lie_\xi \hat{g}_{ab}$ varies slowly, the resulting integral for the generalized force can be expanded in a multipole series in the usual way. One additional subtlety which occurs in the gravitational problem is that even if a particular $\hat{g}_{ab}$ can itself be adequately approximated using a low-order Taylor expansion, the same can not necessarily be said for $\Lie_\xi \hat{g}_{ab}$ when $\xi^a \in K_G( \mathcal{Z}, \{ \fB_s \}; g)$. The generalized Killing fields associated with Dixon's momenta involve the physical metric and all of its attendant difficulties. These difficulties are partially inherited by the generalized Killing fields used to define $P^\mathrm{D}_s$.

One way out of this problem\footnote{It would be more elegant to instead demand that $K_G( \mathcal{Z}, \{ \fB_s \} ; \hat{g})$ and $K_G( \mathcal{Z}, \{ \fB_s \} ; g)$ be identical or otherwise closely related. Such an assumption would restrict possible relations between $g_{ab}$ and $\hat{g}_{ab}$, and is an avenue which has not been explored.} is to choose a bare momentum $P_s (\xi)$ defined by an integral which is structurally identical to \eqref{pDotGR}, but where all $\xi^a$ are elements of $K_G( \mathcal{Z}, \{ \fB_s \} ; \hat{g})$ rather than $K_G( \mathcal{Z}, \{ \fB_s \} ; g)$. Let
\begin{equation}
    P_s (\xi) := \int_{\fB_s} \!\! T^{ab}_\mathrm{body} g_{bc} \xi^c \rmd S_a , \qquad \xi^a \in K_G( \mathcal{Z}, \{ \fB_s \} ; \hat{g}).
    \label{PDefBareGR}
\end{equation}
We take this to be the bare momentum of a self-gravitating mass. The effective metric which appears here is to be regarded at this stage as an additional parameter. The generalized force $\rmd P_s/\rmd s$ follows from a trivial modification of \eqref{pDotGR}. Furthermore, Dixon's momenta are recovered in a test mass limit where $\hat{g}_{ab} \approx g_{b}$. If there exists a $\psi^a \in K_G( \mathcal{Z}, \{ \fB_s \} ; \hat{g})$ such that $\Lie_\psi g_{ab} = 0$, it is evident that $P_s (\psi)$ must be conserved.

\subsubsection{Self-fields and laws of motion}

There are many possible ways to extract an effective metric $\hat{g}_{ab}$ from the physical metric $g_{ab}$. The simplest generalization of the previous discussions involves a two-point tensor field $G_{aba'b'}(x,x')$ which satisfies
\begin{equation}
    G_{aba'b'} = G_{(ab)a'b'} = G_{ab(a'b')}
    \label{GreenSymIndices}
\end{equation}
and
\begin{equation}
  G_{aba'b'}(x,x') = G_{a'b' ab }(x',x).
  \label{GreenSymGR}
\end{equation}
For any such propagator, consider an effective metric $\hat{g}_{ab}$ defined via
\begin{equation}
    g_{ab} = \hat{g}_{ab} + g^\mathrm{S}_{ab},
    \label{gHatDef}
\end{equation}
where
\begin{equation}
  g_{ab}^\mathrm{S}[\mathfrak{R}] := \int_{\mathfrak{R}} \! G_{aba'b'} T^{a'b'}_\mathrm{body} \rmd V'
  \label{gSDef}
\end{equation}
and $g_{ab}^\mathrm{S} = g_{ab}^\mathrm{S}[\fW]$. Note that the volume element in this ``self-field'' is the one associated with $g_{ab}$, not $\hat{g}_{ab}$. Substituting \eqref{gHatDef} into the appropriately-modified form of \eqref{pDotGR} shows that the $s$-integral of the self-field's contribution to $\rmd P_s/ \rmd s$ has the form \eqref{calFdef} with force density
\begin{equation}
    f = \frac{1}{2} T^{ab}_\mathrm{body} T^{a'b'}_\mathrm{body} (\xi^c \nabla_c G_{aba'b'} + 2 \nabla_{a} \xi^c G_{bca'b'} ).
\end{equation}
Applying \eqref{ForceAv} and related results then transforms the law of motion into
\begin{align}
    \frac{\rmd}{\rmd s} \hat{P}_s = \frac{1}{2} \int_{\fB_s} \!\! T^{ab}_\mathrm{body} \Lie_\xi \hat{g}_{ab} \rmd S + \frac{1}{4} \int_{\fB_s} \!\! \rmd S  \int_\fW \! \rmd V' T^{ab}_\mathrm{body} T^{a'b'}_\mathrm{body} \Lie_\xi G_{aba'b'},
    \label{pDotGR2}
\end{align}
where $\hat{P}_s = P_s + E_s$ and
\begin{equation}
    E_s = \frac{1}{4} \left( \int_{\fB_s^+} \!\! T^{ab}_\mathrm{body} \Lie_\xi g_{ab}^\mathrm{S} [ \fB_s^- ] \rmd V - \int_{\fB_s^-} \!\! T^{ab}_\mathrm{body} \Lie_\xi g_{ab}^\mathrm{S} [ \fB_s^+ ] \rmd V \right).
\end{equation}
$E_s$ is a functional of $T^{ab}_\mathrm{body}$ which effectively acts like the momentum of the self-field. If $\Lie_\xi G_{aba'b'}$ is a linear functional of $\Lie_\xi \hat{g}_{ab}$, the term involving $\Lie_\xi G_{aba'b'}$ in \eqref{pDotGR2} renormalizes a body's quadrupole and higher multipole moments. In these cases, a multipole expansion of \eqref{pDotGR2} yields
\begin{equation}
    \frac{\rmd}{\rmd s} \hat{P}_s (\xi) = \frac{1}{2} \sum_{n=2}^\infty \frac{1}{n!} \hat{I}^{c_1 \cdots c_n ab} \Lie_\xi \hat{g}_{ab,c_1 \cdots c_n}.
    \label{pDotGRMultipole}
\end{equation}
The tensor extensions appearing here are extensions of $\hat{g}_{ab}$ in a spacetime with metric $\hat{g}_{ab}$. This means, for example, that $\hat{g}_{ab,c} = 0$ and $\hat{g}_{ab,cd} = \frac{2}{3} \hat{R}_{a(cd)}{}^{f} \hat{g}_{bf}$ [cf. \eqref{MetricExtensions}].

Equation \eqref{pDotGRMultipole} represents not a particular law of motion, but a class of them. This is because many different propagators may be found which satisfy \eqref{GreenSymIndices} and \eqref{GreenSymGR}, and whose Lie derivatives with respect to $\xi^a$ are quasi-local in $\Lie_\xi \hat{g}_{ab}$. Choosing different propagators with these properties leads to different effective metrics, different self-momenta, and different effective multipole moments. Any of these combinations satisfies \eqref{pDotGR2}, and also \eqref{pDotGRMultipole} when the appropriate $\hat{g}_{ab}$ is sufficiently well-behaved. It is of course preferable to choose a propagator such that the associated multipole series may ``typically'' be truncated at low order without significant loss of accuracy. This condition is vague. In the electromagnetic and scalar theories, a (rather imperfect) proxy was the requirement that the effective fields be solutions to the vacuum field equation. The analogous condition in general relativity would be $\hat{R}_{ab} = 0$, where $\hat{R}_{ab}$ denotes the Ricci tensor associated with $\hat{g}_{ab}$. This does not appear to be possible within the currently-considered class of effective metrics. 

We take a pragmatic approach and suppose that $G_{aba'b'}$ is the Detweiler-Whiting S-type Green function associated with
\begin{align}
    \hat{g}^{cd} \hat{\nabla}_c \hat{\nabla}_d G_{aba'b'} - \big[ 2 \hat{g}^{cf} \hat{R}_{f(ab)}{}^{d}  + \frac{1}{2} \big( \hat{g}^{cd} \hat{R}_{ab} + \hat{g}_{ab} \hat{g}^{cf} \hat{g}^{dh} \hat{R}_{fh} \big) \big] G_{cda'b'} 
    \nonumber
    \\
    ~ = - 16 \pi \big( \hat{g}_{ac} \hat{g}_{bd} - \frac{1}{2} \hat{g}_{ab} \hat{g}_{cd} \big) \hat{g}^{c}{}_{(a'} \hat{g}^{d}{}_{b')} \hat{\delta} (x,x').
    \label{EinsteinLin}
\end{align}
This is essentially the prescription suggested in \cite{HarteGR}. In terms of retarded and advanced solutions $G^\pm_{aba'b'}$ to \eqref{EinsteinLin}, the S-type Green function satisfies
\begin{align}
    G_{aba'b'} = \frac{1}{2} ( G^+_{aba'b'} + G^-_{aba'b'} - V_{aba'b'})
\end{align}
for some bitensor $V_{aba'b'}$ which is an appropriate solution to the homogeneous version of \eqref{EinsteinLin}. Somewhat more explicitly,
\begin{equation}
    G_{aba'b'} = 2 \big( \hat{g}_{ac} \hat{g}_{bd} - \frac{1}{2} \hat{g}_{ab} \hat{g}_{cd} \big) \hat{g}^{c}{}_{(a'} \hat{g}^{d}{}_{b')} \hat{\Delta}^{1/2} \delta( \hat{\sigma} ) - \frac{1}{2} V_{aba'b'} \Theta( \hat{\sigma} ).
\end{equation}
Coupling this Green function with \eqref{gHatDef} and \eqref{gSDef} defines $\hat{g}_{ab}$ in terms of $g_{ab}$. Unlike in scalar or electromagnetic theories, the definition of the effective field is highly implicit in general relativity. It reduces to a simple subtraction only in linear perturbation theory. More generally, the Green function itself depends on the field which one intends to find, and must be found by iteration or related methods.

Despite the nonlinearity of the map $g_{ab} \rightarrow \hat{g}_{ab}$, the effective metric does not necessarily satisfy the vacuum Einstein equation. The specific differential equation \eqref{EinsteinLin} is nevertheless inspired by Lorenz-gauge perturbation theory, and has the following desirable properties:
\begin{enumerate}
    \item If $g_{ab}$ is sufficiently close to a background metric $\bar{g}_{ab}$ satisfying the vacuum Einstein equation $\bar{R}_{ab} = 0$, $\hat{g}_{ab}$ satisfies the vacuum Einstein equation linearized about $\bar{g}_{ab}$. 
    
    \item The differential operator is self-adjoint.

    \item The trace of \eqref{EinsteinLin} can be solved independently of the full equation.
\end{enumerate}
The first condition guarantees that the effective metric is reasonable at least in first order perturbation theory. Self-adjointness is useful because it allows the reciprocity condition \eqref{GreenSymGR} to be enforced. Finally, it is important for technical reasons to know the trace $\hat{g}^{ab} G_{aba'b'}$ of $G_{aba'b'}$. The form of \eqref{EinsteinLin} may be used to show that this satisfies 
\begin{equation}
    \hat{g}^{ab} G_{aba'b'} = \mathcal{G} \hat{g}_{a'b'},    
\end{equation}
where $\mathcal{G}$ is an S-type Detweiler-Whiting Green function for the nonminimally-coupled scalar equation
\begin{equation}
    \big( \hat{g}^{ab} \hat{\nabla}_a \hat{\nabla}_b  + \frac{1}{2} \hat{R} \big) \mathcal{G} = 16 \pi \hat{\delta}(x,x').
\end{equation}

\subsubsection{Linear and angular momentum}


Using the Green function associated with \eqref{EinsteinLin} to construct $\hat{g}_{ab}$ and $\hat{P}_s$, linear and angular momenta may be extracted in the usual way. For any $\xi^a \in K_G(\mathcal{Z}, \{ \fB_s \} ; \hat{g} )$ and any $z_s \in \mathcal{Z}$, let
\begin{equation}
    \hat{P}_s ( \xi ) = \hat{p}_a (z_s,s) \xi^a (z_s) + \frac{1}{2} \hat{S}^{a}{}_{b} \hat{\nabla}_a \xi^b (z_s).
    \label{pTopsGR}
\end{equation}
The angular momentum defined in this way is antisymmetric in the sense the $\hat{S}^{(a}{}_{c} \hat{g}^{b)c} = 0$. Differentiating \eqref{pTopsGR} using a covariant derivative associated with $\hat{g}_{ab}$ shows that
\begin{align}
    \frac{ \hat{\rmD} \hat{p}_a }{ \rmd s} = - \frac{1}{2} \hat{R}_{abc}{}^{d} \dot{z}^b_s \hat{S}^{c}{}_{d} + \hat{F}_a, \quad \frac{\hat{\rmD} \hat{S}^{a}{}_{b} }{ \rmd s } = (\hat{g}^{ac} \hat{g}_{bd} - \delta^a_d \delta^c_b) \hat{p}_c \dot{z}^d_s + \hat{N}^{a}{}_{b}, 
    \label{pDotGR3}
\end{align}
where
\begin{align}
    \hat{F}_a = \frac{1}{2} \sum_{n=2}^\infty \frac{1}{n!} \hat{I}^{d_1 \cdots d_n bc} \hat{\nabla}_a \hat{g}_{bc, d_1 \cdots d_n},
\end{align}
and
\begin{align}
    \hat{N}^{a}{}_{c} \hat{g}^{bc} = \sum_{n=2}^\infty \frac{2}{n!} \hat{g}^{f[b} \big( \hat{I}^{|c_1 \cdots c_n|a]d} \hat{g}_{df,c_1 \cdots c_n} + \frac{n}{2} \hat{I}^{a] c_1 \cdots c_{n-1} dh} \hat{g}_{dh,c_1 \cdots c_{n-1} f}  \big) .
    \label{TorqueGR}
\end{align}
$\hat{I}^{c_1 \cdots c_n ab}$ represents the renormalized $2^n$-pole moment of the body's stress-energy tensor. Despite the notation, these are not the same as the moments appearing in the scalar and electromagnetic multipole expansions \eqref{ScalarGenForce} and \eqref{EMmultipole}, which are renormalized differently. In the test body limit where $\hat{g}_{ab} \approx g_{ab}$, \eqref{pDotGR3}-\eqref{TorqueGR} reduce to the multipole expansions derived by Dixon \cite{Dix74}. More generally, they show --- if the multipole expansion is valid --- that with appropriate renormalizations, a self-gravitating body moves instantaneously as though it were a test body in the effective metric $\hat{g}_{ab}$. Note in particular that the derivatives of the momenta which appear in the evolution equations are derivatives associated with $\hat{g}_{ab}$, not $g_{ab}$. This is a consequence of choosing the generalized Killing fields to be constructed using $\hat{g}_{ab}$ instead of $g_{ab}$. 

\subsubsection{Center of mass}

A center of mass frame may be defined by choosing an appropriate foliation together with the worldline $\{ \gamma_s \}$ which guarantees that $\hat{p}_a \hat{S}^{a}{}_{b} = 0$ when $z_s = \gamma_s$. A body's linear momentum is then related to its center of mass velocity via an appropriately ``hatted'' version of \eqref{pCM}. 

\subsubsection{A simple case}

If $\rmd \hat{P}_s/ \rmd s$ is sufficiently small, a body's quadrupole and higher multipole moments might be neglected. In these cases, it follows from \eqref{pDotGR3} that the motion is described by the Mathisson-Papapetrou equations in the effective metric:
\begin{equation}
    \frac{ \hat{\rmD} \hat{p}_a }{ \rmd s} = - \frac{1}{2} \hat{R}_{abc}{}^{d} \dot{z}^b_s \hat{S}^{c}{}_{d} , \qquad \frac{\hat{\rmD} \hat{S}^{a}{}_{b} }{ \rmd s } = (\hat{g}^{ac} \hat{g}_{bd} - \delta^a_d \delta^c_b) \hat{p}_c \dot{z}^d_s .
\end{equation}
Choosing $z_s = \gamma_s$, the squared spin magnitude $\hat{S}^{a}{}_{b} \hat{S}^{b}{}_{a}$ is necessarily conserved. It is therefore consistent to once again consider systems with vanishing spin. Assuming that $\hat{S}^{a}{}_{b} = 0$,
\begin{equation}
    \frac{ \hat{\rmD} }{ \rmd s } \dot{\gamma}_s^a = 0.
    \label{Geodesic}
\end{equation}
Non-spinning masses whose quadrupole and higher interactions may be neglected therefore fall on geodesics associated with $\hat{g}_{ab}$. This generalizes the well-known result that small test bodies in general relativity fall on geodesics associated with the background spacetime. 

Moving beyond the test body limit, the difficult step is to compute $\hat{g}_{ab}$. What is typically referred to as the first order gravitational self-force may nevertheless be derived by considering an appropriate family of successively-smaller extended masses. To lowest nontrivial order, the effective metric looks like the metric of a point particle moving on an appropriate vacuum background $\bar{g}_{ab}$ which is close to $\hat{g}_{ab}$. Using overbars to denote quantities associated with $\bar{g}_{ab}$ and assuming retarded boundary conditions,
\begin{equation}
    \hat{g}_{ab} = \bar{g}_{ab} + \frac{1}{2} \hat{m} \int ( \bar{G}^-_{aba'b'} - \bar{G}^+_{aba'b'} + \bar{V}_{aba'b'} ) \dot{\gamma}^{a'}_s \dot{\gamma}^{b'}_s \rmd s'.
\end{equation}
This is well-behaved even on the body's worldline. So is the connection associated with it, which may be computed using, e.g., methods described in \cite{PoissonRev}. Substituting the result into \eqref{Geodesic} recovers the MiSaTaQuWa equation commonly used to describe the first order gravitational self-force \cite{HarteGR}. Comparisons have not yet been made with second order calculations of the gravitational self-force which have recently been completed using other methods \cite{Pound2ndOrder, Gralla2ndOrder}.

\subsubsection{Future directions}

The formulation of the gravitational problem of motion remains somewhat unsatisfactory. Most importantly, the effective metric which has been adopted here (and in \cite{HarteGR}) is somewhat ad hoc. It is inspired by Lorenz-gauge perturbation theory, but this has no particular significance other than being one way to guarantee hyperbolic field equations. More seriously, the $\hat{g}_{ab}$ defined here does not satisfy the vacuum Einstein equation except in certain limiting cases. This seems unnatural. It would be preferable if the general relativistic laws of motion were completely identical in structure to the laws satisfied by test bodies moving in vacuum backgrounds. A condition like $\hat{R}_{ab} = 0$ would also suggest, at least intuitively, that the associated $\hat{g}_{ab}$ might vary slowly in a wide variety of physical systems. The importance of slow variation to the application of multipole expansions makes it extremely interesting to search for an effective metric which admits a multipole expansion like \eqref{pDotGRMultipole} while also being an exact solution to the vacuum Einstein equation. Although such a metric\footnote{It could also be interesting to consider reformulations where an effective connection is sought instead of an effective metric.} has not yet been found, there are several promising routes by which progress might be made. 

The simplest conceivable modifications of the formalism described here retains the bare momentum \eqref{PDefBareGR} while altering the effective metric $\hat{g}_{ab}$. It is trivial to accomplish this by, for example, modifying the differential equation satisfied by $G_{aba'b'}$ or by introducing $n$-point propagators similar to those in \eqref{PhiSGen}. It can also be useful to alter the functional relation \eqref{gHatDef} between $g_{ab}$, $\hat{g}_{ab}$, and any integrals which may be present. Despite being very simple analytically, relating two metrics to one another via the addition of a second-rank tensor is geometrically rather awkward.

Better-motivated functional relationships between the physical and effective metrics may be more convenient. Geometrically, perhaps the simplest conceivable map between two metrics is a conformal transformation. If $g_{ab} = \Omega^2 \hat{g}_{ab}$ for an appropriate $\Omega$, it is straightforward\footnote{Analyzing the effect of a conformal factor on the laws of motion is similar to considering objects coupled to a particular type of nonlinear scalar field. Despite the nonlinearity, such systems can be understood exactly using only minimal adaptations of the formalism used to analyze the (linear) Klein-Gordon problem.} to obtain an effective metric which exactly satisfies the trace $\hat{R} = 0$ of the vacuum Einstein equation. This is not enough, however. More degrees of freedom are necessary. It may be possible to go further by combining an appropriate conformal factor with a ``generalized Kerr-Schild transformation'' so that
\begin{equation}
    g_{ab} = \Omega^2 ( \hat{g}_{ab} + \ell_{(a} k_{b)} )
    \label{KerrSchild}
\end{equation}
for some 1-forms $\ell_a$ and $k_a$ which are null with respect to $\hat{g}_{ab}$ (and therefore null with respect to $g_{ab}$ as well). Despite the simplicity of this expansion, there is strong evidence that it is very general: Given any analytic $g_{ab}$, $(\Omega, \ell_a, k_a)$ triplets can always be chosen, at least locally, which guarantee that $\hat{g}_{ab}$ is flat \cite{DefThm}. Although it is not known how such choices interact with the laws of motion, the possibility of a flat (or conformally flat) effective metric is intriguing. Among other benefits, it might eliminate the need for generalized Killing vectors\footnote{Quasi-local momenta have recently been proposed in general relativity which use isometric embeddings to lift flat Killing fields into arbitrary spacetimes \cite{Yau1,Yau2}. See also \cite{RQF} for a proposal which allows conformal Killing vectors to be introduced in geometries without symmetries.}.

Combinations of these observations can perhaps be applied to provide one (implicit) map $g_{ab} \rightarrow \hat{g}_{ab}$ with the desired properties. It is likely simpler, however, to instead use them to construct a continuous flow of metrics ${}_\lambda \tilde{g}_{ab}$ which smoothly deforms $g_{ab} = {}_0 \tilde{g}_{ab}$ into an appropriate $\hat{g}_{ab} = {}_{\infty} \tilde{g}_{ab}$. The $\lambda$ parameter here is not necessarily physical, but might be interpreted roughly as the reciprocal of the influence of a body's internal scales. Flows like these have the advantage that individual ``steps'' $\lambda \rightarrow \lambda + \rmd \lambda$ can be viewed as (easily-controlled) linear perturbations. Indeed, it is straightforward to impose differential relations on the $\lambda$-dependence of ${}_\lambda \tilde{g}_{ab}$ which ensure that a flow removes any initial stress-energy as $\lambda \rightarrow \infty$. This requires using a 1-parameter family of Green functions ${}_\lambda G_{aba'b'}$ associated with Einstein's equation linearized about each ${}_\lambda \tilde{g}_{ab}$. Separately, it is also straightforward to construct flows which lead to well-behaved laws of motion. What is more difficult is to find a flow which accomplishes both of these tasks simultaneously. If this were found, varying $\lambda$ would likely vary an object's effective metric, its effective momentum, and its effective multipole moments. While only the $\lambda \rightarrow \infty$ limit might be physical, such variations are highly reminiscent of the running couplings which arise in renormalization group flows. 

Regardless, a great deal of freedom clearly exists and may be exploited to better understand the problem of motion in general relativity. The resulting insights may also shed new light on nonlinear problems more generally.

\section{Discussion}  

The techniques described in this review provide a unified and largely non-perturbative formalism with which to understand how objects move in classical field theories. Although these techniques have thus far been applied only to a handful of specific theories --- Newtonian gravity, Klein-Gordon theory, electromagnetism, and general relativity --- they are easily generalized.

One of the central concepts employed here is what we have called the ``generalized momentum.'' This is used as a convenient observable with which to describe an object's motion in the large, and represents a body's momentum not as a tensor either in the interior of the spacetime or at infinity, but instead as a linear map over a more abstract vector space. This automatically takes into account the nonlocality inherent in the momentum concept and also makes explicit how particular components of the momentum can be ``conjugate to,'' e.g., symmetry-generating vector fields. 

Another important property of the generalized momentum is that it unifies a body's linear and angular momenta into a single object. Given a generalized momentum, linear and angular components can easily be extracted. This process depends, however, on extra information, namely a choice of ``observer'' in the sense of a preferred origin. This origin is arbitrary. It affects the angular momentum even in elementary discussions of Newtonian mechanics, but more generally influences an object's linear momentum as well. This has an interesting physical consequence: \textit{Mathisson-Papapetrou terms arise in the evolution equations governing a body's linear and angular momenta due to the motion of the origin used to extract these components of the generalized momentum}. Mathisson-Papapetrou effects are therefore kinematic in nature, arising from the changing ``character'' of each generalized (or genuine) Killing field at different points.

Once a generalized momentum has been defined as a particular linear map on a particular vector space, stress-energy conservation may be used to derive its rate of change. The resulting generalized force is another linear map on the same vector space. Letting $\xi^a$ be a particular element of that space, generalized forces typically have the form
\begin{equation}
  F = \int_{\fB_s} \!\! \rho \Lie_\xi \phi \rmd S,
  \label{GenericInt}
\end{equation}
where $\phi$ represents some (not necessarily scalar) long-range field and $\rho$ its source. $\fB_s$ is an appropriate 3-volume and $\rmd S$ an associated volume element. Integrals like these can be difficult to evaluate directly, so it is important to seek approximations in practical problems. The simplest such approximations involve some combination of test body and smallness conditions which guarantee that $\Lie_\xi \phi$ ``varies slowly'' throughout $\fB_s$. A multipole expansion can then be performed to express $F$ in terms of $\phi$ and its derivatives as well as the multipole moments of $\rho$ computed on a (largely arbitrary) worldline $\{ z_s \}$: $F = q(z_s,s) \Lie_\xi \phi(z_s) + q^a(z_s,s) \Lie_\xi \nabla_a \phi(z_s) + \ldots$ .

It is far more difficult to obtain useful multipole expansions when an object's self-field can no longer be ignored. The potentially-complicated nature of $\rho$ is then inherited by $\phi$ via the field equation, and there is typically no sense in which $\Lie_\xi \phi$ can be approximated using Taylor-like expansions inside $\fB_s$. Coping with this is perhaps the main theoretical problem associated with self-interaction in the classical theory of motion.

It is dealt with here by considering methods which alter the integrand in \eqref{GenericInt} without affecting the integral as a whole (or which alter the integral only via terms which can be interpreted as renormalizations). Particularly useful for this purpose are nonlocal deformations $\phi \rightarrow \hat{\phi}$ generated by appropriate classes of propagators. Although the cases discussed here have used 2-point Green functions associated with the physical field equations, other types of propagators can be more useful in other contexts. Regardless, the large variety of possible deformations may be tailored to optimally simplify whichever problem is at hand.

In particular, it is often possible to find a deformation $\phi \rightarrow \hat{\phi}$ such that $\Lie_\xi \hat{\phi}$ varies slowly even when $\Lie_\xi \phi$ does not. Appropriately-modified multipole expansions can then be applied much more generally than might have been expected. This leads to the main physical principle which dictates motion in each of the theories we have considered: \textit{Laws governing the motion of self-interacting masses are structurally identical to laws governing the motion of test bodies}. The fields appearing in these laws are nontrivial, however. Objects generally act as though they were accelerated not by the physical field (i.e., $\phi$), but instead by the ``effective field'' $\hat{\phi}$.

The use of effective fields to understand problems of motion is not new per se. Standard formulations of Newtonian celestial mechanics heavily rely, for example, on external gravitational potentials which are distinct from the physical potentials. What has been  been stressed here is that generalizing the external field concept (where the ``external'' label is replaced by the more appropriate ``effective'') is similarly essential for a simple understanding of motion even in highly non-Newtonian regimes. All classical results on the self-force can easily be recovered, for example, once the appropriately-formulated laws of motion have been derived. Even point particle limits of these laws are well-defined precisely as stated; they require no independent postulates or regularizations.

The standard deformation $\phi \rightarrow \hat{\phi}$ of the physical Newtonian gravitational potential into its effective counterpart leaves forces and torques completely unaffected: The Newtonian self-force and self-torque both vanish. Other theories are not so simple. Writing generalized forces in terms of effective fields generally requires the introduction of compensating counterterms. It is only when these counterterms have a particularly simple form that the associated effective field is likely to be useful. Indeed, we have considered systems where these terms act only to make a body's momenta or other multipole moments appear to be shifted from those moments which might have been deduced using knowledge of a body's internal structure. The details of this structure are rarely known in practice, in which case it is natural to ``remove'' residual forces and torques by appropriately redefining an object's momenta or other multipole moments. These are renormalizations. They affect generic extended objects, and are always finite in this context. Considerable effort has been devoted here to identifying renormalizations and interpreting them physically. 

The resulting techniques have shown that a large variety of renormalizations are possible even in simple theories. The effective 4-momentum of an electric charge may differ from its bare momentum not only in length (i.e., mass), but also in direction. Spins and center of mass positions can be renormalized as well. Adding the additional complication of a curved spacetime, even the quadrupole and higher multipole moments associated with a body's stress-energy tensor may be dynamically shifted via the forces exerted by its self-field.

Two general mechanisms have been shown responsible for these effects. Both of these are associated with generalizations of --- or failures to generalize --- Newton's third law. One mechanism relates from a direct violation of this law, while the other arises from an inability to fully take advantage of ``action-reaction cancellations.'' The second of these is simpler and affects a body's linear and angular momenta. It is associated with self-fields which are nonlocal in time, in which case forces are sourced in four dimensions but act on matter only in three-dimensional slices. If the propagators associated with these statements satisfy certain minimal constraints, the inability to construct action-reaction pairs in this context conspires to dynamically shift an object's momenta. Such effects are essentially universal in relativistic theories, but can also be relevant for some non-relativistic systems. 

The second renormalization mechanism discussed here stems from more direct violations of Newton's third law. Mathematically, it is related to the behavior of the relevant propagators under Lie dragging. If, say, a self-field is defined in terms of a propagator $G$, and $\Lie_\xi G$ depends only on $\Lie_\xi \phi$ for some field $\phi$, the multipole moments coupling to $\phi$ are renormalized by the self-force. In the cases considered here, $\phi$ was the metric and the relevant moments were those associated with a body's stress-energy tensor. The same mechanism applied to, e.g., a nonlinear scalar theory would instead renormalize a body's charge moments.

Despite the generality of these results, much remains to be learned. Besides the various technical details which remain open --- some of which have been mentioned in the text --- it would also be interesting to understand how the techniques developed here can be applied in new ways. It may be possible, for example, to adapt these techniques to systems where long-range fields couple to an object's surface instead of its volume. Such problems arise when considering the motion of solid objects through fluids, among other cases. More generally, it might be possible to investigate problems which are not related to motion at all. Quantities similar to \eqref{GenericInt} occur in many fields of physics and mathematics, as do various types of regularizations and renormalizations.
It appears likely that the methods developed here can be applied to better understand at least some of these systems. Such speculations have only just begun to be explored.

\bibliographystyle{unsrt}

\bibliography{Harte_eom_proceedings_2013}

\end{document}